\PassOptionsToPackage{prologue,dvipsnames}{xcolor}
\documentclass{article}
\usepackage{ulem}
\usepackage{amsmath}
\usepackage{amssymb}
\usepackage{geometry}
\counterwithin{figure}{section}
\counterwithin{table}{section}
\usepackage{subfigure}
\usepackage{amsfonts,amsthm,bm}
\newtheorem{theorem}{Theorem}
\usepackage{mathrsfs}
\usepackage[mathscr]{euscript}
\usepackage{BOONDOX-cal}
\usepackage{tikz}
\usetikzlibrary{positioning} 
\usepackage[dvipsnames]{xcolor} 
\usepackage{multirow}
\definecolor{mint}{RGB}{102,249,207}
\definecolor{Lead orchid}{RGB}{181,215,227}
\definecolor{moon}{RGB}{255,255,191}
\definecolor{DM}{RGB}{243,83,58}
\definecolor{BH}{RGB}{250,159,66}
\definecolor{RF}{RGB}{138,216,121}
\definecolor{TN}{RGB}{90,207,201}
\definecolor{FT}{RGB}{0,150,0}	
\usepackage{wallpaper}
\usepackage{extarrows}
\allowdisplaybreaks[4]
\numberwithin{equation}{section}
\labelformat{section}{{\color{red}{Section #1}}}
\labelformat{subsection}{{\color{red}{Section #1}}}
\labelformat{subsubsection}{{\color{red}{Section #1}}}
\labelformat{equation}{{\color{blue}{Equation #1}}}
\labelformat{figure}{{\color{FT}{Figure #1}}}
\labelformat{table}{{\color{FT}{Table #1}}}
\title{A New Method for Time Domain Displacement Response due to Concentrated Vertical Force on Free Surface of Elastic Half-space}
\author{$\text{Tian Lin}^{\#}$ and $\text{Hengshan Hu}^{*}$}
\date{}
\begin{document}
	\maketitle
	\begin{itemize}
		\item[$\#$] Department of Aerospace Science and Mechanics, Harbin Institute of Technology, 150001, Harbin, China. E-mail: 21b918011@stu.hit.edu.cn
		\item[*] Corresponding author. Department of Aerospace Science and Mechanics, Harbin Institute of Technology, 150001, Harbin, China. E-mail: hhs@hit.edu.cn
	\end{itemize}
	\begin{abstract}		
		Since 1904, numerous researchers, including Cagniard, Pinney, Pekeris, Chao, de Hoop, Johnson, Kausel, Feng and Zhang have contributed to the Lamb's problem, developing the Cagniard-de Hoop method for obtaining solutions to Lamb's problem and simplifying the solutions. Focusing on the problems of the solutions acquired by the Cagniard-de Hoop method: limitations of Poisson's ratio in existing simplified solutions, difficulty in decomposing the complex expressions of whole wave-field, this article introduces a novel metod to handle these problems. This method is named as Huygens method by us, which was inspired by Huygens–Fresnel principle. The problem where the Huygens method is applied is a special case of the Lamb's problem about displacement in time domain due to a vertical force on the free surface. In this article, we expand the Green's function to an equivalent form, which is named as Time-Green function by us. The displacements we provide in this new form are without singularity of existing works and hold for arbitrary Poisson's ratio. Based on the results acquired by the Huygens method, we decompose the wave-field of displacements into three component-waves with different generation mechanisms, direct wave, surface wave, Huygens wave, and analyze the temporal and spatial properties of each component-wave. The direct wave shares the same form with the solutions in full-space and includes two kinds of body waves, the longitudinal and transverse wave. The surface wave contains the Rayleigh wave and a wave group, while a unified expression for the velocity of the Rayleigh wave is provided, which solves the limitations of Poisson's ratio and is more concise than the existing segmented expression. The Huygens wave is a new wave we difined from the Huygens method. It is genenrated by the Huygens effect that the surface wave generate the direct wave. Therefore, in this problem, there are four kinds of Huygens waves, H-R-L, H-R-T, H-G-L, H-G-T, where R, G, L and T mean Rayleigh wave, wave group, longitudinal wave and transverse wave separately. In the transverse wave generated by the wave group (H-G-T), an aspherical wavefront is discovered. In the horizontal plane with a fixed depth, there existing a circle with radius linearly depending on the depth, the H-G-T propagates with a fixed velocity of the longitudinal wave out of the circle. This phenomenon is named as {\it variation} in this article. In the H-R-L component of the axial displacement $u_z$ underground, except the head wave, another waveform is found following the head wave, which is named as {\it tail flame}. Tail flame is a visual effect caused by plural mutagenic points with long enough distances of Green's functions, which is look like a different wave but shares the same mechanism with head wave. This visual effect gives H-R-L a two-stage attenuation, that the head wave attenuates with the increase of depth slowly and the tail flame attenuates slowly with the increase of diameter. This good detection property is not found in the other component-waves, for they are always sensitive to the depth or the diameter. Furthermore, we provide the approximations of the displacements near the axis, which are all in the form of elementary functions and reduce the algorithmic complexity to $O(1)$ or $O(n)$.\\
		\\
		\noindent{\textbf{Key word: }Lamb's problem; Green's function; Analytical solutions; Rayleigh waves; Acoustics; Seismology; Mechanics of solid}	
	\end{abstract}
	\thispagestyle{empty}
	\newpage
	\setcounter{page}{1}
	\section{Introduction}
	Lamb's problem (1904, \cite{1954Propagation}) delves into the elastic displacements that arise from a point force applied to a half-space. Cagniard's sophisticated approach (1939, \cite{cagniard1939reflexion}), utilizing Laplace transforms with respect to time, yielded solutions within the time domain. Pinney (1954, \cite{pinney1954surface}) derived results expressed through finite integrals on surfaces with underground sources. Pekeris (1955a, \cite{pekeris1955seismica}; 1955b, \cite{pekeris1955seismicb}) provided a closed-form solution for a vertical point source on the surface with a Poisson’s ratio of 0.25, as well as exact integral solutions for the displacements generated by a point pressure impulse underground. Pekeris and Lifson (1957, \cite{pekeris1957motion}) provided exact integral solutions for the displacements resulting from a vertical force. Chao (1960, \cite{chao1960dynamical}) offered a closed-form solution for a tangential point force on the surface with a Poisson’s ratio of 0.25. The approach underwent subsequent modification by de Hoop (1960, \cite{de1960modification}), often termed the Cagniard-de Hoop method. Kawasaki, Suzuki, and Sato (1972a, \cite{1972Seismica}; 1972b, \cite{1972Seismicb}) along with Sato (1972, \cite{sato1972seismic}) have addressed the surface displacements arising from a double-couple source in a half-space. Johnson (1974, \cite{johnson1974green}) synthesized the contributions of Cagniard (1939, \cite{cagniard1939reflexion}), Dix (1954, \cite{dix1954method}), Pinney (1954, \cite{pinney1954surface}), Pekeris (1955a, \cite{pekeris1955seismica}; 1955b, \cite{pekeris1955seismicb}), Pekeris and Lifson (1957, \cite{pekeris1957motion}), de Hoop (1961, \cite{de1961theoretical}), and Aggarwal and Ablow (1967, \cite{aggarwal1967solution}), presenting displacement solutions in a uniform notation. Mooney (1974, \cite{mooney1974some}) extended the studies of Pekeris (1955a, \cite{pekeris1955seismica}) and Chao (1960, \cite{chao1960dynamical}) to encompass an arbitrary Poisson’s ratio, albeit excluding the radial component. Richards (1979, \cite{richards1979elementary}) derived a comprehensive set of formulas for investigating spontaneous crack propagation, although no proofs or derivations were provided. Kausel (2013, \cite{kausel2013lamb}) expanded upon Richards's work, offering further details and discussing special conditions for surface displacements caused by a surface point load. Building on Johnson's groundwork, Feng and Zhang conducted significant mathematical transformations, simplifying Johnson's results into elliptic functions for a wide range of Poisson's ratios ($0\sim0.2631$). Their contributions include the exact closed-form solution for displacement at the surface of an elastic half-space induced by a buried point source (2018, \cite{feng2018exact}), and the exact closed-form solution for displacement within the interior of an elastic half-space due to a buried point force with a Heaviside step function time history (2021, \cite{feng2021exact}). Pan (2019, \cite{pan2019green}) collectted the works in half-plane by forces and dislocations in his review. Currently, these works on exact solutions for Lamb's problem primarily rely on the Cagniard-de Hoop method.\\
	\\
	The solutions obtained through the Cagniard-de Hoop method present three primary challenges: limitations imposed by the range of Poisson's ratios, the absence of a unified expression for the velocity of Rayleigh waves, and complexities in decomposition, often involving complex singularities. This article introduces a novel approach, the Huygens method, aimed at addressing these challenges comprehensively. The core focus of this article lies in providing analytical displacement solutions for a half-space subjected to a vertical concentrated force on the free surface, employing the Huygens method instead of the conventional Cagniard-de Hoop method. This problem, while a subset of Lamb's problem, is investigated as a central aspect, with its solutions serving as fundamental components for constructing solutions to the broader Lamb's problem within the framework of the Huygens method.\\
	\\
	The starting point shared by both the Huygens method and the Cagniard-de Hoop method is the solutions in the transform domain, as \ref{2} of this article, reminiscent of the relevant content in Johnson's seminal work (1974, \cite{johnson1974green}). The Huygens method, unlike the Cagniard-de Hoop method, imposes no constraints on the medium's properties (such as isotropy) and is applicable in linear scenarios. It is devised as a tool to decompose the entire wave into component-waves with distinct generation mechanisms. The outcomes of the Huygens method manifest in a novel form derived from and equivalent to the Green's function (referred to as Time-Green function herein). Several noteworthy results presented in this article may capture the interest of readers, including a unified algebraic formula for Rayleigh wave velocity, solutions devoid of singularities, clear separation and generation mechanisms of waves, and approximations employing elementary functions to describe displacements near the axis.\\
	\\
	To enhance readability, here is a concise preview of the upcoming sections. In \ref{2}, we provide mathematical descriptions of the problem and solutions in the transform domain. Diverging from conventional approaches that treat the concentrated force as a non-homogeneous term of body force, our work regards the source as a boundary condition, offering equivalence between these two descriptions. \ref{3} delves into the detailed steps of the Huygens method, presenting expressions for displacements and numerical verification of the results obtained using this method. Lastly, \ref{4} elucidates the generation mechanisms of each component-wave, alongside discussions on their temporal and spatial properties. Additionally, we explore approximations for displacements near the axis.\\
	\section{Displacements in frequency and wave-number domain}\label{2}
	This section delineates the half-space problem by constructing the mathematical model (\ref{2.1}), and succinctly outlines the procedure for deriving displacements in both the frequency and wave-number domains (\ref{2.2}). The outcomes presented herein serve as the foundational premise for both the Huygens method and the Cagniard-de Hoop method employed in this paper.\\
	\subsection{Problem discription}\label{2.1}
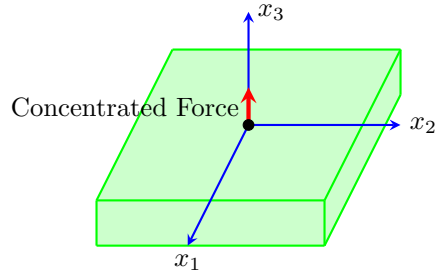
\begin{figure}[h!]
		\centering
		\begin{tikzpicture}[scale = 2]
			\fill[green!20] (-1,-0.5)--(0.5,-0.5)--(0.5,-0.8)--(-1,-0.8);
			\fill[green!20] (-1,-0.5)--(0.5,-0.5)--(1,0.5)--(-0.5,0.5);
			\fill[green!20] (0.5,-0.5)--(1,0.5)--(1,0.2)--(0.5,-0.8);
			\draw[green,thick] (-1,-0.5)--(0.5,-0.5);
			\draw[green,thick] (-1,-0.5)--(-0.5,0.5);
			\draw[green,thick] (-1,-0.5)--(-1,-0.8);
			\draw[green,thick] (-1,-0.8)--(0.5,-0.8);
			\draw[green,thick] (0.5,-0.8)--(0.5,-0.5);
			\draw[green,thick] (0.5,-0.8)--(1,0.2);
			\draw[green,thick] (1,0.2)--(1,0.5);
			\draw[green,thick] (-0.5,0.5)--(1,0.5);
			\draw[green,thick] (0.5,-0.5)--(1,0.5);
			\draw[blue,-stealth,thick] (0,0)--(1,0);
			\draw[blue,-stealth,thick] (0,0)--(-0.4,-0.8);
			\draw[blue,-stealth,thick] (0,0)--(0,0.75);
			\draw[red,-stealth,ultra thick] (0,0)--(0,0.25);
			\filldraw[fill=black] (0,0) circle (1pt);
			\node [above left] at (0,0) {Concentrated Force};
			\node [below] at (-0.4,-0.8) {$x_1$};
			\node [right] at (1,0) {$x_2$};
			\node [right] at (0,0.75) {$x_3$};
		\end{tikzpicture}
		\caption{Concentrated force in vertical direction acts on free surface in half-space.}
		\label{13th-Sept.-2023-P001}
	\end{figure}

	The subject of investigation is an isotropic elastic half-space with a free surface subjected to a vertical concentrated force. The elastic medium obeys Hooke's law,\\
	\begin{equation}
		\tau_{ij}=\lambda\delta_{ij}+\mu\left(\varepsilon_{ij}+\varepsilon_{ji}\right).
		\label{13th-Sept.-2023-E001}
	\end{equation}
	In \ref{13th-Sept.-2023-E001}, $\tau_{ij}$ is the stress tensor, while $\varepsilon_{ij}$ is the strain tensor. $\lambda$ signifies the Lamé constant, and $\mu$ stands for the shear modulus. Hereinafter, all the variables with indices hereafter adhere to the Einstein summation convention.\\
	\\
	In situations involving small deformations, the strain $\varepsilon_{ij}$ satisfying the following relation with the displacements $u_i$.\\
	\begin{equation}
		\varepsilon_{ij}=\frac{1}{2}\left(\frac{\partial u_i}{\partial x_j}+\frac{\partial u_j}{\partial x_i}\right).
		\label{13th-Sept.-2023-E002}
	\end{equation} 
	\ref{13th-Sept.-2023-E001} can be reformulated as follows,\\
	\begin{equation}
		\tau_{ij}=\lambda\delta_{ij}u_{k,k}+\mu\left(u_{i,j}+u_{j,i}\right).
		\label{13th-Sept.-2023-E003}
	\end{equation}
	By combining \ref{13th-Sept.-2023-E003} with the momentum theorem, we can derive the dynamic equations\\
	\begin{equation}
		\mu u_{i,jj}+\left(\lambda+\mu\right)u_{j,ji}+\rho f_i=\rho \ddot{u}_i
		\label{13th-Sept.-2023-E004},
	\end{equation}
	where $\rho$ represents the density of the medium and $f_i$ denotes the accelerations generated by external body forces. It should be noted that $\dot{Q}=Q_{,t}$ and $\ddot{Q}=Q_{,tt}$.\\
	\\
	We establish a Cartesian coordinate system (\ref{13th-Sept.-2023-P001}) with origin coincide with point where the concentrated force acts. \ref{13th-Sept.-2023-E004}, boundary conditions and initial conditions, constitute a partial differential equation problem (hereinafter referred to as a PDE problem).\\
	\\
	The specificity of this problem lies in the placement of the concentrated force on the boundary. Consequently, two sets of PDE problems can describe this physical situation. The first set treats the concentrated force as an external body force, where $f_i=\delta_{i3}\delta(t)\delta(\mathbf{x})$.\\
	\\
	\begin{subequations}
		\begin{align}
			\textbf{SET 1 } & & \notag \\
			& \text{Non-homogeneous equations: } & \mu u_{i,jj}+\left(\lambda+\mu\right)u_{j,ji}+\rho\delta_{i3}\delta(t)\delta(\mathbf{x})=\rho \ddot{u}_i;\\
			& \text{Homogeneous boundary conditions: } & \sqrt{x_1^2+x_2^2+x_3^2}\to\infty,\quad u_i\to0
			\label{13th-Sept.-2023-E005a};\\
			&  & x_3=0,\quad \tau_{i3}=0
			\label{13th-Sept.-2023-E005b};\\
			& \text{Homogeneous initial conditions: } & t=0,\quad u_i=\dot{u}_i=0.
			\label{13th-Sept.-2023-E006}
		\end{align}
	\end{subequations}
	The second set regards the concentrated force as stress on the surface, implying the absence of a free surface but rather a surface with a stress distribution, where $f_i=0$.\\
	\begin{subequations}
		\begin{align}
			\textbf{SET 2 } & & \notag \\
			& \text{Homogeneous equations: } & \mu u_{i,jj}+(\lambda+\mu)u_{j,ji}=\rho \ddot{u_i}
			\label{17th-Jan.-2024-E002a};\\
			& \text{Non-homogeneous boundary conditions: } & \sqrt{x_1^2+x_2^2+x_3^2}\to\infty,\quad u_i\to0\label{17th-Jan.-2024-E002b};\\
			&  & x_3=0,\quad \tau_{i3}=\rho\delta_{i3}\delta(x_1)\delta(x_2)
			\label{13th-Sept.-2023-E009};\\
			& \text{Homogeneous initial conditions: } & t=0,\quad u_i=\dot{u}_i=0.
		\end{align}
	\end{subequations}
	The solutions of \textbf{SET 1} and \textbf{SET 2} exist within a specific interval,\\
	\begin{equation}
		(t,x_1,x_2,x_3)\in\left(0,\infty\right)\times\mathbb{R}^2\times\left(-\infty,0\right).
		\label{13th-Sept.-2023-E007}
	\end{equation}
	These two descriptions are entirely equivalent in terms of their results, yet they convey the notion that the boundary acts as a source and can be interchanged. This concept forms the basis for constructing the complete solutions of Lamb's problem and various stress boundary problems in a half-space, even in situations beyond isotropy. This fundamental distinction is what sets apart the Huygens method from the Cagniard-de Hoop method.\\
	\subsection{Solution to problem}\label{2.2}
	This article selects \textbf{SET 2} to describe the problem. Utilizing the Helmholtz decomposition theorem and considering the axisymmetric condition (the problem exhibits axisymmetry), the displacements in the cylindrical coordinate system are determined as follows,\\
		\begin{subequations}
		\begin{align}
			u_r&=\frac{\partial \varphi}{\partial r}-\frac{\partial \psi}{\partial z},\label{18th.Jan.-2024-E002a}\\
			u_z&=\frac{\partial \varphi}{\partial z}+\frac{\partial\psi}{\partial r}+\frac{\psi}{r},\label{18th.Jan.-2024-E002b}\\
			u_{\theta}&=0.
		\end{align}
	\label{2nd-Apr.-2024-E001}
	\end{subequations}
	The stress on the free surface can be expressed as follows by \ref{13th-Sept.-2023-E003},\\
	\begin{subequations}
		\begin{align}		
			\tau_{zz}&=\lambda\left(\frac{\partial^2\varphi}{\partial r^2}+\frac{1}{r}\frac{\partial\varphi}{\partial r}+\frac{\partial^2\varphi}{\partial z^2}\right)+2\mu\left(\frac{\partial^2\varphi}{\partial z^2}+\frac{\partial^2\psi}{\partial r\partial z}+\frac{1}{r}\frac{\partial\psi}{\partial z}\right),\label{18th.-Jan.-2024-E001a}\\
			\tau_{rz}&=\mu\left(2\frac{\partial^2\varphi}{\partial r\partial z}+\frac{\partial^2\psi}{\partial r^2}+\frac{1}{r}\frac{\partial\psi}{\partial r}-\frac{\psi}{r^2}-\frac{\partial^2\psi}{\partial z^2}\right).\label{18th.-Jan.-2024-E001b}
		\end{align}
	\end{subequations}
	With this decomposition, \ref{17th-Jan.-2024-E002a} can be transformed into the wave equations concerning displacement potential functions, $\varphi$ and $\psi$. The similar work can be found in Achenbach's book (2012, \cite{achenbach2012wave}).\\
	\begin{subequations}
		\begin{align}
			(\frac{\partial^2}{\partial r^2}+\frac{1}{r}\frac{\partial}{\partial r}+\frac{\partial^2}{\partial z^2})\varphi&=\frac{1}{c_L^2}\frac{\partial^2\varphi}{\partial t^2},\label{17th-Jan.-2024-E001a}\\
			(\frac{\partial^2}{\partial r^2}+\frac{1}{r}\frac{\partial}{\partial r}+\frac{\partial^2}{\partial z^2}-\frac{1}{r^2})\psi&=\frac{1}{c_T^2}\frac{\partial^2\psi}{\partial t^2}\label{17th-Jan.-2024-E001b}.
		\end{align}
	\end{subequations}
	After performing Laplace transformation on \ref{17th-Jan.-2024-E001a} and \ref{17th-Jan.-2024-E001b}, and subsequently employing the variable separation method, we derive the general solutions in the frequency and wave-number domain,\\
		\begin{subequations}
		\begin{align}
			\Phi(s,r,z)&=\left(A_{\Phi}(s,k){\rm J}_0(kr)+B_{\Phi}(s,k){\rm N}_0(kr)\right)\left(C_{\Phi}(s,k)\exp\left(z\sqrt{\frac{s^2}{c_L^2}+k^2}\right)+D_{\Phi}(s,k)\exp\left(-z\sqrt{\frac{s^2}{c_L^2}+k^2}\right)\right),
			\label{14th-Sept.-2023-E014a}\\
			\Psi(s,r,z)&=\left(A_{\Psi}(s,k){\rm J}_1(kr)+B_{\Psi}(s,k){\rm N}_1(kr)\right)\left(C_{\Psi}(s,k)\exp\left(z\sqrt{\frac{s^2}{c_T^2}+k^2}\right)+D_{\Psi}(s,k)\exp\left(-z\sqrt{\frac{s^2}{c_T^2}+k^2}\right)\right).
			\label{14th-Sept.-2023-E014b}
		\end{align}
	\end{subequations}
	Detailed steps are provided in \ref{A.1}.\\
	\\
	In \ref{14th-Sept.-2023-E014a} and \ref{14th-Sept.-2023-E014b}, $\Phi(s,r,z)$ and $\Psi(s,r,z)$ represent the Laplace transformations of $\varphi$ and $\psi$, respectively. $A$, $B$, $C$, and $D$ are pending coefficients. ${\rm J}_n$ denotes the Bessel function, and ${\rm N}_n$ represents the Neumann function. Additionally, $s$ is the variable of Laplace transformation, $k$ is the parameter used for variable separation method, $r=\sqrt{x_1^2+x_2^2}$ and $z=x_3$ are the space coordinates in the axis coordinate system. $c_L$ is the velocity of longitudinal waves, and $c_T$ is the velocity of transverse waves.\\
	\begin{subequations}
		\begin{align}
			c_L&=\left(\lambda+2\mu\right)/\rho,
			\label{14th-Sept.-2023-E005a}\\
			c_T&=\mu/\rho.
			\label{14th-Sept.-2023-E005b}
		\end{align}
	\end{subequations}
	Due to \ref{17th-Jan.-2024-E002b}, the coefficients of the Neumann function are zero because the Neumann function ${\rm N}_n(kr)$ diverges as $r\to 0$ and $r\to\infty$. Therefore, $\Phi$ and $\Psi$ only retain the Bessel function ${\rm J}_n(kr)$.\\
	\\
	The function $f(s,k;z)=\exp\left(z\sqrt{s^2/c^2+k^2}\right)$ is multivalued, which means we need to select one of its branches for calculation. Regardless of the chosen branch, we must ensure that the imaginary part of $\sqrt{s^2/c^2+k^2}$ remains greater than zero to satisfy \ref{17th-Jan.-2024-E002b}. Let's assume that the branch we select yields an effective solution for $\exp\left(z\sqrt{s^2/c^2+k^2}\right)$. The details of how to analyze the multivalued function and select the appropriate branch are outlined in \ref{A.2}. Subsequently, the general solutions can be simplified accordingly.\\
	\begin{subequations}
		\begin{align}
			\Phi&=A(s,k){\rm J}_0(kr)e^{k_Lz},
			\label{15th-Sept.-2023-E005a}\\
			\Psi&=B(s,k){\rm J}_1(kr)e^{k_Tz},
			\label{15th-Sept.-2023-E005b}
		\end{align}
	\end{subequations}
	where $k_L=\sqrt{s^2/c_L^2+k^2}$ and $k_T=\sqrt{s^2/c_T^2+k^2}$.\\
	\\
	Perform Laplace transformation on \ref{13th-Sept.-2023-E009}, and then utilize the Fourier-Bessel formula,\\
	\begin{equation}
		\delta(r-r_0)=r\int_{0}^{\infty}k{\rm J}_m(kr){\rm J}_m(kr_0){\rm d}k
		\label{15th-Sept.-2023-E009},
	\end{equation}
	to expend $\delta(x_1)\delta(x_2)$.\\
	\begin{equation}
		\delta(x_1)\delta(x_2)=\frac{\delta(r)}{2\pi r}=\frac{1}{2\pi}\int_{0}^{\infty}k{\rm J}_0(kr){\rm d}k \text{(in axisymmetric case)}.
		\label{15th-Sept.-2023-E010}
	\end{equation}
	With \ref{18th.-Jan.-2024-E001a}, \ref{18th.-Jan.-2024-E001b}, \ref{15th-Sept.-2023-E005a}, and \ref{15th-Sept.-2023-E005b}, the boundary conditions on the surface form a system of linear equations involving $A$ and $B$,\\
	\begin{equation}
		\left(\begin{array}{cc}
			\rho s^2+2\mu k^2 & 2\mu kk_T\\
			2\mu kk_L & \rho s^2+2\mu k^2\\
		\end{array}\right)
		\left(\begin{array}{c}
			A\\
			B\\
		\end{array}\right)=
		\left(\begin{array}{c}
			\frac{\rho k}{2\pi}\\
			0\\
		\end{array}\right).
		\label{15th-Sept.-2023-E011}
	\end{equation}
	Solutions of \ref{15th-Sept.-2023-E011} can be obtained using Cramer's rule.\\
	\begin{subequations}
		\begin{align}
			A&=\frac{\rho k}{2\pi}\frac{\rho s^2+2\mu k^2}{(\rho s^2+2\mu k^2)^2-4\mu^2k^2k_Lk_T},
			\label{15th-Sept.-2023-E012a}\\
			B&=-\frac{\rho k}{2\pi}\frac{2\mu kk_L}{(\rho s^2+2\mu k^2)^2-4\mu^2k^2k_Lk_T}.
			\label{15th-Sept.-2023-E012b}
		\end{align}
	\end{subequations}
	The Laplace transformations of the displacement potential functions are represented by the following equations. \\
	\begin{subequations}
		\begin{align}
			\Phi(s;r,z)&=\int_{0}^{\infty}A(s,k){\rm J}_0(kr)e^{k_Lz}{\rm d}k,
			\label{15th-Sept.-2023-E013a}\\
			\Psi(s;r,z)&=\int_{0}^{\infty}B(s,k){\rm J}_1(kr)e^{k_Tz}{\rm d}k.
			\label{15th-Sept.-2023-E013b}
		\end{align}
	\end{subequations}
	Finally, we obtain the displacements in the form of frequency and wave-number integrals.\\
	\begin{subequations}
		\begin{align}
			u_r(t,r,z)&=-\frac{1}{2\pi\mathbf{i}}\int_{\sigma-\mathbf{i}\infty}^{\sigma+\mathbf{i}\infty}\int_0^{\infty}\left(kA(s,k)e^{k_Lz}+k_TB(s,k)e^{k_Tz}\right){\rm J}_1(kr)e^{st}{\rm d}k{\rm d}s
			\label{15th-Sept.-2023-E014a}\\
			u_z(t,r,z)&=\frac{1}{2\pi\mathbf{i}}\int_{\sigma-\mathbf{i}\infty}^{\sigma+\mathbf{i}\infty}\int_0^{\infty}\left(k_LA(s,k)e^{k_Lz}+kB(s,k)e^{k_Tz}\right){\rm J}_0(kr)e^{st}{\rm d}k{\rm d}s
			\label{15th-Sept.-2023-E014b}
		\end{align}
	\end{subequations}
	\section{Huygens method and numerical verification}\label{3}
	In this section, the Huygens method is employed to solve the integrals in the frequency and wave-number domain, and its accuracy is verified through comparison with numerical integration. This section encompasses solutions to the integrals of displacement potential functions in the frequency and wave-number domain and displacements calculated by \ref{2nd-Apr.-2024-E001} (\ref{3.1}), the method for transforming the Green's function to the Time-Green function (\ref{3.2}), and numerical verifications of the results presented in this article (\ref{3.3}).\\
	\\
	The presentation of the Huygens method draws inspiration from the Huygens–Fresnel principle, which posits that every point on a wavefront can be regarded as a secondary source, and the wavefront is the envelope of all these secondary sources. We envision that the three-dimensional wave-field underground is the body wave (the propagation term in \ref{3.1}, abbreviated as the P-term) generated by the response of the low-dimensional boundary to the source (the distribution term in \ref{3.1}, abbreviated as the D-term). This phenomenon is the Huygens effect we called below.\\
	\subsection{Solutions to the integrals}\label{3.1}
	In this section, the process to solve the integrals of $\varphi$ and $\psi$ in the frequency and wave-number domain is divided into two steps: \\
	\begin{itemize}
		\item[1.] The separation of integrals by the convolution property (\ref{3.1.1}).\\
		\item[2.] The computation of D-terms (\ref{3.1.2}) and P-terms (\ref{B.1}).
	\end{itemize}
	Finally, the Green's functions are provided (\ref{3.1.3}).\\
	\subsubsection{Sepration of D-term and P-term}\label{3.1.1}
	We define a symbolic operation known as the three-dimensional convolution,\\
	\begin{equation}
		\left<f(t,x_1,x_2),g(t,x_1,x_2)\right>:=\iint_{\mathbb{R}^2}\int_{0}^{\infty}f(\tau,\xi_1,\xi_2)g(t-\tau,\xi_1-x_1,\xi_2-x_2){\rm d}\tau{\rm d}\xi_1{\rm d}\xi_2.
		\label{16th-Sept.-2023-E001}\\
	\end{equation}
	The displacement potential functions defined by the Helmholtz decomposition to the displacements (\ref{18th.Jan.-2024-E002a} and \ref{18th.Jan.-2024-E002b}) can be expressed in the following form,\\
	\begin{subequations}
		\begin{align}
			\varphi&=\rho\frac{\partial}{\partial z}\left<L,P_L\right>,
			\label{16th-Sept.-2023-E002a}\\
			\psi&=\rho\frac{\partial}{\partial r}\left<T,P_T\right>.
			\label{16th-Sept.-2023-E002b}
		\end{align}
	\end{subequations}
	Here, $L$ and $T$ are the responses of the low-dimensional boundary to the source. $P_L$ and $P_T$ are the body waves generated by that. The detailed process to obtain \ref{16th-Sept.-2023-E002a} and \ref{16th-Sept.-2023-E002b} is outlined in \ref{B.1}.\\
	\\
	In the above equations, we refer to $L$ and $T$ as distribution terms (D-terms),\\
	\begin{subequations}
		\begin{align}
			L=\frac{1}{4\pi^2\mathbf{i}}\int_{\sigma-\mathbf{i}\infty}^{\sigma+\mathbf{i}\infty}\int_{0}^{\infty}\frac{\rho s^2+2\mu k^2}{(\rho s^2+2\mu k^2)^2-4\mu^2k^2k_Lk_T}k{\rm J}_0(kr)e^{st}{\rm d}k{\rm d}s,
			\label{16th-Sept.-2023-E003a}\\
			T=\frac{1}{4\pi^2\mathbf{i}}\int_{\sigma-\mathbf{i}\infty}^{\sigma+\mathbf{i}\infty}\int_{0}^{\infty}\frac{2\mu k_Lk_T}{(\rho s^2+2\mu k^2)^2-4\mu^2k^2k_Lk_T}k{\rm J}_0(kr)e^{st}{\rm d}k{\rm d}s.
			\label{16th-Sept.-2023-E003b}
		\end{align}
	\end{subequations}
	while $P_L$ and $P_T$ are denoted as propagation terms (P-terms),\\
	\begin{subequations}
		\begin{align}
			P_L&=\frac{1}{2\pi R}\delta(t-\frac{R}{c_L}),
			\label{16th-Sept.-2023-E004a}\\
			P_T&=\frac{1}{2\pi R}\delta(t-\frac{R}{c_T}).
			\label{16th-Sept.-2023-E004b}
		\end{align}
	\end{subequations}
	\subsubsection{Computation of D-term}\label{3.1.2}
	In this section, the computation of $L$, as shown in \ref{16th-Sept.-2023-E003a}, serves as an example of how to calculate the D-terms. This computation consists of three parts:
	\begin{itemize}
		\item[1.] Changing the order of the integral of $k$ and $\omega$, where $\omega$ is the variable of the Laplace transform after performing a substitution, $s=\sigma+\mathbf{i}\omega$. 
		\item[2.] Designing the path of the $\omega$-integral.
		\item[3.] Calculating the $k$-integral.
	\end{itemize}
	The validity of the operation of changing integral order and the intricacies of the computation are elaborated in \ref{B.2}.\\
	\\
	\underline{\textbf{Change of integral-order}} The distribution term $L$ can be expressed in the following form,\\
	\begin{equation}
		L=\frac{e^{\sigma t}}{4\rho\pi^2}\int_{0}^{\infty}\int_{-\infty}^{\infty}\left((\sigma+\mathbf{i}\omega)^2+2c_T^2k^2\right)\frac{F_2(\sigma+\mathbf{i}\omega)}{F(\sigma+\mathbf{i}\omega)}e^{\mathbf{i}\omega t}{\rm d}\omega k{\rm J}_0(kr){\rm d}k.\\
		\label{17th-Sept.-2023-E001}
	\end{equation}
	Functions in the above formula are defined below,\\
	\begin{subequations}
		\begin{align}
			F(x)&:=F_1(x)F_2(x)=x^2(x^2+a^2c_T^2k^2)(x^2-b^2c_T^2k^2)(x^2-c^2c_T^2k^2).
			\label{17th-Sept.-2023-E002c}\\
			F_1(x)&:=(x^2+2c_T^2k^2)^2-4c_T^2k^2\sqrt{x^2+c_T^2k^2}\sqrt{\eta x^2+c_T^2k^2},
			\label{17th-Sept.-2023-E002a}\\
			F_2(x)&:=(x^2+2c_T^2k^2)^2+4c_T^2k^2\sqrt{x^2+c_T^2k^2}\sqrt{\eta x^2+c_T^2k^2},
			\label{17th-Sept.-2023-E002b}		
		\end{align}
	\end{subequations} 
	The parameter $\eta$ is the square of the ratio of $c_T$ and $c_L$.\\
	\begin{equation}
		\eta=\frac{c_T^2}{c_L^2}=\frac{\mu}{\lambda+2\mu},\quad\eta\in(0,0.75).\quad\text{(For the Poisson's ratio from -0.5 to 0.5.)}
		\label{17th-Sept.-2023-E003}
	\end{equation}
	The other parameters $a$, $b$, and $c$ pertain to the roots of $F(x)$ and are solely associated with the parameters of the medium. (Exact forms are detailed in \ref{B.2}.)\\
	\\
	\underline{\textbf{Path of $\omega$-integral}} We begin by addressing the inner frequency integral of \ref{17th-Sept.-2023-E001} first. (The detailed process is presented in \ref{B.3}.) The integral in the frequency domain can be regarded as a path integral along the real axis in the complex plane of $\omega$.\\
	\begin{equation}
		\begin{aligned}
			I&:=\int_{-\infty}^{\infty}\left((\sigma+\mathbf{i}\omega)^2+2c_T^2k^2\right)\frac{F_2(\sigma+\mathbf{i}\omega)}{F(\sigma+\mathbf{i}\omega)}e^{\mathbf{i}\omega t}{\rm d}\omega=\int_{R.A.}\left((\sigma+\mathbf{i}\omega)^2+2c_T^2k^2\right)\frac{F_2(\sigma+\mathbf{i}\omega)}{F(\sigma+\mathbf{i}\omega)}e^{\mathbf{i}\omega t}{\rm d}\omega.\\
		\end{aligned}
		\label{19th-Sept.-2023-E001}
	\end{equation}
	Simply denote the integral as $\int_{R.A.}$.\\
	\\
	To calculate the integral, we construct an integral contour $C$ (\ref{19th-Sept.-2023-P001}): the outer contour encompasses the real axis ($R.A.$) and an infinite semicircular arc $A_{\infty}$ enclosing the upper plane; the inner contour comprises four branch-cut paths $L_{\text{Branch-cut}}$ (two branch cuts, each with upper and lower paths) and four arcs surrounding four branch points $A_{\text{Branch-point}}$.\\
	\begin{figure}[h!]
		\begin{center}
			\begin{tikzpicture}
				\draw  [->](-6,0)--(6,0);
				\node[below left] at (6,0) {${\rm Re}[\omega]$};
				\draw  [->](0,-1)--(0,4);
				\node[below left] at (0,4) {${\rm Im}[\omega]$};
				\draw[blue,dashed,ultra thick,->] (-5,0)--(5,0);
				\draw[blue,dashed,ultra thick,->] (5,0)arc(0:180:5);
				\draw  [green](-4,2)--(-2.5,2);
				\draw  [green](4,2)--(2.5,2);
				\filldraw [red] (0,2) circle (2pt);
				\node [above right] at (0,2) {$0+\mathbf{i}\sigma(2)$};
				\filldraw [red] (-1,2) circle (2pt);
				\node [below] at (-1,2) {$-ac_Tk+\mathbf{i}\sigma$};
				\filldraw [red] (1,2) circle (2pt);
				\node [below] at (1,2) {$ac_Tk+\mathbf{i}\sigma$};
				\filldraw [green] (-4,2) circle (2pt);
				\node[below] at(-3.75,2) {$-c_Lk+\mathbf{i}\sigma$};
				\filldraw [green] (-2.5,2) circle (2pt);
				\node[above] at(-2.5,2) {$-c_Tk+\mathbf{i}\sigma$};
				\filldraw [green] (4,2) circle (2pt);
				\node[below] at(4,2) {$c_Lk+\mathbf{i}\sigma$};
				\filldraw [green] (2.5,2) circle (2pt);
				\node[above] at(2.5,2) {$c_Tk+\mathbf{i}\sigma$};
			\end{tikzpicture}
		\end{center}
		\caption{Integral circuit and distributaion of inner singularities.}
		\label{19th-Sept.-2023-P001}
	\end{figure}
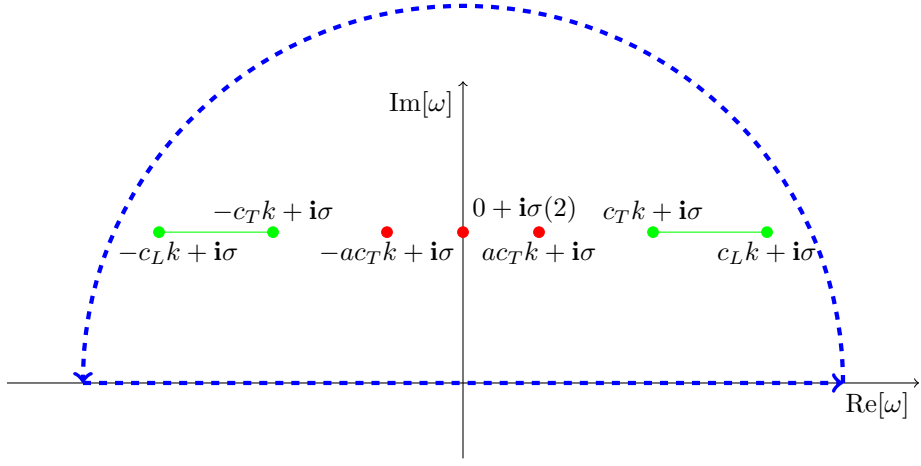\\
	In the multiply connected domain surrounded by the integral contour, there is one second-order pole at $\omega=\mathbf{i}\sigma$ (red point in \ref{19th-Sept.-2023-P001}), two first-order poles at $\omega=\pm ac_Tk+\mathbf{i}\sigma$ (red points), four branch points at $\omega=\pm c_Tk+\mathbf{i}\sigma,\pm c_Lk+\mathbf{i}\sigma$ (green points), and two branch cuts (green lines).\\
	\\
	Cauchy's theorem applies in this domain.\\
	\begin{equation}
		\int_{R.A.}+\int_{A_{\infty}}+\int_{L_{Branch-cut}}+\int_{A_{Branch-point}}=2\pi\mathbf{i}\left(Res[\mathbf{i}\sigma]+Res[-ac_Tk+\mathbf{i}\sigma]+Res[ac_Tk+\mathbf{i}\sigma]\right).
		\label{19th-Sept.-2023-E002}
	\end{equation}
	Via the Jordan's lemma and the residue theorem, the $\omega$-integral can be written as\\
	\begin{equation}
		I=I_{pole}-I_{Branch-cut},
	\end{equation}
	where\\
	\begin{subequations}
		\begin{align}
			&I_{pole}=2\pi\mathbf{i}\left(Res[\mathbf{i}\sigma]+Res[-ac_Tk+\mathbf{i}\sigma]+Res[ac_Tk+\mathbf{i}\sigma]\right)=4\rho\pi^2e^{-\sigma t}\left(M_1\frac{t}{2\pi}-N_1\frac{ac_T\sin (ac_Ttk)}{k}\right),\label{19th-Sept.-2023-E004}\\
			&I_{Branch-cut}=\frac{16}{c_T^2k}e^{-\sigma t}\int_{1}^{1/\sqrt{\eta}}W_L(x)xc_T\sin(xc_Ttk){\rm d}x,\\
			&M_1=\frac{16}{a^2 b^2 c^2},\quad N_1=\frac{(2-a^2)^3}{c_T^2\pi a^4(a^2+b^2)(a^2+c^2)},\quad W_L(x)=\frac{\left(2-x^2\right)\sqrt{x^2-1}\sqrt{1-\eta x^2}}{x^3(x^2-a^2)(x^2+b^2)(x^2+c^2)}.
			\notag
		\end{align}
	\end{subequations}
	\underline{\textbf{Calculation of $k$-integral}} $L$ can be written as \\
	\begin{equation}
		L=\frac{e^{\sigma t}}{4\rho\pi^2}\int_{0}^{\infty}\left(I_{pole}-I_{Branch-cut}\right)k{\rm J}_0(kr){\rm d}k.
		\label{19th-Sept.-2023-E008}
	\end{equation}
	Introduce a integral formula of Bessel function,\\
	\begin{equation}
		\int_0^{\infty}{\rm J}_0(ax)\sin (bx){\rm d}x=
		\left\{\begin{array}{cc}
			0&,\quad0<b<a\\
			\frac{1}{\sqrt{b^2-a^2}}&,\quad0<a<b\\
		\end{array}\right..
		\label{19th-Sept.-2023-E009}
	\end{equation}\\
	Use \ref{19th-Sept.-2023-E009} to rewrite \ref{19th-Sept.-2023-E008}. Similarly, we can obtain the form of $T$. In conclusion, the D-terms can be expressed as\\
	\begin{subequations}
		\begin{align}
			L&=\frac{1}{\rho}\left(Mt\delta(\mathbf{r})-N_1\frac{{\rm H}(t-r/(ac_T))}{\sqrt{t^2-r^2/(ac_T)^2}}-C_1\int_{1}^{1/\sqrt{\eta}}W_L(x)\frac{{\rm H}(t-r/(xc_T))}{\sqrt{t^2-r^2/(xc_T)^2}}{\rm d}x\right),\label{19th-Sept.-2023-E010a}\\
			T&=\frac{1}{\rho}\left(Mt\delta(\mathbf{r})-N_2\frac{{\rm H}(t-r/(ac_T))}{\sqrt{t^2-r^2/(ac_T)^2}}-C_2\int_{1}^{1/\sqrt{\eta}}W_T(x)\frac{{\rm H}(t-r/(xc_T))}{\sqrt{t^2-r^2/(xc_T)^2}}{\rm d}x\right),
			\label{19th-Sept.-2023-E010b}
		\end{align}
	\end{subequations}
	where ${\rm H}$(x) is the Heaviside step function, with the parameters and functions being defined as below
	\begin{equation}
		\begin{aligned}
			&M=M_1=\frac{16}{a^2 b^2 c^2}=\frac{1}{1-\eta},\\
			&N_2=\frac{(2-a^2)^4}{2c_T^2\pi a^4(a^2+b^2)(a^2+c^2)},\\
			&C_1=\frac{4}{c_T^2\pi^2},\quad C_2=\frac{2}{c_T^2\pi^2},\\
			&W_T(x)=\frac{\left(2-x^2\right)^2\sqrt{x^2-1}\sqrt{1-\eta x^2}}{x^3(x^2-a^2)(x^2+b^2)(x^2+c^2)}.\\
		\end{aligned}
	\end{equation}
	\subsubsection{Green's function}\label{3.1.3}
	In this section, certain special functions are defined, and a geometrical discussion regarding the form of these special functions is conducted for the arrival time and the over-peak time. Finally, Green's functions are represented using these special functions. Detailed calculations elucidating the intricate expressions of the special functions are provided in \ref{C.1}.\\
	\\
	In the preceding paragraph, we provided the exact expressions of the distribution terms: \ref{19th-Sept.-2023-E010a} and \ref{19th-Sept.-2023-E010b}, and the propagation terms: \ref{16th-Sept.-2023-E004a} and \ref{16th-Sept.-2023-E004b}. Now, what we need to consider are $<L,P_L>$ and $<T,P_T>$. The main components of these 3D convolutions are \\
	\begin{subequations}
		\begin{align}
			\mathscr{A}(t,r,z;c)&:=\left<t\delta(\mathbf{r}),\frac{1}{R}\delta(t-\frac{R}{c})\right>,
			\label{20th-Sept.-2023-E002a}\\
			\mathscr{H}_v(t,r,z;p,c;m,n)&:=\left<\frac{{\rm H}(t-r/p)}{\sqrt{t^2-r^2/p^2}},\left(\frac{x_1}{r}\right)^v\frac{r^{m}}{R^{n}}\delta(t-\frac{R}{c})\right>, \label{20th-Sept.-2023-E002b}\\
		\end{align}
	\end{subequations}
	where $p$ retrieves a value in $\{ac_T\}\cup\left(c_T,c_L\right)$ and $c$ retrieves a value in $\{c_T,c_L\}$.\\
	\begin{subequations}
		\begin{align}
			&\begin{aligned}
				\left<L,P_L\right>=\frac{1}{2\pi\rho}\left(M\mathscr{A}(t,r,z;c_L)-N_1\mathscr{H}_0(t,r,z;ac_T,c_L;0,1)-C_1\int_{1}^{1/\sqrt{\eta}}W_L(x)\mathscr{H}_0(t,r,z;xc_T,c_L;0,1){\rm d}x\right),
			\end{aligned}\\
			&\begin{aligned}
				\left<T,P_T\right>=\frac{1}{2\pi\rho}\left(M\mathscr{A}(t,r,z;c_T)-N_2\mathscr{H}_0(t,r,z;ac_T,c_T;0,1)-C_2\int_{1}^{1/\sqrt{\eta}}W_T(x)\mathscr{H}_0(t,r,z;xc_T,c_T;0,1){\rm d}x\right).
			\end{aligned}		
		\end{align}
	\end{subequations}
	We name $\mathscr{A}$ as direct wave-field which is in the same form of body waves, and $\mathscr{H}_v$ as Huygens wave-fields which is the mathematical expressions of Huygens effect. \ref{20th-Sept.-2023-E002a} is easy to calculate.\\
	\begin{equation}
		\begin{aligned}
			\mathscr{A}(t,r,z;c)&=\iint_{\mathbb{R}^2}\int_{0}^{\infty}\frac{\tau\delta(\bm{\xi})}{\sqrt{|\mathbf{r}-\bm{\xi}|^2+z^2}}\delta(t-\tau-\frac{\sqrt{|\mathbf{r}-\bm{\xi}|^2+z^2}}{c}){\rm d}\tau{\rm d}A\\
			&=\iint_{\mathbb{R}^2}\frac{\delta(\bm{\xi})}{\sqrt{|\mathbf{r}-\bm{\xi}|^2+z^2}}\left(t-\frac{\sqrt{|\mathbf{r}-\bm{\xi}|^2+z^2}}{c}\right){\rm H}(t-\frac{\sqrt{|\mathbf{r}-\bm{\xi}|^2+z^2}}{c}){\rm d}A\\
			&=\left(\frac{t}{R}-\frac{1}{c}\right){\rm H}(t-\frac{R}{c}).\\
		\end{aligned}
		\notag
	\end{equation}
	The functions of Huygens wave-fields are more complex to calculate. Eventually, their forms will be simplified to definite integrals. The details of obtaining them and their exact expressions are presented in \ref{C}. Subsequent work will not focus on their exact forms, and they will only be involved in specific calculations or illustrations. Here, we discuss the physical insights they provide. These functions share a similar integral form,\\
	\begin{equation}
		\begin{aligned}
			&\left<\frac{{\rm H}(t-r/p)}{\sqrt{t^2-r^2/p^2}},\frac{A}{R^n}\delta(t-\frac{R}{c})\right>=\left<\frac{A}{R^n}\delta(t-\frac{R}{c}),\frac{{\rm H}(t-r/p)}{\sqrt{t^2-r^2/p^2}}\right>,\\
			&=\int_{0}^{\infty}\int_{0}^{2\pi}\int_0^{\infty}\frac{\xi A(\tau,\xi,\theta)}{(\sqrt{\xi^2+z^2})^n\sqrt{(t-\tau)^2-|\mathbf{r}-\bm{\xi}|^2/p^2}}\delta(\tau-\frac{\sqrt{\xi^2+z^2}}{c}){\rm H}(t-\tau-\frac{|\mathbf{r}-\bm{\xi}|}{p}){\rm d}\xi{\rm d}\theta{\rm d}\tau.\\
		\end{aligned}
		\label{22th-Sept.-2023-E001}
	\end{equation}
	The parameters $p$ and $c$ determine the time properties of wave propagation. The integral region in \ref{22th-Sept.-2023-E001} depends on $\delta(\tau-\sqrt{\xi^2+z^2}/c)$ and ${\rm H}(t-\tau-|\mathbf{r}-\bm{\xi}|/p)$. Geometrically, this represents the intersection between a hyperbolic surface and a cone. As time $t$ increases, they intersect and pass through each other. The arrival time corresponds to the first instance of their intersection. \ref{22th-Sept.-2023-P001} shows this process.\\
	\begin{figure}[h!]
		\centering
		\begin{tikzpicture}
			\filldraw[blue!20] (1.5,0.5)--(4,-2)--(-1,-2);
			\draw[->] (-4,0) --(4,0) node[right] {$\xi_1$};
			\draw[->] (0.85,0.5)--(-1.7,-1) node[below]{$\xi_2$};
			\draw[->] (0,-2) --(0,4) node[above] {$\tau$};
			\draw[->,dashed,thick] (1.5,-1) --(1.5,3); 
			\draw[domain =-3:3,red,ultra thick] plot (\x ,{sqrt(\x*\x+1)}) node[black,right] {$\tau=\sqrt{\xi^2+z^2}/c$};
			\draw[dash dot,domain =-1:1.5,blue,ultra thick] plot (\x ,\x-1);
			\node [below,black] at (1.5,-1) {$\tau\leq t-\frac{|\mathbf{r}-\bm{\xi}|}{p}$};
			\draw[dash dot,domain =1.5:4,blue,ultra thick] plot (\x ,-\x+2);
			\filldraw [red] (0,1) circle (2pt);
			\node[below left] at (0,1) {$\left(0,|z|/c\right)$};
			\filldraw [blue] (1.5,0.5) circle (2pt);
			\node[right] at (1.5,0.5) {$\left(r,t\right)$};
		\end{tikzpicture}
		\caption{The integral area}
		\label{22th-Sept.-2023-P001}
	\end{figure}
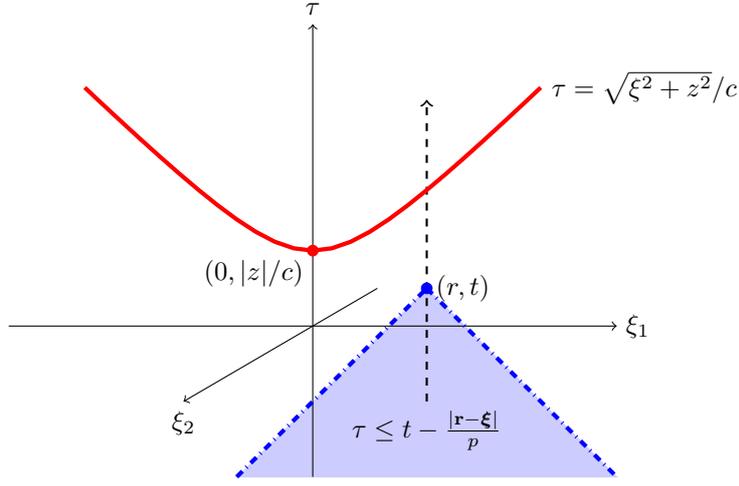\\
	When $rp\leq Rc$, the top of the cone will initially touch the hyperbolic surface as time increases. Before this intersection occurs, the integral region represents an empty set, indicating that there is no response during this period.\\
	\begin{equation}
		t_{arrival}=\frac{R}{c}.
		\notag
	\end{equation}
	When $rp > Rc$, the cone is wider and will first touch the hyperbolic surface tangentially before its apex reaches it. This scenario only occurs when $c = c_T$.\\
	\begin{equation}
		\frac{R}{c_L}<t_{arrival}=\frac{1}{p}\left(r+|z|\sqrt{\frac{p^2}{c^2}-1}\right)<\frac{R}{c_T}.
		\notag
	\end{equation}
	This phenomenon can only be observed in the branch cut integral of $\left<T,P_T\right>$. We refer to it as "variation," and a discussion on its formation mechanisms is provided in the next section.\\
	\begin{equation}
		t_{arrival}(p,c)=\left\{\begin{array}{ll}
			\frac{R}{c}&,\quad rp\leq Rc\\
			\frac{1}{p}\left(r+|z|\sqrt{\frac{p^2}{c^2}-1}\right)&,\quad rp>Rc.\\
		\end{array}\right.
		\label{22th-Sept.-2023-E002}
	\end{equation}
	In addition to $t_{\text{arrival}}$, another significant time point worth attention is the over-peak time $t_{\text{peak}}$. Geometrically, this is the time when the peak of the hyperbolic surface touches the surface of the cone. It holds significance in the change of amplitude of the Green's function.\\
	\begin{equation}
		t_{peak}(p,c)=\frac{r}{p}+\frac{|z|}{c}.
		\label{22th-Sept.-2023-E003}
	\end{equation}
	The Green's functions can be expressed as combinations of $\mathscr{A}$ and $\mathscr{H}_0$ and their derivatives.\\
	\begin{subequations}
		\begin{align}
			&\begin{aligned}
				G_r^{(z)}&=\frac{M}{2\pi}\left(\frac{3rz}{R^5}t\left({\rm H}(t-\frac{R}{c_L})-{\rm H}(t-\frac{R}{c_T})\right)+\frac{rz}{R^3}\left(\frac{1}{c_L^2}\delta(t-\frac{R}{c_L})-\frac{1}{c_T^2}\delta(t-\frac{R}{c_T})\right)\right)\\
				&-\frac{N_1}{2\pi}\frac{\partial^2}{\partial z\partial r}\mathscr{H}_0(t,r,z;ac_T,c_L;0,1)+\frac{N_2}{2\pi}\frac{\partial^2}{\partial r\partial z}\mathscr{H}_0(t,r,z;ac_T,c_T;0,1)\\
				&-\frac{1}{2\pi}\int_{1}^{1/\sqrt{\eta}}\left(C_1W_L(x)\frac{\partial^2}{\partial z\partial r}\mathscr{H}_0(t,r,z;xc_T,c_L;0,1)-C_2W_T(x)\frac{\partial^2}{\partial r\partial z}\mathscr{H}_0(t,r,z;xc_T,c_T;0,1)\right){\rm d}x,
			\end{aligned}\\
			&\begin{aligned}
				G_z^{(z)}&=\frac{M}{2\pi}\left(\frac{2z^2-r^2}{R^5}t\left({\rm H}(t-\frac{R}{c_L})-{\rm H}(t-\frac{R}{c_T})\right)+\frac{1}{R^3}\left(\frac{z^2}{c_L^2}\delta(t-\frac{R}{c_L})+\frac{r^2}{c_T^2}\delta(t-\frac{R}{c_T})\right)\right)\\
				&-\frac{N_1}{2\pi}\frac{\partial^2}{\partial z^2}\mathscr{H}_0(t,r,z;ac_T,c_L;0,1)-\frac{N_2}{2\pi}\nabla_r^2\mathscr{H}_0(t,r,z;ac_T,c_T;0,1)\\
				&-\frac{1}{2\pi}\int_{1}^{1/\sqrt{\eta}}\left(C_1W_L(x)\frac{\partial^2}{\partial z^2}\mathscr{H}_0(t,r,z;xc_T,c_L;0,1)+C_2W_T(x)\nabla_r^2\mathscr{H}_0(t,r,z;xc_T,c_T;0,1)\right){\rm d}x.
			\end{aligned}
		\end{align}
	\end{subequations}
	where $\mathscr{A}$ and its derivatives are represented in exact closed forms and $\nabla_r$ is the nabla operator in radial plane,\\
	\begin{equation}
		\nabla_r^2:=\frac{\partial^2}{\partial x_1^2}+\frac{\partial^2}{\partial x_2^2}=\frac{\partial^2}{\partial r^2}+\frac{1}{r}\frac{\partial}{\partial r}\text{ (in axisymmetric case)}.
	\end{equation}
	\subsection{Displacements in form of Time-Green fuction}\label{3.2}
	In this section, the method for obtaining the Time-Green functions from the Green's functions is explained, and the displacements in the form of the Time-Green functions are provided (\ref{3.2.1}). Additionally, the displacements under special situations ($r=0$ or $z=0$) are discussed (\ref{3.2.2}).\\ 
	\subsubsection{Time-Green fuction}\label{3.2.1}
	In general, a Green's function is a tensor representing the displacements induced by a impulse at a single point in many physical problems. Green's functions are typically utilized in the following manner,
	\begin{equation}
		\begin{aligned}
			&u_{i}=\iiint_{\mathbb{R}}\int_0^{t}G_{ij}(\tau,\bm{\xi})F_{j}(t-\tau,\mathbf{x}-\bm{\xi}){\rm d}\tau{\rm d}\xi_1{\rm d}\xi_2{\rm d}\xi_3.\\
		\end{aligned}
		\label{20th-Sept.-2023-E001}
	\end{equation}
	where $G_{ij}$ represents the displacement in the $i$ direction induced by a impulse at a single point in the $j$ direction; $F_j$ denotes a distribution in the time-space domain of the force in the $j$ direction, while $u_i$ represents the displacement field in the $i$ direction induced by this distributed force. In this article, our result is $G_{i3}$, and we denote it as a vector $\mathbf{G}^{(z)}$. $\mathbf{G}^{(z)}$ in the cylindrical coordinate system can be represented as\\
	\begin{equation}
		\mathbf{G}^{(z)}=\left(\begin{array}{c}
			G_{r}^{(z)} \\
			G_{z}^{(z)} \\
			G_{\theta}^{(z)} 
		\end{array}\right)=\rho\left(\begin{array}{c}
			\frac{\partial^2}{\partial r\partial z}\left(\left<L,P_L\right>-\left<T,P_T\right>\right) \\
			\frac{\partial^2}{\partial z^2}\left<L,P_L\right>+\left(\frac{\partial^2}{\partial r^2}+\frac{1}{r}\frac{\partial}{\partial r}\right)\left<T,P_T\right> \\
			0 
		\end{array}\right).
		\label{24th-Sept.-2023-E005}
	\end{equation}
	Problems in reality often manifest as distributed forces rather than situations described by Green's functions. However, the importance of Green's functions lies in \ref{20th-Sept.-2023-E001}, which indicates that we can obtain all solutions to real situations via Green's functions. The theorem associated with this conclusion is known as the Duhamel principle.\\
	\\
	Since the impulse at a single point is a mathematical singularity, Green's functions also possess this property, which complicates the derivation process when simplifying Green's functions. Here, this article introduces a new form of Green's function, which extends the traditional Green's function tensor. We refer to this new form as the \textit{\textbf{Time-Green function}} $\mathbf{G}^{(T)}$. The new form is still a Green's function but is more easily applicable to specific situations and provides a more direct analysis of physical phenomena. it is important to note that the form of the Time-Green function is not unique, and what this article presents is a form without singularities, which holds significant value in engineering applications.\\
	\\
	Time-Green fuction is builded on the derivative transfer property of convolution. Denote a symbolic operation as\\
	\begin{equation}
		[f*g](t):=\int_0^tf(\tau)g(t-\tau){\rm d}\tau.
		\label{24th-Sept.-2023-E001}
	\end{equation}
	The derivative transfer property is\\
	\begin{equation}
		\frac{{\rm d}^n}{{\rm d}t^n}[f*g](t)=\left[\frac{{\rm d}^nf}{{\rm d}t^n}*g\right](t)=\left[f*\frac{{\rm d}^ng}{{\rm d}t^n}\right](t).
		\label{24th-Sept.-2023-E003}
	\end{equation}
	When the source is vertical, concentrated and time-continued (not a impulse), which means the $F_j$ in \ref{20th-Sept.-2023-E001} is\\
	\begin{equation}
		F_j=F(t)\delta(\mathbf{r})\delta(z)\delta_{3j}.
		\label{24th-Sept.-2023-E002}
	\end{equation}
	where $\delta_{3j}$ is the Kronecker delta symbol.\\
	\\
	Via \ref{20th-Sept.-2023-E001} the displacements under the activation of \ref{24th-Sept.-2023-E002} are\\
	\begin{equation}
		\begin{aligned}
			u_{i}&=\iiint_{\mathbb{R}}\int_0^{t}G_{i}^{(z)}(\tau,\bm{\xi})F(t-\tau)\delta(\mathbf{r}-\xi_1\mathbf{e}_1-\xi_2\mathbf{e}_2)\delta(z-\xi_3){\rm d}\tau{\rm d}\xi_1{\rm d}\xi_2{\rm d}\xi_3\\
			&=\int_0^{t}G_{i}^{(z)}(\tau,\mathbf{r},z)F(t-\tau)\delta(\mathbf{r})\delta(z){\rm d}\tau=\left[G_{i}^{(z)}*F\right](t),\quad i=r,z,\theta.
		\end{aligned}
		\label{26th-Sept.-2023-E001}
	\end{equation}
	In \ref{24th-Sept.-2023-E005}, there are partial derivatives with respect to position of $<L,P_L>$ and $<T,P_T>$ in the components of $\mathbf{G}^{(z)}$. As discussed in the section on pre-computation of some convolutions, we know that there are integrals in the representations of $<L,P_L>$ and $<T,P_T>$, which makes deriving these functions challenging, increases the complexity of calculation programs, and limits their engineering applications. If partial derivatives with respect to position can be transformed into partial derivatives with respect to time, the complex calculations for the partial derivatives in $\mathbf{G}^{(z)}$ will be simplified into a straightforward process of deriving the known source function $F(t)$. Hence, it is necessary to represent $\mathbf{G}^{(z)}$ as a linear combination of some functions and their time derivatives.\\
	\begin{equation}
		\mathbf{G}^{(z)}=\mathbf{f}_0(t) +\frac{\partial\mathbf{f}_1 }{\partial t}(t)+\frac{\partial^2\mathbf{f}_2 }{\partial t^2}(t)+\mathbf{g}_0(t)+\frac{\partial\mathbf{g}_1 }{\partial t}(t)+\frac{\partial^2\mathbf{g}_2 }{\partial t^2}(t)+...
		\notag
	\end{equation}
	Then \ref{26th-Sept.-2023-E001} can be written in another form,\\
	\begin{equation}
		\mathbf{u}=\mathbf{G}^{(z)}*F=\left(\mathbf{f}_0+\mathbf{g}_0\right)*F+\left(\mathbf{f}_1+\mathbf{g}_1\right)*\frac{{\rm d}F}{{\rm d}t}+\left(\mathbf{f}_2+\mathbf{g}_2\right)*\frac{{\rm d}^2F}{{\rm d}t^2}+...
		\notag
	\end{equation}
	which can be written as the form of Time-Green fuction as well.\\
	\begin{equation}
		\mathbf{u}=\mathbf{G}^{(T)}*\mathbf{F}^{(T)}.
		\label{28th-Sept.-2023-E003}
	\end{equation}
	where\\
	\begin{equation}
		\mathbf{G}^{(T)}=\left(\begin{array}{ccc}
			G_{0-r}^{(T)} & G_{1-r}^{(T)} & G_{2-r}^{(T)}\\
			G_{0-z}^{(T)} & G_{1-z}^{(T)} & G_{2-z}^{(T)}\\
			0 & 0 & 0\\
		\end{array}	
		\right),\label{26th-Sept.-2023-E002}
	\end{equation}
	and\\
	\begin{equation}
		\mathbf{F}^{(T)}:=\left(F(t),\frac{{\rm d}F}{{\rm d}t},\frac{{\rm d}^2F}{{\rm d}t^2}\right)^{\mathbf{T}}.
	\end{equation}
	In fact, the position variables $r$ and $z$ are related to the time variable $t$ in $\mathscr{H}_0$, which is also the cause of wavefront surfaces. Because the variables $(r,t)$ and $(z,t)$ are combined as a denominator in $\mathscr{H}_0$, the partial derivatives with respect to position can be transformed into partial derivatives with respect to time. The details of how this transformation works are outlined in \ref{C.2}. The exact forms of the components of \ref{26th-Sept.-2023-E002} are below, where the variables $(t,r,z)$ are omitted in the Huygens functions $\mathscr{H}_v(t,r,z;p,c;m,n)$ and the exact expressions are \ref{9th-Nov.-2023-E001} and \ref{11th-Mar.-2024-E001},\\
	\begin{subequations}
		\begin{align}
			&\begin{aligned}
				G_{0-r}^{(T)}&=\frac{Mrz}{2\pi R^3}\left(\frac{1}{c_L^2}\delta(t-\frac{R}{c_L})-\frac{1}{c_T^2}\delta(t-\frac{R}{c_T})\right)+\frac{3Mrz}{2\pi R^5}t\left({\rm H}(t-\frac{R}{c_L})-{\rm H}(t-\frac{R}{c_T})\right)\\
				&-\frac{3N_1z}{2\pi}\mathscr{H}_1(;ac_T,c_L;1,5)+\frac{3N_2z}{2\pi}\mathscr{H}_1(;ac_T,c_T;1,5)\\
				&-\frac{3C_1z}{2\pi}\int_{1}^{1/\sqrt{\eta}}W_L(x)\mathscr{H}_1(;xc_T,c_L;1,5){\rm d}x+\frac{3C_2z}{2\pi}\int_{1}^{1/\sqrt{\eta}}W_T(x)\mathscr{H}_1(;xc_T,c_T;1,5){\rm d}x,
			\end{aligned}\label{28th-Sept.-2023-E001a}\\
			&\begin{aligned}
				G_{1-r}^{(T)}&=-\frac{3N_1z}{2\pi c_L}\mathscr{H}_1(;ac_T,c_L;1,4)+\frac{3N_2z}{2\pi c_T}\mathscr{H}_1(;ac_T,c_T;1,4)\\
				&-\frac{3C_1z}{2\pi c_L}\int_{1}^{1/\sqrt{\eta}}W_L(x)\mathscr{H}_1(;xc_T,c_L;1,4){\rm d}x+\frac{3C_2z}{2\pi c_T}\int_{1}^{1/\sqrt{\eta}}W_T(x)\mathscr{H}_1(;xc_T,c_T;1,4){\rm d}x,
			\end{aligned}\label{28th-Sept.-2023-E001b}\\
			&\begin{aligned}
				G_{2-r}^{(T)}&=-\frac{N_1z}{2\pi c_L^2}\mathscr{H}_1(;ac_T,c_L;1,3)+\frac{N_2z}{2\pi c_T^2}\mathscr{H}_1(;ac_T,c_T;1,3)\\
				&-\frac{C_1z}{2\pi c_L^2}\int_{1}^{1/\sqrt{\eta}}W_L(x)\mathscr{H}_1(;xc_T,c_L;1,3){\rm d}x+\frac{C_2z}{2\pi c_T^2}\int_{1}^{1/\sqrt{\eta}}W_T(x)\mathscr{H}_1(;xc_T,c_T;1,3){\rm d}x,
			\end{aligned}\label{28th-Sept.-2023-E001c}\\
			&\begin{aligned}
				G_{0-z}^{(T)}&=\frac{M}{2\pi R^3}\left(\frac{z^2}{c_L^2}\delta(t-\frac{R}{c_L})+\frac{r^2}{c_T^2}\delta(t-\frac{R}{c_T})\right)+\frac{M(2z^2-r^2)}{2\pi R^5}t\left({\rm H}(t-\frac{R}{c_L})-{\rm H}(t-\frac{R}{c_T})\right)\\
				&-\frac{N_1}{2\pi}\left(3z^2\mathscr{H}_0(;ac_T,c_L;0,5)-\mathscr{H}_0(;ac_T,c_L;0,3)\right)\\
				&-\frac{N_2}{2\pi}\left(3\mathscr{H}_0(;ac_T,c_T;2,5)-2\mathscr{H}_0(;ac_T,c_T;0,3)\right)\\
				&-\frac{C_1}{2\pi}\int_{1}^{1/\sqrt{\eta}}W_L(x)\left(3z^2\mathscr{H}_0(;xc_T,c_L;0,5)-\mathscr{H}_0(;xc_T,c_L;0,3)\right){\rm d}x\\
				&-\frac{C_2}{2\pi}\int_{1}^{1/\sqrt{\eta}}W_T(x)\left(3\mathscr{H}_0(;xc_T,c_T;2,5)-2\mathscr{H}_0(;xc_T,c_T;0,3)\right){\rm d}x,
			\end{aligned}\label{28th-Sept.-2023-E001d}\\
			&\begin{aligned}
				G_{1-z}^{(T)}&=-\frac{N_1}{2\pi c_L}\left(3z^2\mathscr{H}_0(;ac_T,c_L;0,4)-\mathscr{H}_0(;ac_T,c_L;0,2)\right)\\
				&-\frac{N_2}{2\pi c_T}\left(3\mathscr{H}_0(;ac_T,c_T;2,4)-2\mathscr{H}_0(;ac_T,c_T;0,2)\right)\\
				&-\frac{C_1}{2\pi c_L}\int_{1}^{1/\sqrt{\eta}}W_L(x)\left(3z^2\mathscr{H}_0(;xc_T,c_L;0,4)-\mathscr{H}_0(;xc_T,c_L;0,2)\right){\rm d}x\\
				&-\frac{C_2}{2\pi c_T}\int_{1}^{1/\sqrt{\eta}}W_T(x)\left(3\mathscr{H}_0(;xc_T,c_T;2,4)-2\mathscr{H}_0(;xc_T,c_T;0,2)\right){\rm d}x,
			\end{aligned}\label{28th-Sept.-2023-E001e}\\
			&\begin{aligned}
				G_{2-z}^{(T)}&=-\frac{N_1z^2}{2\pi c_L^2}\mathscr{H}_0(;ac_T,c_L;0,3)-\frac{N_2}{2\pi c_T^2}\mathscr{H}_0(;ac_T,c_T;2,3)\\
				&-\frac{C_1z^2}{2\pi c_L^2}\int_{1}^{1/\sqrt{\eta}}W_L(x)\mathscr{H}_0(;xc_T,c_L;0,3){\rm d}x-\frac{C_2}{2\pi c_T^2}\int_{1}^{1/\sqrt{\eta}}W_T(x)\mathscr{H}_0(;xc_T,c_T;2,3){\rm d}x.
			\end{aligned}\label{28th-Sept.-2023-E001f}
		\end{align}
	\label{28th-Sept.-2023-E001}
	\end{subequations}
	\subsubsection{Special conditions}\label{3.2.2}
	The impulse at a single point is placed at the origin, which results in the formulas we obtained earlier exhibiting singularities when $r=0$ or $z=0$. It is necessary to conduct specific discussions on these two conditions.\\
	\\
	\underline{\textbf{Displacements on axis ($r=0$)}} Considering the condition where $r\to0$, we note that all terms related to refracted wave fields of $G^{(T)}_{0-r}$, $G^{(T)}_{1-r}$, and $G^{(T)}_{2-r}$ include the function $\mathscr{H}_1(;p,c;1,n)$. This function vanishes when $r=0$,\\
	\begin{equation}
		\begin{aligned}
			\mathscr{H}_0(t,0,z;p,c;0,0,n_3)&=\int_{\tau_{near}}^{\tau_{far}}{\color{red}{\int_{-\pi}^{\pi}}}\int_{0}^{\infty}\frac{\xi^2{\color{red}{\cos\theta}}}{(\sqrt{\xi^2+z^2})^{n_3}\sqrt{(t-\tau)^2-\xi^2/p^2}}\delta(\tau-\frac{\sqrt{\xi^2+z^2}}{c}){\rm d}\xi{\color{red}{{\rm d}\theta}}{\rm d}\tau=0,\\
		\end{aligned}
		\notag
	\end{equation}
	which means points on the source axis, $r=0$, have zero radial displacement, $u_r(r=0)=0$. This is consistent with \ref{15th-Sept.-2023-E014a} above. When $r=0$, ${\rm J}_1(kr)=0$, and thus \ref{15th-Sept.-2023-E014a} vanishes.\\
	\\
	The vertical displacement $u_z$ does not have a singularity on the axis. It can be calculated normally. Furthermore, \ref{28th-Sept.-2023-E001d}, \ref{28th-Sept.-2023-E001e}, and \ref{28th-Sept.-2023-E001f} can be simplified when $r=0$. Notice that all terms related to refracted wave fields of $G^{(T)}_{0-z}$, $G^{(T)}_{1-z}$, and $G^{(T)}_{2-z}$ include $\mathscr{H}_0(;p,c;0,n)$ and $\mathscr{H}_0(;p,c;2,n)$, which can be represented by elementary functions (See \ref{C.3}), making it convenient for application.\\
	\\
	\underline{\textbf{Displacements on surface ($z=0$)}} The detailed process is outlined in \ref{C.4}. Consider the condition where $z\to0$. In this condition, there is degeneracy of weak function.\\
	\begin{equation}
		\lim_{|z|\to0}\frac{|z|}{r^2+z^2}=\delta(r).
		\label{28th-Sept.-2023-E002}
	\end{equation}
	See \ref{28th-Sept.-2023-E001a}, \ref{28th-Sept.-2023-E001b}, and \ref{28th-Sept.-2023-E001c}. When $z=0$, $G^{(T)}_{0-r}=0$, $G^{(T)}_{1-r}=0$, and $G^{(T)}_{2-r}=0$, which implies $u_r=0$ on the free surface. However, this is not consistent with \ref{15th-Sept.-2023-E014a}. The reason for this inconsistency is due to degeneracy of weak function occurring in the representations of $\mathbf{G}^{(T)}$. Therefore, instead of \ref{28th-Sept.-2023-E003} to calculate $u_r$ when $z=0$, we directly calculate and simplify \ref{15th-Sept.-2023-E014a} to obtain the Green's function. This process is similar to the aforementioned approach.\\
	\begin{equation}
		\begin{aligned}
			z=0,G_r&=\rho\frac{\partial }{\partial r}\left(L-T\right)
			\\
			&=-(N_1-N_2)\frac{\partial }{\partial r}\frac{{\rm H}(t-r/(ac_T))}{\sqrt{t^2-r^2/(ac_T)^2}}-\int_{1}^{1/\sqrt{\eta}}\left(C_1W_L(x)-C_2W_T(x)\right)\frac{\partial }{\partial r}\frac{{\rm H}(t-r/(xc_T))}{\sqrt{t^2-r^2/(xc_T)^2}}{\rm d}x.\\
		\end{aligned}
		\label{30th-Sept.-2023-E001}
	\end{equation}
	Thus the representations of Time-Green functions are\\
	\begin{subequations}
		\begin{align}
			z=0,\quad&G_{0-r}^{(T)}=0,\\
			&G_{1-r}^{(T)}=-(N_1-N_2)S_1(t,r;ac_T)-\int_{1}^{1/\sqrt{\eta}}\left(C_1W_L(x)-C_2W_T(x)\right)S_1(t,r;xc_T){\rm d}x,\\
			&G_{2-r}^{(T)}=-(N_1-N_2)S_2(t,r;ac_T)-\int_{1}^{1/\sqrt{\eta}}\left(C_1W_L(x)-C_2W_T(x)\right)S_2(t,r;xc_T){\rm d}x.
		\end{align}
	\label{1st-Apr.-2024-E001}
	\end{subequations}\\
	Here, integration by parts is utilized to eliminate the singularity of the Green's function and obtain an equivalent form without singularity, known as the Time-Green functions. A formal sound source function $F(t)$ needs to satisfy\\
	\begin{equation}
		F(0)=\frac{{\rm d}F}{{\rm d}t}(0)=0,\notag
	\end{equation}
	If the source function is not in a formal form, the real displacement under it will have an additional term that includes a part of the singularity.\\
	\begin{equation}
		u_r=\mathbf{G}^{(T)}*\mathbf{F}^{(T)}\cdot\mathbf{e}_r{\color{red}{-(N_1-N_2)ET(t,r;ac_T)-\int_{1}^{1/\sqrt{\eta}}\left(C_1W_L(x)-C_2W_T(x)\right)ET(t,r;xc_T){\rm d}x}}.
	\end{equation}
	where\\
	\begin{subequations}
		\begin{align}
			&S_1(t,r;p)=-\frac{\sqrt{t^2-r^2/p^2}}{rt}{\rm H}(t-\frac{r}{p})\label{9th.-Jan.-2024-E003a},\\
			&S_2(t,r;p)=-\frac{1}{p}\arccos(\frac{r}{pt}){\rm H}(t-\frac{r}{p})\label{9th.-Jan.-2024-E003b},\\
			&ET(t,r;p)=-\left[\left(\frac{r}{p^2\sqrt{t^2-r^2/p^2}}+\frac{\sqrt{t^2-r^2/p^2}}{r}\right)\frac{{\color{red}{F(0)}}}{t}+\frac{1}{p}\arccos(\frac{r}{pt}){\color{red}{F'(0)}}\right]{\rm H}(t-\frac{r}{p}).
		\end{align}
	\end{subequations}
	Furthermore, the Time-Green functions of $u_z$ also exist in different forms.\\
	\begin{subequations}
		\begin{align}
			z=0,\quad&G^{(T)}_{0-z}=\frac{M}{4\pi c_T^2r}\delta(t-\frac{r}{c_T}),\\
			&G^{(T)}_{1-z}=0,\\
			&G^{(T)}_{2-z}=-\frac{1}{4\pi c_T^2}\left(N_2\mathscr{H}_0(t,r,0;ac_T,c_T;0,1)+C_2\int_1^{1/\sqrt{\eta}}W_T(x)\mathscr{H}_0(t,r,0;xc_T,c_T;0,1){\rm d}x\right).
		\end{align}
	\label{1st-Apr.-2024-E002}
	\end{subequations}
	\subsection{Numerical verification}\label{3.3}
\begin{figure}[h!]
		\centering
		\subfigure[Source function.]{\includegraphics[scale=0.8]{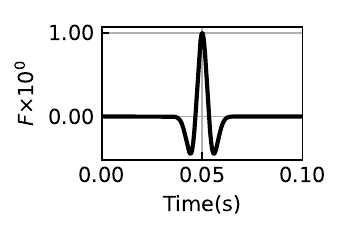}}
		\subfigure[First-order derivative of source function.]{\includegraphics[scale=0.8]{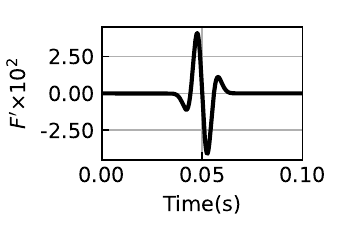}}
		\subfigure[Second-order derivative of source function.]{\includegraphics[scale=0.8]{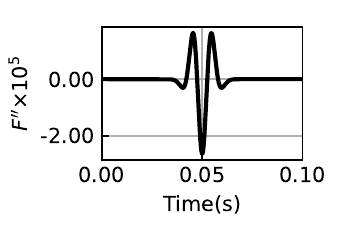}}
		\caption{Source vector $\mathbf{F}^{(T)}$.}\label{8th-Nov.-2023-P001}
	\end{figure}

\begin{figure}[h!]
		\centering
		\subfigure[Radial displacement $u_r$ of situation \textbf{a}.]{\includegraphics[scale=0.8]{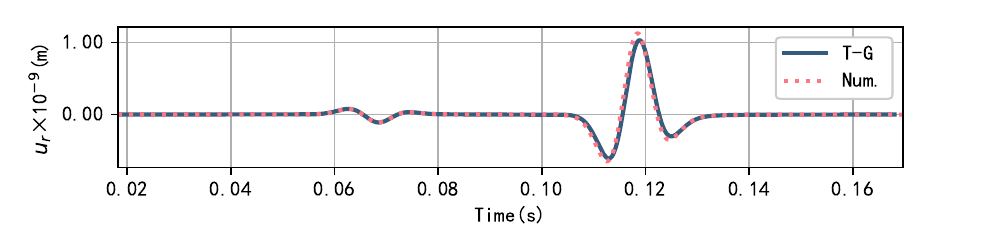}}\\
		\subfigure[Axial displacement $u_z$ of situation \textbf{a}.]{\includegraphics[scale=0.8]{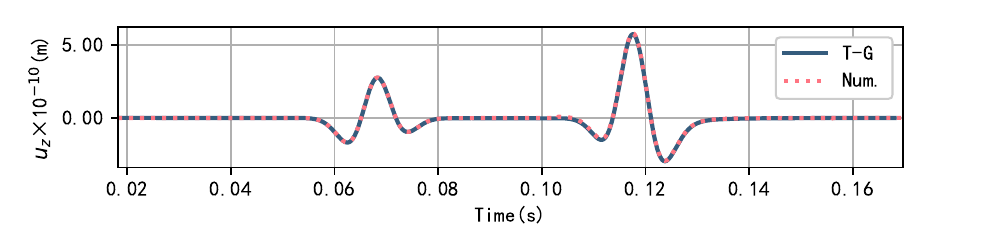}}\\
		\subfigure[Radial displacement $u_r$ of situation \textbf{b}.]{\includegraphics[scale=0.8]{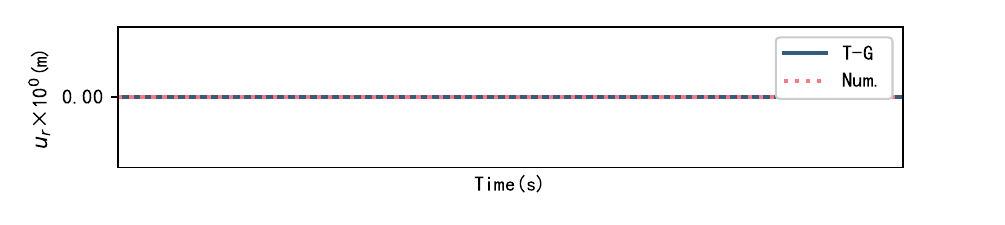}}\\
		\subfigure[Axial displacement $u_z$ of situation \textbf{b}.]{\includegraphics[scale=0.8]{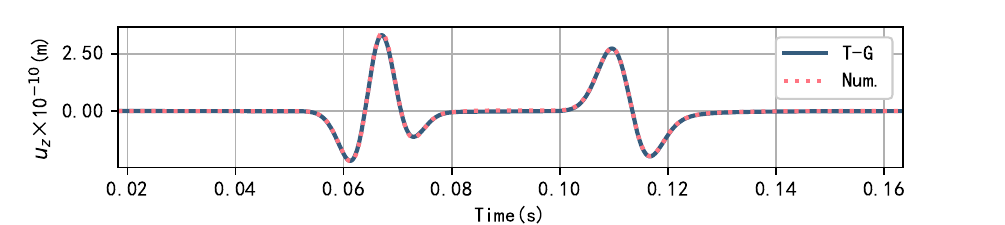}}\\
		\subfigure[Radial displacement $u_r$ of situation \textbf{c}.]{\includegraphics[scale=0.8]{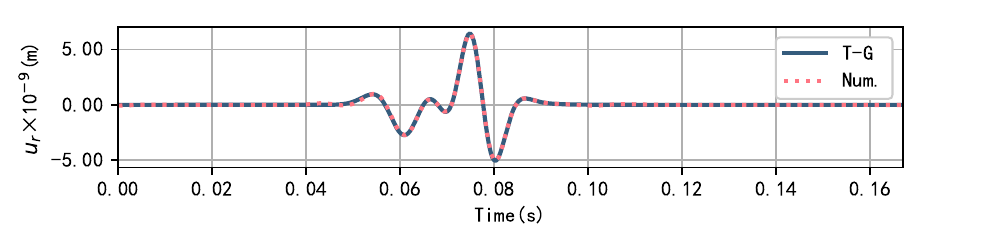}}\\
		\subfigure[Axial displacement $u_z$ of situation \textbf{c}.]{\includegraphics[scale=0.8]{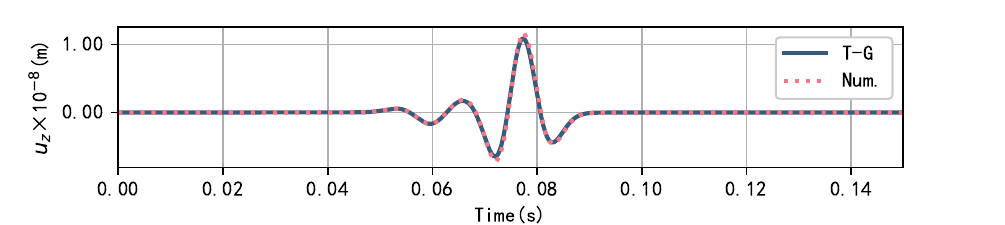}}
		\caption{Comparison between the displacements calculated by the Time-Green function and numerical integral.}\label{21th-Mar.-2024-P001}
	\end{figure}

	To demonstrate the correctness of the results, we select three positions and compare the full-wave responses obtained using our results with those obtained using \ref{15th-Sept.-2023-E014a} and \ref{15th-Sept.-2023-E014b} by numerical integral:
	\begin{itemize}
		\item[a.] Common situation ($r=20$m and $z=-50$m).
		\item[b.] Point on the axis ($r=0$m and $z=-50$m).
		\item[c.] Point on the free surface ($r=20$m and $z=0$m).
	\end{itemize}
	Here, we use the Ricker wavelet (1953, \cite{ricker1953form}) as the source function,\\
	\begin{equation}
		F(t)=\left(1-\frac{1}{2}\omega_s^2(t-T_s)^2\right)\exp\left(-\frac{1}{4}\omega_s^2(t-T_s)^2\right)
		\label{30th-Sept.-2023-E002}
	\end{equation}
	where $\omega_s$ controls the bandwidth and $T_s$ controls the arrival time of the peak value.\\
	\\
	The parameters used in this section are listed in \ref{8th-Nov.-2023-T001}.\\
	\begin{table}[h!]
		\begin{center}
			\begin{tabular}{llrl}
				\hline\hline\noalign{\smallskip}
				\textbf{Symbol} & \textbf{Explanation} & \textbf{Value} & \textbf{Unit}\\
				\hline
				$\rho$ & Density of medium & 1000 &${\rm kg}/{\rm m}^3$\\
				$c_L$ & Longitudinal wave velocity & 3000 &${\rm m}/{\rm s}$ \\
				$c_T$ & Transverse wave velocity & 800 &${\rm m}/{\rm s}$ \\
				$\omega_s$ & Bandwidth & $2\pi/0.015$ & rad/s\\
				$T_s$ & Arrival time of peak value & 50 &ms\\
				\noalign{\smallskip}\hline
			\end{tabular}\caption{Parameters of medium and source.}\label{8th-Nov.-2023-T001}
		\end{center}
	\end{table}\\
	\ref{8th-Nov.-2023-P001} is the schema of $\mathbf{F}^{(T)}$.\\
	\\
	After calculations of $u_r$ and $u_z$ on the given points, the comparison of results given by Time-Green functions and results given by numerical integral are shown in \ref{21th-Mar.-2024-P001}, where solid line represents the results of Time-Green functions and dotted line represents the results of numerical integral. The curves of all the situations computed by the Time-Green functions are identical to the numerical results. Any small mismatches are attributed to errors in numerical integration.\\
	\\
	\section{Physical explanation and mathematical analysis to results}\label{4}
	In this section, the results presented in \ref{3.2} are analyzed both physically and mathematically. The main elements include:
	\begin{itemize}
		\item[1.] Classification of waves by different generation mechanisms (\ref{4.1}). Discussions about the Rayleigh wave are provided in \ref{4.1.2}, while discussions about the Huygens effect are outlined in \ref{4.1.3}.
		\item[2.] Discussions about the time domain properties of each component-wave (\ref{4.2}).
		\item[3.] Attenuation of displacements in different component-wave modes activated by the Ricker wavelet with changes in spatial coordinates (\ref{4.3}). Approximations of displacements underground near the axis are provided in \ref{4.3.2}.
	\end{itemize}
	The displacements in this section are splitted into two parts,\\
	\begin{equation}
		\mathbf{u}=\mathbf{u}^{direct}+\mathbf{u}^{Huygens}
	\end{equation}
	\subsection{Generation mechanisms of waves}\label{4.1}
	In this section, the entire wave-field is divided into three components: the direct wave (\ref{4.1.1}), the surface wave (\ref{4.1.2}), and the Huygens wave (\ref{4.1.3}).\\
	\subsubsection{Direct wave}\label{4.1.1}
\begin{figure}[h!]
		\centering
		\subfigure[$u_r^{direct}(r=20\text{ m},z=-50\text{ m})$]{\includegraphics[scale=0.8]{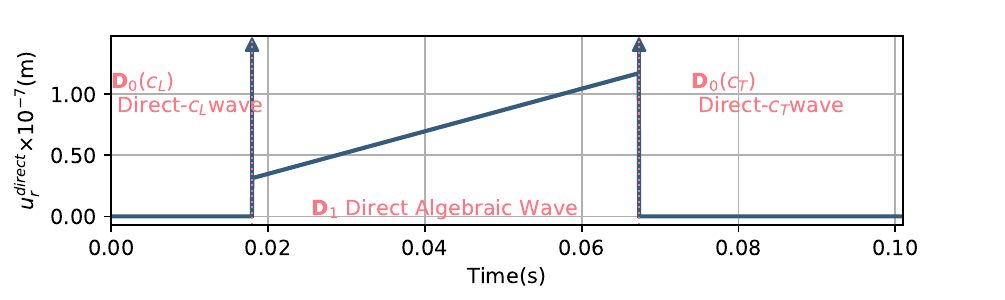}}\\
		\subfigure[$u_z^{direct}(r=20\text{ m},z=-50\text{ m})$]{\includegraphics[scale=0.8]{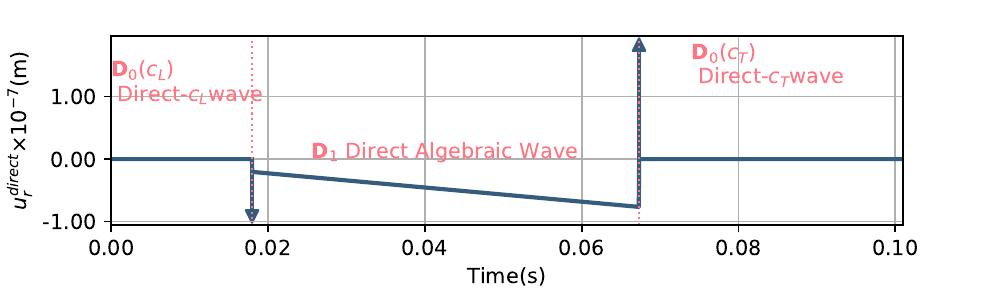}}
		\caption{Direct waves of displacements at $r=20\text{ m}$ and $z=-50\text{ m}$.}\label{21th-Mar.-2024-P002}
	\end{figure}

	There are two modes of direct wave,\\
	\begin{subequations}
		\begin{align}
			&\mathbf{D}_0(c)=\frac{1}{R}\delta(t-\frac{R}{c}),\label{16th-Nov.-2023-E001a}\\
			&\mathbf{D}_1=\frac{t}{R^3}({\rm H}(t-\frac{R}{c_L})-{\rm H}(t-\frac{R}{c_T})).\label{16th-Nov.-2023-E001b}
		\end{align}
	\end{subequations}
	We designate \ref{16th-Nov.-2023-E001a} as the \emph{direct-$c$ wave} (D-c) and \ref{16th-Nov.-2023-E001b} as the \emph{direct algebraic wave} (D-A). In \ref{21th-Mar.-2024-P002}, these two kinds of waves are drawn and the arrow means the $\delta$ functions to the infinite. The components of direct waves comprise common longitudinal waves and transverse waves. Utilizing \ref{28th-Sept.-2023-E001c} and \ref{28th-Sept.-2023-E001f}, the constituent parts of direct waves are\\
	\begin{subequations}
		\begin{align}
			&\begin{aligned}
				u^{direct}_r&=\frac{2}{1-\eta}\left(\frac{rz}{4\pi R^3}\left(\frac{1}{c_L^2}\delta(t-\frac{R}{c_L})-\frac{1}{c_T^2}\delta(t-\frac{R}{c_T})\right)+\frac{3rz}{4\pi R^5}t\left({\rm H}(t-\frac{R}{c_L})-{\rm H}(t-\frac{R}{c_T})\right)\right)\\
				&=\frac{2}{1-\eta}\left(\frac{rz}{4\pi R^2}\left(\frac{1}{c_L^2}\mathbf{D}_0(c_L)-\frac{1}{c_T^2}\mathbf{D}_0(c_T)\right)+\frac{3rz}{4\pi R^2}\mathbf{D}_1\right),
			\end{aligned}\\
			&\begin{aligned}
				u^{direct}_z&=\frac{2}{1-\eta}\left(\frac{1}{4\pi R^3}\left(\frac{z^2}{c_L^2}\delta(t-\frac{R}{c_L})+\frac{r^2}{c_T^2}\delta(t-\frac{R}{c_T})\right)+\frac{(2z^2-r^2)}{4\pi R^5}t\left({\rm H}(t-\frac{R}{c_L})-{\rm H}(t-\frac{R}{c_T})\right)\right)\\
				&=\frac{2}{1-\eta}\left(\frac{1}{4\pi R^2}\left(\frac{z^2}{c_L^2}\mathbf{D}_0(c_L)+\frac{r^2}{c_T^2}\mathbf{D}_0(c_T)\right)+\frac{(2z^2-r^2)}{4\pi R^2}\mathbf{D}_1\right).
			\end{aligned}			
		\end{align}
	\end{subequations}
	The Green's functions in full-space are (\cite{achenbach2012wave})\\
	\begin{equation}
		\begin{aligned}
			G^{full}_{ij}&=\frac{1}{4\pi c_L^2R}\gamma_i\gamma_j\delta(t-\frac{R}{c_L})+\frac{1}{4\pi c_T^2R}(\delta_{ij}-\gamma_i\gamma_j)\delta(t-\frac{R}{c_T})+\frac{1}{4\pi R^3}(3\gamma_i\gamma_j-\delta_{ij})t({\rm H}(t-\frac{R}{c_L})-{\rm H}(t-\frac{R}{c_T})).
		\end{aligned}
	\end{equation}
	where $\gamma_i=x_i/R$. Obviously,\\
	\begin{equation}
		u_r^{direct}=\frac{2}{1-\eta}G^{full}_{r3},\quad u_z^{direct}=\frac{2}{1-\eta}G^{full}_{z3},\notag
	\end{equation}
	This implies that the direct waves share the same physical properties as waves in the full-space; they are spherical waves directly activated by the source.\\
	\\
	However, it is counterintuitive that two half-space problems cannot simply be added up to obtain the full-space problem. They are two entirely different problems, which is why the coefficient is not 2 but a parameter related to the elastic constant. If we split the full-space into two by the plane $z=0$, a fact will be found: the shear stress and tangential displacement on the surface are not zero,
	\begin{equation}
		\lim_{z\to0}\tau^{full}_{rz}\neq0,\quad \lim_{z\to0}G^{full}_{rz},\neq0\notag
	\end{equation}
	This implies that additional energy input (or output) is required for the half-space we remove. The extra energy is provided by the other half-space. Consider $\eta=0$, which means $\mu=0$. In this situation, even if we split the full-space, there is no energy exchange between the two half-spaces as long as $\tau_{rz}\equiv0$ at all times. Furthermore, in the general situation, the transfer of this extra energy is prohibited, which would cause the boundary to tremble and ultimately create the Huygens waves.\\
	\subsubsection{Surface wave}\label{4.1.2}
\begin{figure}[h!]
		\centering
		\subfigure[$Q^{surface}_L(r=20\text{ m})$]{\includegraphics[scale=0.8]{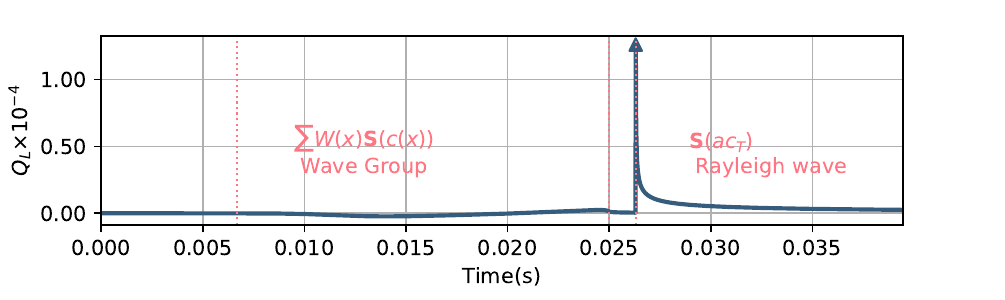}}
		\subfigure[$Q^{surface}_T(r=20\text{ m})$]{\includegraphics[scale=0.8]{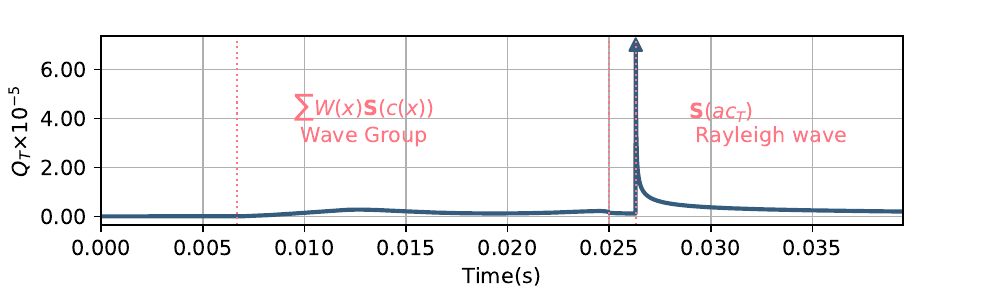}}
		\caption{Physical quantities of surface wave mode at $r=20\text{ m}$.}\label{21th-Mar.-2024-P003}
	\end{figure}

\begin{figure}[h!]
		\centering
		\includegraphics[scale=0.8]{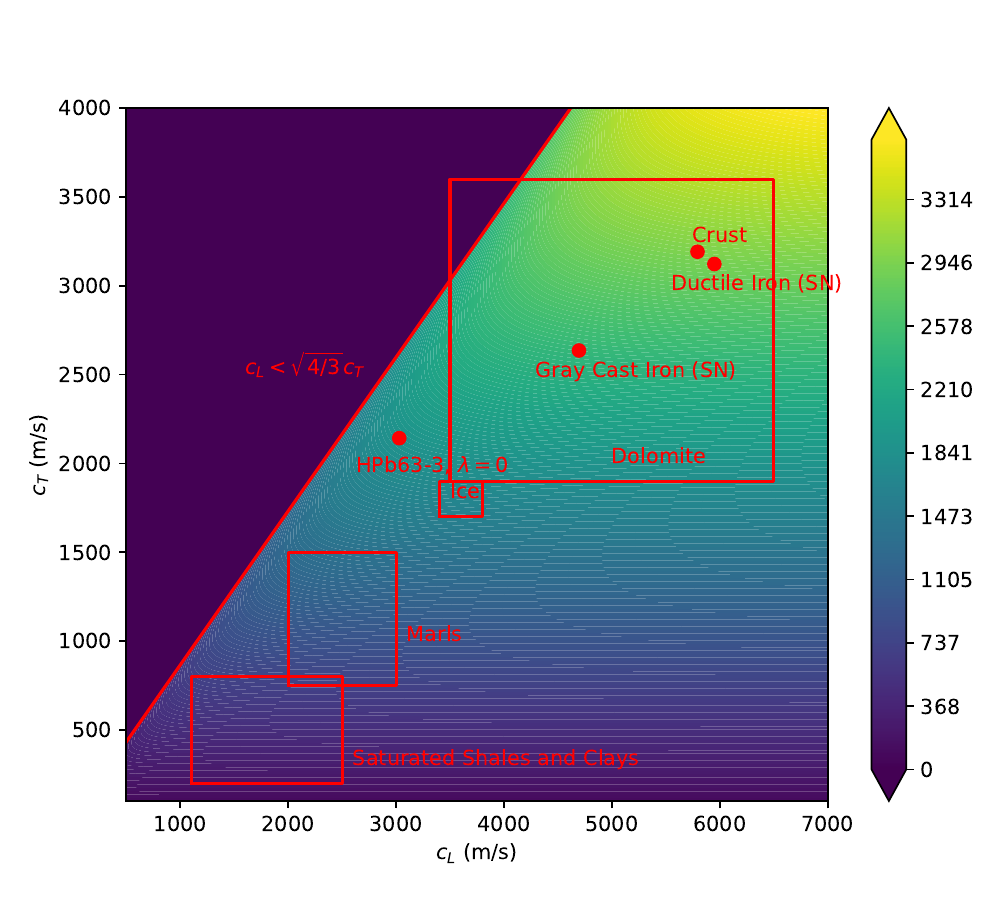}
		\caption{$c_R$ varies with $c_L$ and $c_T$.}\label{20th-Nov.-2023-P001}
	\end{figure}

\begin{figure}[h!]
		\centering
		\includegraphics[scale=0.8]{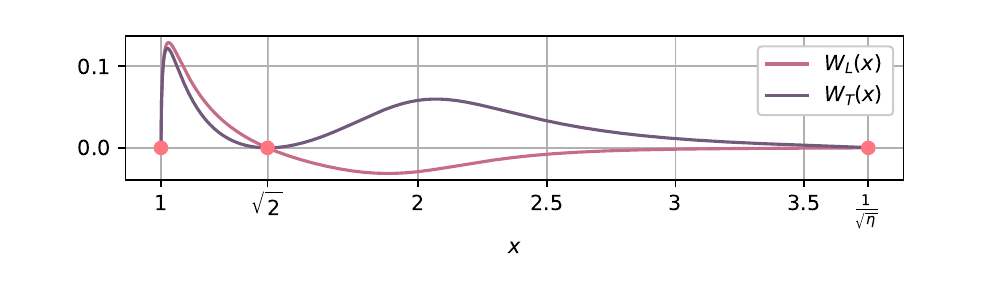}
		\caption{The schema of weught functions.}\label{21th-Mar.-2024-P004}
	\end{figure}

	Surface waves are not equivalent to the actual wave-fields on the surface but rather serve as precursors to generate Huygens waves. They manifest in certain physical quantities only on the surface and when the Huygens effect disappears. From the discussion about $z=0$ in \ref{3.2.2}, we understand that the surface waves are fully evident in radial displacement $G_r$, whereas axial displacement $G_z$ represents Huygens waves in the final expressions. Fortunately, due to the new method introduced in this article, we are able to extract the surface waves from the expressions without letting them be overridden.\\
	\\
	The distribution terms $L$ and $T$ in \ref{3.1.1} provide expressions for surface waves, which include the \emph{Rayleigh wave} (S-R) $ac_T$ (slightly slower than the transverse wave) and a \emph{wave group} (S-G) whose velocity is distributed in the range $(c_T,c_L)$ (lacking both longitudinal and transverse waves). This is consistent with the existing consensus.\\
	\begin{subequations}
		\begin{align}
			Q^{surface}_L&=N_1\frac{{\rm H}(t-r/(ac_T))}{\sqrt{t^2-r^2/(ac_T)^2}}+C_1\int_{1}^{1/\sqrt{\eta}}W_L(x)\frac{{\rm H}(t-r/(xc_T))}{\sqrt{t^2-r^2/(xc_T)^2}}{\rm d}x,\\
			Q^{surface}_T&=N_2\frac{{\rm H}(t-r/(ac_T))}{\sqrt{t^2-r^2/(ac_T)^2}}+C_2\int_{1}^{1/\sqrt{\eta}}W_T(x)\frac{{\rm H}(t-r/(xc_T))}{\sqrt{t^2-r^2/(xc_T)^2}}{\rm d}x.
		\end{align}
	\end{subequations}
	As explained earlier, surface waves do not actually exist except in some special situations. Therefore, it is unnecessary to study a specific physical quantity like $G_r$, but rather analyze the form as follows,\\
	\begin{equation}
		N\frac{{\rm H}(t-r/(ac_T))}{\sqrt{t^2-r^2/(ac_T)^2}}+C\int_{1}^{1/\sqrt{\eta}}W(x)\frac{{\rm H}(t-r/(xc_T))}{\sqrt{t^2-r^2/(xc_T)^2}}{\rm d}x,\quad W(1)=W(\sqrt{2})=W(\frac{1}{\sqrt{\eta}})=0,\notag
	\end{equation}
	or surface waves mode in the simplified form
	\begin{equation}
		\mathbf{S}(p)=\frac{{\rm H}(t-r/p)}{\sqrt{t^2-r^2/p^2}}.
	\end{equation}
	A "wave" is simply a term used to denote a specific type of motion mode. All physical quantities exhibiting the same mode of motion can be considered as part of the same wave. In \ref{21th-Mar.-2024-P003}, two kinds of surface waves: Rayleigh wave with velocity $ac_T$ and wave groups are shown, where the arrow means the value toward infinity.\\
	\\
	Another valuable piece of information is the expression for the velocity of the Rayleigh wave, different form a segmented expression in Vinh’s work (2004 \cite{vinh2004formulas}). In \ref{10th.Jan.-2024-E001}, the square root operation is defined as $Z^{1/2}=|Z|^{1/2}\exp(\mathbf{i}\theta/2)$, and the cube root operation is defined as $Z^{1/3}=|Z|^{1/3}\exp(\mathbf{i}\theta/3)$, which naturally results in the coefficient behind $c_T$ being a real number.\\
	\begin{equation}
		c_{R}=c_T\sqrt{\frac{8}{3}+\frac{1}{6}\frac{8-48\eta}{\left(17 - 45\eta + 3\sqrt{-192\eta^3 + 321\eta^2 - 186\eta + 33}\right)^{\frac{1}{3}}}-\frac{2}{3}\left(17 - 45\eta + 3\sqrt{-192\eta^3 + 321\eta^2 - 186\eta + 33}\right)^{\frac{1}{3}}}.\label{10th.Jan.-2024-E001}
	\end{equation}
	In \ref{10th.Jan.-2024-E001}, $c_R$ depends on the combined effect of $c_L$ and $c_T$. In \ref{20th-Nov.-2023-P001}, materials having Poisson's ratio between -0.5 and 0.5 are considered. Certain metal and geological materials are highlighted. In fact, the ratio $a$ changes very little with $\eta$ to most natural materials ($\lambda\geq0$, $\eta$ varies from 0 to 0.5), ranging from 0.874 to 0.955, almost as a constant. However, to some artificial materials ($\lambda<0$, $\eta$ varies from 0.5 to 0.75), the ratio $a$ changes obviously, ranging from 0.689 to 0.874. The equation yielding $c_R$ (the imaginary part of the root of \ref{17th-Sept.-2023-E002a} and \ref{17th-Sept.-2023-E002c}) is identical to the velocity equation of the Rayleigh wave found in Achenbach's book (\cite{achenbach2012wave}).\\
	\begin{equation}
		\left(2-\frac{c_R^2}{c_T^2}\right)^2-4\left(1-\frac{c_R^2}{c_L^2}\right)^{1/2}\left(1-\frac{c_R^2}{c_T^2}\right)^{1/2}=0.\label{1st-Apr.-2024-E003}
	\end{equation}
	 \\
	The wave group emerges from the branch-cut integral, which can be conceived as a collection of surface waves amalgamated together with a weighting function.\\
	\begin{equation}
		\int_{1}^{1/\sqrt{\eta}}W(x)\frac{{\rm H}(t-r/(xc_T))}{\sqrt{t^2-r^2/(xc_T)^2}}{\rm d}x=\sum_{c(x)\in(c_T,\sqrt{2}c_T)\cup(\sqrt{2}c_T,c_L)}W(x)\mathbf{S}(c(x))\Delta x,\quad c(x)=xc_T.\notag
	\end{equation}
	Aside from the absence of longitudinal and transverse waves within the wave group, there's also the absence of waves with a velocity of $\sqrt{2}c_T$. In \ref{21th-Mar.-2024-P004}, two weight functions are drawn and three zero points $x=1,\sqrt{2},1/\sqrt{\eta}$ are marked.\\
	 \\
	The distinction between surface waves and other types of waves lies in their confinement to the 2D boundary rather than propagating through the entire 3D domain. In contrast, direct waves and Huygens waves propagate throughout the entire medium. Additionally, the presence of a ridge edge could potentially give rise to a new type of wave, known as line waves, which would differ significantly from both body waves and surface waves. Notably, the concept of a "point wave" emerges in the form of $t\delta(\mathbf{r})$ within $L$ and $T$, contributing to the generation of direct waves via the Huygens effect within the 3D domain. Despite being labeled as "point waves," they are still considered a direct response to the source and hold equal significance to surface waves. For better understanding, \ref{21th-Mar.2024-T001} gives the classifications of the above wave modes from two perspectives.\\
	\begin{table}[h!]
		\centering
		\subtable[Waves in their stages of fluctuation.]{
			\begin{tabular}{clllc}
				\hline\hline
				\textbf{activation} & \textbf{Boundary} & \textbf{1st. responsiveness} & \textbf{2nd. responsiveness} & \textbf{Final responsiveness}\\
				\hline
				$\delta(t)\delta(\mathbf{r})$ & 1D ridge edge & 1D line waves & Huygens waves  & Direct waves\\
				& 2D surface & 2D surface waves & genrated & and Huygens waves\\
				& & (3D) Direct waves & by 1D and 2D waves &\\
				\hline
		\end{tabular}}\\
		\subtable[Classification of waves based on dimensions.]{
			\begin{tabular}{llll}
				\hline\hline
				\textbf{Low dimensional waves} & \textbf{3D waves} & \textbf{Invisible precursors} & \textbf{Real wave-field}\\
				\hline
				1D line waves &  Direct waves & Low dimensional waves & 3D waves\\
				2D surface waves & Huygens waves &  &  \\
				\hline
		\end{tabular}}
		\caption{Relations between waves.}\label{21th-Mar.2024-T001}
	\end{table}
	\subsubsection{Huygens wave}\label{4.1.3}
\begin{figure}[h!]
		\centering
		\subfigure[Components of $\mathbf{u}_r^{Huygens}$.]{\includegraphics[scale=0.8]{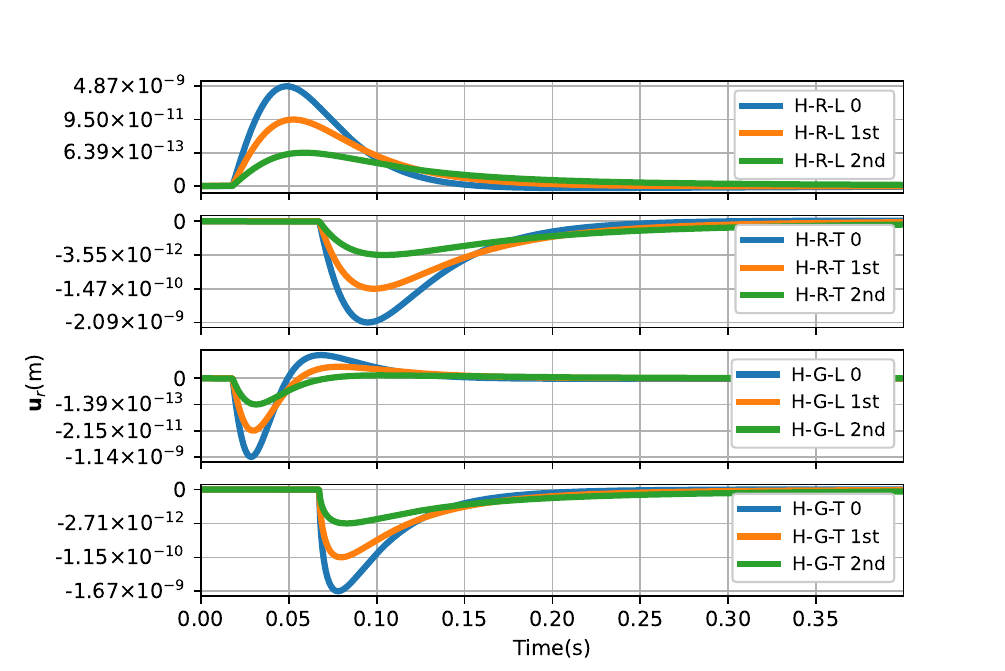}}
		\subfigure[Components of $\mathbf{u}_z^{Huygens}$.]{\includegraphics[scale=0.8]{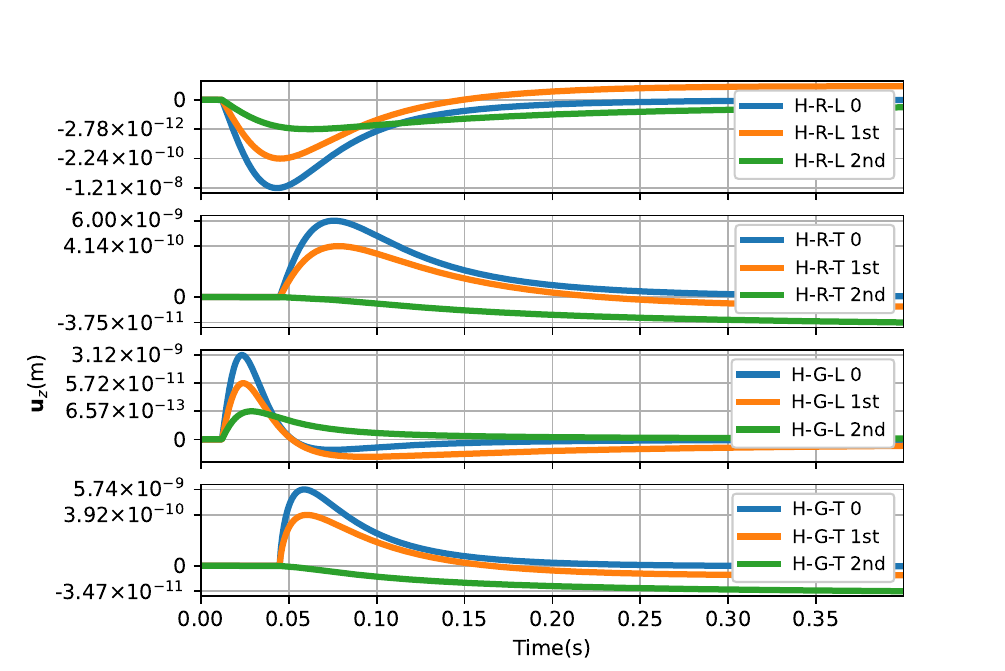}}
		\caption{Huygens waves in displacements.}\label{23th-Nov.-2023-P001}
	\end{figure}

\begin{figure}[h!]
		\centering
		\subfigure[Component-waves in $\mathbf{u}_r^{Huygens}$.]{\includegraphics[scale=0.8]{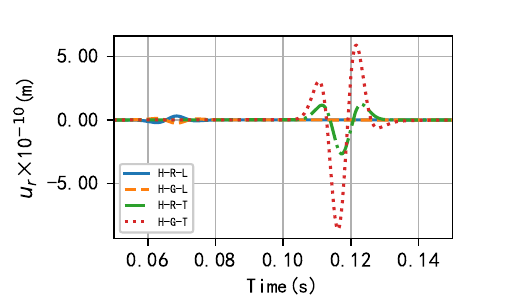}}
		\subfigure[Component-waves in $\mathbf{u}_z^{Huygens}$.]{\includegraphics[scale=0.8]{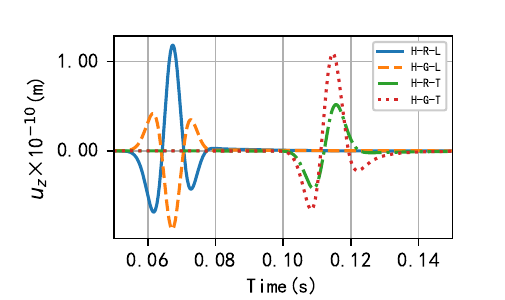}}
		\caption{Component-waves in displacements.}\label{23th-Jan.-2024-P001}
	\end{figure}

	In this article, the phenomenon where surface waves induce 3D waves in the medium is referred to as the Huygens effect. This term is chosen because this phenomenon aligns with the description of Huygens' principle, as evidenced by the expressions for $\mathscr{H}_{v}$. It is worth noting that Huygens' principle also applies to 3D waves, but treating it as the response to a point load yields no discernible difference. Therefore, the focus of this article is on waves generated by the Huygens effect of boundary waves.\\
	\\
	As explained in \ref{4.1.2}, waves generated by boundary conditions can be conceptualized as the vibration of the boundary itself. The mathematical description of this process is as follows,\\
	\begin{equation}
		\left<f_S(t,\mathbf{r})\mathbf{S}(p),f_D(t,\mathbf{R})\mathbf{D}_0(c)\right>,\label{1st-Apr.-2024-E004}
	\end{equation}
	where $f_S(t,\mathbf{r})\mathbf{S}(p)$ represents the surface wave, and $f_D(t,\mathbf{R})\mathbf{D}_0(c)$ is the body wave generated by this surface wave.\\
	\\
	In \ref{3.2.1}, Time-Green functions in common conditions include these kinds of Huygens waves:\\
	\begin{subequations}
		\begin{align}
			&\begin{aligned}
				&\mathbf{u}_r^{Huygens}=\mathbf{u}_r^{(c_R,c_L)H}+\mathbf{u}_r^{(c_R,c_T)H}+\sum\mathbf{u}_r^{(c(x),c_L)H}+\sum\mathbf{u}_r^{(c(x),c_T)H}\\
				&=\frac{1}{2\pi}\left(-N_1\mathbf{B}_r(c_R,c_L)+N_2\mathbf{B}_r(c_R,c_T)-C_1\int_1^{1/\sqrt{\eta}}W_L(x)\mathbf{B}_r(xc_T,c_L){\rm d}x+C_2\int_1^{1/\sqrt{\eta}}W_T(x)\mathbf{B}_r(xc_T,c_T){\rm d}x\right),\\
			\end{aligned}\\
			&\begin{aligned}
				&\mathbf{u}_z^{Huygens}=\mathbf{u}_z^{(c_R,c_L)H}+\mathbf{u}_z^{(c_R,c_T)H}+\sum\mathbf{u}_z^{(c(x),c_L)H}+\sum\mathbf{u}_z^{(c(x),c_T)H}\\
				&=-\frac{1}{2\pi}\left(N_1\mathbf{B}_z(c_R,c_L)+N_2\mathbf{B}_z(c_R,c_T)+C_1\int_1^{1/\sqrt{\eta}}W_L(x)\mathbf{B}_z(xc_T,c_L){\rm d}x+C_2\int_1^{1/\sqrt{\eta}}W_T(x)\mathbf{B}_z(xc_T,c_T){\rm d}x\right).\\
			\end{aligned}
		\end{align}
	\end{subequations}
	where $\mathbf{B}_r(p,c)$ and $\mathbf{B}_z(p,c)$ are the Huygens wave modes which are three dimensional vectors.\\
	\begin{subequations}
		\begin{align}
			&\begin{aligned}
				&\mathbf{B}_r(p,c)=\left[\left<\mathbf{S}(p),\frac{3zx_1}{R^4}\mathbf{D}_0(c)\right>,\left<\mathbf{S}(p),\frac{3zx_1}{cR^3}\mathbf{D}_0(c)\right>,\left<\mathbf{S}(p),\frac{zx_1}{c^2R^2}\mathbf{D}_0(c)\right>\right]\\
				&=\left[3z\mathscr{H}_1(;p,c;1,5),\frac{3z}{c}\mathscr{H}_1(;p,c;1,4),\frac{z}{c^2}\mathscr{H}_1(;p,c;1,3)\right],
			\end{aligned}\\
			&\begin{aligned}
				&\mathbf{B}_z(p,c_L)=\left[\left<\mathbf{S}(p),\left(\frac{3z^2}{R^4}-\frac{1}{R^2}\right)\mathbf{D}_0(c_L)\right>,\left<\mathbf{S}(p),\left(\frac{3z^2}{c_LR^3}-\frac{1}{c_LR}\right)\mathbf{D}_0(c_L)\right>,\left<\mathbf{S}(p),\frac{z^2}{c_L^2R^2}\mathbf{D}_0(c_L)\right>\right]\\
				&=\left[\left(3z^2\mathscr{H}_0(;p,c_L;0,5)-\mathscr{H}_0(;p,c_L;0,3)\right),\frac{1}{c_L}\left(3z^2\mathscr{H}_0(;p,c_L;0,4)-\mathscr{H}_0(;p,c_L;0,2)\right),\frac{z^2}{c_L^2}\mathscr{H}_0(;p,c_L;0,3)\right],\\	
			\end{aligned}\\
			&\begin{aligned}
				&\mathbf{B}_z(p,c_T)=\left[\left<\mathbf{S}(p),\left(\frac{3r^2}{R^4}-\frac{2}{R^2}\right)\mathbf{D}_0(c_T)\right>,\left<\mathbf{S}(p),\left(\frac{3r^2}{c_TR^3}-\frac{2}{c_TR}\right)\mathbf{D}_0(c_T)\right>,\left<\mathbf{S}(p),\frac{r^2}{c_T^2R^2}\mathbf{D}_0(c_T)\right>\right]\\
				&=\left[\left(3\mathscr{H}_0(;p,c_T;2,5)-2\mathscr{H}_0(;p,c_T;0,3)\right),\frac{1}{c_T}\left(3\mathscr{H}_0(;p,c_T;2,4)-2\mathscr{H}_0(;p,c_T;0,2)\right),\frac{1}{c_T^2}\mathscr{H}_0(;p,c_T;2,3)\right].
			\end{aligned}
		\end{align}
	\end{subequations}
	The types of direct waves are depended on medium. In isotropic medium, there are two types, longitudinal wave and transverse wave. As shown in \ref{22th-Mar.-2024-P001}, Huygens wave propagates in the 3D domain with velocity $c$ generated by surface waves with velocity $p$ is called as \emph{$(p,c)$-Huygens wave} (H-p-c). $\mathbf{S}(p)$ comes from the distributaion terms $L$ and $T$ and $\mathbf{D}_0(c)$ comes from the propagation terms $P_L$ and $P_T$. In fact, due to the continuity in time domain of surface waves, the secondary point soureces are distributed in a circle and moving all the time.\\
	\begin{figure}[h!]
		\centering
		\begin{tikzpicture}
			\draw[-, line width=3pt, draw = brown](-8,0) --(8,0);
			
			\node[above] at (0,0) {Source (also secondary)};
			\node[above] at (-7,0) {$\mathbf{S}(p)$};
			\node[above] at (7,0) {$\mathbf{S}(p)$};
			\node[above] at (-4,0) {Secondary source};
			\node[above] at (4,0) {Secondary source};
			\node[above] at (2,-1) {$\mathbf{D}_0(c)$};
			\draw[->,dash dot,draw=blue,line width=3pt] (0,0)--(-7,0);
			\draw[->,dash dot,draw=blue,line width=3pt] (0,0)--(7,0);
			\filldraw[fill=black] (0,0) circle (3pt);
			\filldraw[fill=red] (-3,0) circle (3pt);
			\filldraw[fill=red] (-5,0) circle (3pt);
			\filldraw[fill=red] (3,0) circle (3pt);
			\filldraw[fill=red] (5,0) circle (3pt);
			\draw[red,line width=3pt] (-0.9,0) arc(-180:0:0.9);
			\draw[red,line width=3pt] (-0.6,0) arc(-180:0:0.6);
			\draw[red,line width=3pt] (-0.3,0) arc(-180:0:0.3);
			\draw[red,line width=3pt] (-3.6,0) arc(-180:-80:0.6);
			\draw[red,line width=3pt] (-3.3,0) arc(-180:-80:0.3);
			\draw[red,line width=3pt] (-5.3,0) arc(-180:-80:0.3);
			\draw[red,line width=3pt] (3.6,0) arc(0:-100:0.6);
			\draw[red,line width=3pt] (3.3,0) arc(0:-100:0.3);
			\draw[red,line width=3pt] (5.3,0) arc(0:-100:0.3);
		\end{tikzpicture}
		\caption{Huygens effect generated by surface wave.}\label{22th-Mar.-2024-P001}
	\end{figure}
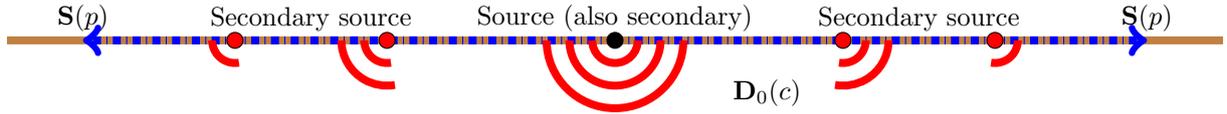\\
	In \ref{23th-Nov.-2023-P001}, four kinds of Huygens waves with different generated mechanisms are shown and the ordinate of the figures are scaled to have a better perception.\\
	\\
	For a clearer grasp of the Huygens effect, let's draw an analogy with a scenario involving a person and a submarine in the sea. Imagine a man who can only swim on the water's surface and a submarine that can navigate both above and below the sea. The man's swimming speed represents the velocity of the surface wave, while the submarine's speed symbolizes that of the body wave. Consequently, we often observe Huygens waves propagating at the speed of body waves, aligning with the least time principle (Fermat's principle). However, certain circumstances, such as wave-fields on the surface and the {\it variation} phenomenon discussed in \ref{22th-Sept.-2023-E002} (to be elucidated in \ref{4.2.1}), can complicate the propagation of Huygens waves. In \ref{23th-Jan.-2024-P001}, a comprehensive illustration of each component-wave (H-R-L, H-R-T, H-G-L, H-G-T) in displacements at $r=20$ m and $z=-50$ m under the influence of a Ricker wavelet is presented in a single image.\\
	\\
	On the surface, the Huygens effect vanishes in radial displacement, leaving only $(p,c_L)$-Huygens waves in axial displacement. Drawing from the analogy above, it is akin to the man's submarine being rendered inoperable. In radial displacement, both the $c_L$ and $c_T$ submarines malfunction, forcing the man to swim unaided. This elucidates why surface waves manifest directly in the real wave-field. In \ref{3.2.2}, the issue of degeneracy of weak function (\ref{28th-Sept.-2023-E002}) can be likened to a flat tire, where the propagation terms degrade to a point.\\
	\begin{equation}
		\lim_{z\to0}\frac{\partial}{\partial z}P=\delta(t)\delta(\mathbf{r}).\notag
	\end{equation}
	For axial displacement, as elucidated in \ref{30th-Oct.-2023-E001}, only $P_T$ is present in the expressions.\\
	\subsection{Temporal analysis to waves}\label{4.2}
	In this section, we analyze the arrival time (\ref{4.2.1}), over-peak time (\ref{4.2.2}), and the end time (\ref{4.2.3}) of each component-wave.\\
	\subsubsection{Wave front and Fermat principle}\label{4.2.1}
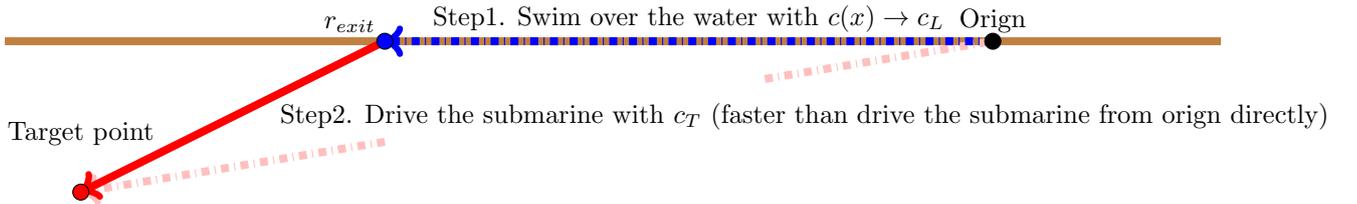
\begin{figure}[h!]
		\centering
		\begin{tikzpicture}
			\draw[-, line width=3pt, draw = brown](-8,0) --(8,0);
			
			\node[above] at (5,0) {Orign};
			\node[above] at (1,0) {Step1. Swim over the water with $c(x)\to c_L$};
			\node[above left] at (-3,0) {$r_{exit}$};
			\draw[dash dot,draw=pink,line width=3pt] (2,-1/2)--(5,0);
			\draw[->,dash dot,draw=pink,line width=3pt] (-3,-4/3)--(-7,-2);
			\node[right] at (-4.5,-1) {Step2. Drive the submarine with $c_T$ (faster than drive the submarine from orign directly)};
			\draw[->,dash dot,draw=blue,line width=3pt] (5,0)--(-3,0);
			\draw[->,draw=red,line width=3pt] (-3,0)--(-7,-2);
			\node[above] at (-7,-1.5) {Target point};
			\filldraw[fill=black] (5,0) circle (3pt);
			\filldraw[fill=blue] (-3,0) circle (3pt);
			\filldraw[fill=red] (-7,-2) circle (3pt);
		\end{tikzpicture}
		\caption{Maximum speed program to the target.}\label{22th-Mar.-2024-P002}
	\end{figure}

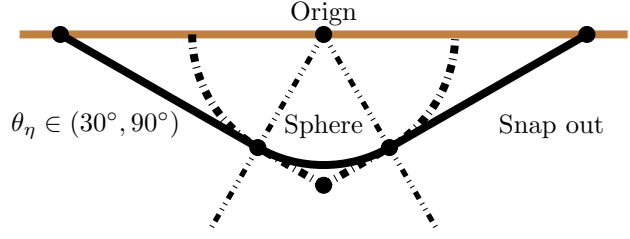
\begin{figure}[h!]
		\centering
		\begin{tikzpicture}
			\draw[-, line width=3pt, draw = brown](-4,0) --(4,0);
			
			\node[above] at (0,0) {Orign};
			\node[above] at (-3,-1.5) {$\theta_{\eta}\in(30^{\circ},90^{\circ})$};
			\node[above] at (3,-1.5) {Snap out};
			\node[above] at (0,-1.5) {Sphere};
			\draw[-, line width=3pt](-3.464,0) --(-0.866,-1.5);
			\draw[-, line width=3pt](3.464,0) --(0.866,-1.5);
			\draw[dash dot, line width=3pt](0,-2) --(-0.866,-1.5);
			\draw[dash dot, line width=3pt](0,-2) --(0.866,-1.5);
			\draw[line width=3pt] (-0.866,-1.5) arc(-120:-60:1.73);
			\draw[dash dot, line width=3pt] (-1.73,0) arc(-180:-120:1.73);
			\draw[dash dot, line width=3pt] (0.866,-1.5) arc(-60:0:1.73);
			\draw[dash dot dot, line width=2pt](0,0) --(-1.5,-2.6);
			\draw[dash dot dot, line width=2pt](0,0) --(1.5,-2.6);
			\filldraw[fill=black] (0,0) circle (3pt);
			\filldraw[fill=black] (-3.464,0) circle (3pt);
			\filldraw[fill=black] (3.464,0) circle (3pt);
			\filldraw[fill=black] (-0.866,-1.5) circle (3pt);
			\filldraw[fill=black] (0.866,-1.5) circle (3pt);
			\filldraw[fill=black] (0,-2) circle (3pt);
		\end{tikzpicture}
		\caption{The wave front of H-G-T.}\label{22th-Mar.-2024-P003}
	\end{figure}

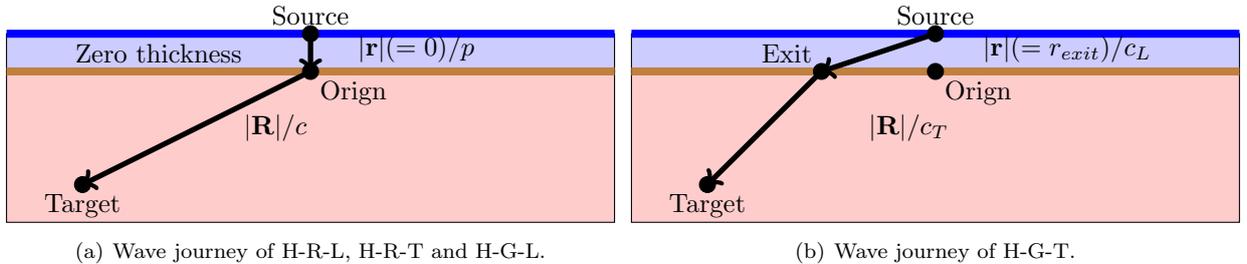
\begin{figure}[h!]
		\centering
		\subfigure[Wave journey of H-R-L, H-R-T and H-G-L.]{\begin{tikzpicture}
				\filldraw[fill=blue!20] (-4,0) --(4,0)--(4,0.5) --(-4,0.5)--(-4,0);
				\filldraw[fill=red!20] (-4,0) --(4,0)--(4,-2) --(-4,-2)--(-4,0);
				\draw[-, line width=3pt, draw = brown](-4,0) --(4,0);
				\draw[-, line width=3pt, draw = blue](-4,0.5) --(4,0.5);
				\node[below right] at (0,0) {Orign};
				\node[above] at (0,0.5) {Source};
				\node[above] at (-2,0) {Zero thickness};
				\node[right] at (-1,-0.75) {$|\mathbf{R}|/c$};
				\node[below] at (-3,-1.5) {Target};
				\node[above right] at (0.5,0) {$|\mathbf{r}|(=0)/p$};
				\filldraw[fill=black] (0,0) circle (3pt);
				\filldraw[fill=black] (0,0.5) circle (3pt);
				\filldraw[fill=black] (-3,-1.5) circle (3pt);
				\draw[->, line width=2pt](0,0.5) --(0,0);
				\draw[->, line width=2pt](0,0) --(-3,-1.5);
		\end{tikzpicture}}
		\subfigure[Wave journey of H-G-T.]{\begin{tikzpicture}
				\filldraw[fill=blue!20] (-4,0) --(4,0)--(4,0.5) --(-4,0.5)--(-4,0);
				\filldraw[fill=red!20] (-4,0) --(4,0)--(4,-2) --(-4,-2)--(-4,0);
				\draw[-, line width=3pt, draw = brown](-4,0) --(4,0);
				\draw[-, line width=3pt, draw = blue](-4,0.5) --(4,0.5);
				\node[below right] at (0,0) {Orign};
				\node[above] at (0,0.5) {Source};
				\node[above left] at (-1.5,0) {Exit};
				\node[right] at (-1,-0.75) {$|\mathbf{R}|/c_T$};
				\node[below] at (-3,-1.5) {Target};
				\node[above right] at (0.5,0) {$|\mathbf{r}|(=r_{exit})/c_L$};
				\filldraw[fill=black] (0,0) circle (3pt);
				\filldraw[fill=black] (-1.5,0) circle (3pt);
				\filldraw[fill=black] (0,0.5) circle (3pt);
				\filldraw[fill=black] (-3,-1.5) circle (3pt);
				\draw[->, line width=2pt](0,0.5) --(-1.5,0);
				\draw[->, line width=2pt](-1.5,0) --(-3,-1.5);
		\end{tikzpicture}}
		\caption{Optical explanation of Huygens waves.}\label{22th-Mar.-2024-P004}
	\end{figure}

	The arrival time refers to the moment when a point in space begins to move from a state of rest, and it is solely determined by spatial coordinates.\\
	\begin{equation}
		t_{arrival}(\mathbf{x}):=\inf_{\mathbf{u}(\tau,\mathbf{x})\neq0}\tau,\quad\mathbf{x}\in\mathbb{R}^2\times(-\infty,0).\label{15th-Nov.-2023-E002}
	\end{equation}
	When the $t_{arrival}(\mathbf{x})=\text{const.}$, it represents a surface $\Sigma_{W.F.}(t)$ in space which is the so called wave front at time equal this constant.\\
	\begin{equation}
		\Sigma_{W.F.}(t):t_{arrival}(\mathbf{x})=t.\label{6th.-Jan.-2024-E001}
	\end{equation}
	The wave front, also known as the head wave, forms a closed surface in space together with the boundary. In this section, the arrival times and wave fronts of various types of waves are discussed, including direct longitudinal waves (D-L), direct transverse waves (D-T), direct algebraic waves (D-A), Rayleigh waves (S-R), surface wave groups (S-G), and Huygens waves generated by four different mechanisms: $(c_R,c_L)$, $(c_R,c_T)$, $(c(x),c_L)$, and $(c(x),c_T)$ (H-R-L, H-R-T, H-G-L, H-G-T). The variation (mentioned in \ref{3.1.3}) phenomenon only occurs in the $(c(x),c_T)$-Huygens wave (H-G-T), \\
	\begin{subequations}
		\begin{align}
			&t^{D-L}_{arrival}=t^{D-A}_{arrival}=t^{H-R-L}_{arrival}=t^{H-G-L}_{arrival}=\frac{R}{c_L},\label{15th-Nov.-2023-E001a}\\
			&t^{D-T}_{arrival}=t^{H-R-T}_{arrival}=\frac{R}{c_T},\label{15th-Nov.-2023-E001b}\\
			&t^{H-G-T}_{arrival}=\left\{\begin{array}{ll}
				\frac{R}{c_T}&,\quad r/R\leq \sqrt{\eta}\\
				\frac{1}{c_L}\left(r+|z|\sqrt{\frac{1}{\eta}-1}\right)&,\quad r/R>\sqrt{\eta}\\
			\end{array}\right.\label{15th-Nov.-2023-E001c},\\
			&t^{S-R}_{arrival}=\frac{r}{c_R}\label{15th-Nov.-2023-E001d},\\
			&t^{S-G}_{arrival}=\frac{r}{c_L}\label{15th-Nov.-2023-E001e}.
		\end{align}
	\end{subequations}
	The wave fronts of \ref{15th-Nov.-2023-E001a} and \ref{15th-Nov.-2023-E001b} are common spherical waves, while the wave front of \ref{15th-Nov.-2023-E001d} is a circular wave, \\
	\begin{subequations}
		\begin{align}
			&\Sigma_{W.F.}^{D-L,D-A,H-R-L,H-G-L}(t):R=c_Lt\label{6th.-Jan.-2024-E003a},\\
			&\Sigma_{W.F.}^{D-T,H-R-T}(t):R=c_Tt\label{6th.-Jan.-2024-E003b},\\
			&\Sigma_{W.F.}^{S-R}(t):r=c_Rt.
		\end{align}	
	\end{subequations}
	Here, \ref{15th-Nov.-2023-E001c} and \ref{15th-Nov.-2023-E001e} will be analyzed separately. \\
	\\
	$(c(x),c_T)$-Huygens waves are generated by the surface wave group. Every wave in the group has a faster velocity than the direct-transverse wave, which means the group of men swim faster than the submarine. As shown in \ref{22th-Mar.-2024-P002}, in some specific domain, the fastest plan to reach the point from the origin is not for the man to drive the $c_T$ submarine from the origin to the target, but for the man to swim over the sea with a velocity infinitely close to $c_L$ firstly, and then drive the $c_T$ submarine to the point.\\
	\\
	So the wave front is a part of a taper whose bottom is on the surface,\\
	\begin{equation}
		\Sigma_{W.F.}^{H-G-T}(t):r+|z|\sqrt{\frac{1}{\eta}-1}=c_Lt,\notag
	\end{equation}
	in domain:\\
	\begin{equation}
		\{\mathbf{x}|r/R>\sqrt{\eta}\}\notag=\text{Half-space}-\text{Cone with origin as top point and $\sqrt{\eta}$ as gradient}.
	\end{equation}
	In the other domain, the wave front is still a part of sphere.\\
	\begin{equation}
		\Sigma_{W.F.}^{H-G-T}(t):\left\{\begin{array}{ll}
			R=c_Tt&,\quad r/R\leq \sqrt{\eta}\\
			r+|z|\sqrt{\frac{1}{\eta}-1}=c_Lt&,\quad r/R>\sqrt{\eta}\\
		\end{array}\right.\label{6th.-Jan.-2024-E004}.
	\end{equation}
	The wave front of $(c(x),c_T)$-Huygens wave resembles a part of a taper combined with a part of a sphere, resembling the surface of a shuttlecock (as shown in \ref{22th-Mar.-2024-P003}). The generatrix of the taper is tangent to the sphere. $(c(x),c_T)$-Huygens wave is faster than the common transverse wave, a phenomenon referred to as variation in this article.\\ 
	\\
	The exit point of an exact point of coordinate $(r,z,\theta)$ can be comupted by\\
	\begin{equation}
		f(x):=x-r+\sqrt{\frac{(r-x)^2+z^2}{\eta}}-|z|\sqrt{\frac{1}{\eta}-1}=0,\quad 0<x<r
	\end{equation}
	and the suitable solution is\\
	\begin{equation}
		r_{exit}:=r-\frac{\sqrt{\eta}}{\sqrt{2}+\sqrt{\eta+1}}|z|.
	\end{equation}
	Obviously, $\theta_{exit}=\theta$ and the coordinate of the exit point is $(r_{exit},0,\theta)$.\\
	\\
	Finally, let's discuss the surface wave group. As indicated in \ref{4.1.2}, there is no component of longitudinal wave in the wave group, but there exists a wave with a velocity infinitely close to $c_L$. To define the arrival time under this condition, the infimum is utilized in \ref{15th-Nov.-2023-E002}.\\
	\begin{equation}
		t^{S-G}_{arrival}(\mathbf{x})=\inf_{{\rm H}(\tau-r/c(x))\neq0}\tau=\sup_{c(x)\in(c_T,\sqrt{2}c_T)\cup(\sqrt{2}c_T,c_L)}\frac{r}{c(x)}=\frac{r}{c_L}\notag.
	\end{equation}
	Different from D-L, D-T, D-A, H-R-L, H-R-T, H-G-L, and S-R, which have solid wave fronts indicating waves occurring at $t \geq t_{\text{arrival}}$, H-G-T and S-G have hollow wave fronts (or a part of), indicating that waves do not really propagate at $t = t_{\text{arrival}}$. \\
	\\
	In conclusion, the phenomenon of wave fronts described above can be explained by Fermat's principle, which states that waves always propagate in a way that minimizes the travel time. As shown in \ref{22th-Mar.-2024-P004}, viewing the surface as a very thin medium-1, different from the body medium-2, the Huygens effect can be seen as the refraction between medium-1 and medium-2. The velocities of these two media can be viewed as speeds of light in different media. The wave always seeks to travel more of its journey in the fast medium when the start and endpoint are fixed. That is why the wave fronts of H-R-L, H-R-T, and H-G-L have a straight-line wave journey, while H-G-T's has a polyline wave journey. \\
	\\
	Thus, we can also define the $t_{arrival}$ of wave front by Fermat principle.\\
	\begin{equation}
		t_{arrival}^{H-p-c}(\mathbf{x}):=\min\left(\frac{|\mathbf{r}|}{p}+\frac{|\mathbf{R}|}{c}\right),\quad\mathbf{r}\in Surface,\quad\mathbf{R}\in Body.
	\end{equation} 
	Although acoustics and optics are distinct disciplines, the notion of wave fronts in acoustics bears resemblance to the behavior of light in optics. This parallel prompts us to explore further similarities between these two fields.\\
	\subsubsection{Over-peak time and peak value of waves}\label{4.2.2}
\begin{figure}[h!]
		\centering
		\subfigure[$u_r^{(c_R,c_L)H}$.]{\includegraphics[scale=0.8]{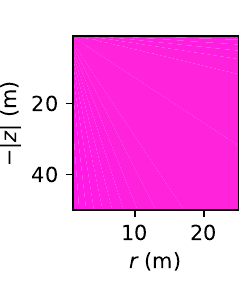}}
		\subfigure[$u_r^{(c_R,c_T)H}$.]{\includegraphics[scale=0.8]{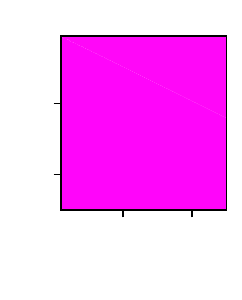}}
		\subfigure[$u_r^{(c(x),c_L)H}$.]{\includegraphics[scale=0.8]{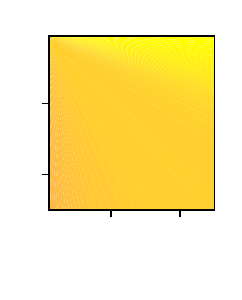}}
		\subfigure[$u_r^{(c(x),c_T)H}$.]{\includegraphics[scale=0.8]{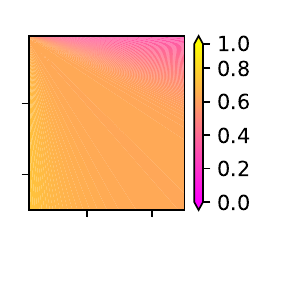}}\\
		\subfigure[$u_z^{(c_R,c_L)H}$.]{\includegraphics[scale=0.8]{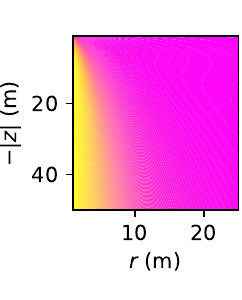}}
		\subfigure[$u_z^{(c_R,c_T)H}$.]{\includegraphics[scale=0.8]{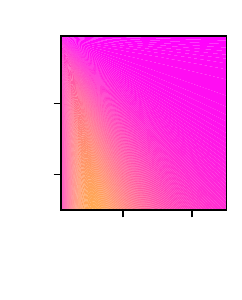}}
		\subfigure[$u_z^{(c(x),c_L)H}$.]{\includegraphics[scale=0.8]{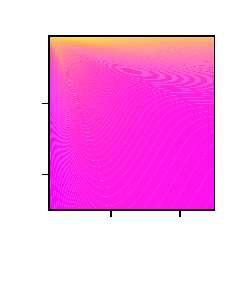}}
		\subfigure[$u_z^{(c(x),c_T)H}$.]{\includegraphics[scale=0.8]{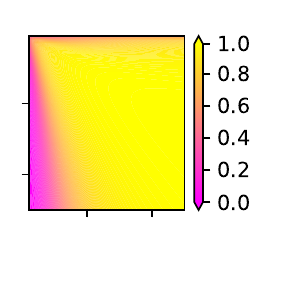}}
		\caption{The distribuation of Huygens waves with different generation mechanisms in displacements.}\label{10th.-Jan.-2024-P001}
	\end{figure}

	As defined in \ref{3.1.3}, the over-peak time refers to the moment when the peak of $\delta(t-R/c)$ intersects with the surface defined by ${\rm H}(t-r/p)$. This concept is particularly valuable when discussing Huygens waves.\\
	\begin{equation}
		t_{peak}(p,c)=\frac{r}{p}+\frac{|z|}{c}.
		\notag
	\end{equation}
	However, achieving the over-peak time does not imply that the wave reaches its maximum amplitude. Determining the exact time of peak amplitude is more complex than identifying $t_{peak}$. This article selects $t_{peak}$ as a marker to signify the peak value, as no other quantity has a clear association with the peak amplitude. $t_{peak}$ serves as a precursor for predicting the arrival of the peak value. The amplitudes of waves at $t_{peak}$ serve as a criterion for determining the primary focus of this section. For the physical quantity $Q$ in this context, four types of Huygens waves ($Q_{H-R-L}$, $Q_{H-R-T}$, $Q_{H-G-L}$, and $Q_{H-G-T}$) are involved. Utilizing the metric outlined below, we will analyze each component of Huygens waves in displacements as they vary with spatial coordinates.\\
	\begin{subequations}
		\begin{align}
			\mathbf{M}(Q)=\left(\frac{|Q_{H-R-L}|}{Q_H},\frac{|Q_{H-R-T}|}{Q_H},\frac{|Q_{H-G-L}|}{Q_H},\frac{|Q_{H-G-T}|}{Q_H}\right),
		\end{align}
	\end{subequations}
	In this section, we delve into $\mathbf{M}(G^{(T)}_{2-r})$ and $\mathbf{M}(G^{(T)}_{2-z})$ in detail. The precise expressions for each Huygens wave mode in $G^{(T)}_{2-r}$ and $G^{(T)}_{2-z}$ are provided in \ref{7th.-Dec.-2023-T001}. The component values at $t=t_{peak}$ are utilized to represent the amplitude of each mode. We define $Q_H=\sqrt{Q_{H-R-L}^2+Q_{H-R-T}^2+Q_{H-G-L}^2+Q_{H-G-T}^2}$.\\
	\begin{table}[h!]
		\begin{center}
			\begin{tabular}{c|cc}
				\hline\hline\noalign{\smallskip}
				& $G_{2-r}$ & $G_{2-z}$\\ 
				\hline
				\textbf{H-R-L} & $-\frac{N_1z}{2\pi c_L^2}\mathscr{H}_1(t_{peak};ac_T,c_L;1,3)$ & $-\frac{N_1z^2}{2\pi c_L^2}\mathscr{H}_0(t_{peak};ac_T,c_L;0,3)$\\
				\textbf{H-R-T} & $\frac{N_2z}{2\pi c_T^2}\mathscr{H}_1(t_{peak};ac_T,c_T;1,3)$ & $-\frac{N_2}{2\pi c_T^2}\mathscr{H}_0(t_{peak};ac_T,c_T;2,3)$\\
				\textbf{H-G-L} & $-\frac{C_1z}{2\pi c_L^2}\int_{1}^{1/\sqrt{\eta}}W_L(x)\mathscr{H}_1(t_{peak};xc_T,c_L;1,3){\rm d}x$ & $-\frac{C_1z^2}{2\pi c_L^2}\int_{1}^{1/\sqrt{\eta}}W_L(x)\mathscr{H}_0(t_{peak};xc_T,c_L;0,3){\rm d}x$\\
				\textbf{H-G-T} & $\frac{C_2z}{2\pi c_T^2}\int_{1}^{1/\sqrt{\eta}}W_T(x)\mathscr{H}_1(t_{peak};xc_T,c_T;1,3){\rm d}x$ & $-\frac{C_2}{2\pi c_T^2}\int_{1}^{1/\sqrt{\eta}}W_T(x)\mathscr{H}_0(t_{peak};xc_T,c_T;2,3){\rm d}x$\\
				\noalign{\smallskip}\hline
			\end{tabular}\caption{Expressions of $\mathbf{M}(G^{(T)}_{2-r})$ and $\mathbf{M}(G^{(T)}_{2-z})$.}\label{7th.-Dec.-2023-T001}
		\end{center}
	\end{table}\\
	In \ref{10th.-Jan.-2024-P001}, the contribution of each type of Huygens wave to $u_r$ and $u_z$ is illustrated. In $u_r$, the Huygens waves generated by the wave group consistently dominate, particularly in regions where $r/R>\sqrt{\eta}$. Notably, H-G-L surpasses H-G-T significantly and becomes the predominant component of displacement. The amplitude of H-G-T decreases sharply due to the variation phenomenon. In $u_z$, the situation is more intricate. Each Huygens wave component plays a crucial role under specific conditions. Near the axis with increasing depth, H-R-L emerges as the primary component. Near the surface, H-R-T takes precedence. In a narrow range of angles near the axis, H-G-L dominates. Across most of the area, H-G-T emerges as the primary contributor.\\
	\\
	\subsubsection{Attenuation and end of fluctuation with time}\label{4.2.3}
\begin{figure}[h!]
		\centering
		\subfigure[2nd-order term of H-R-T]{\includegraphics[scale=0.8]{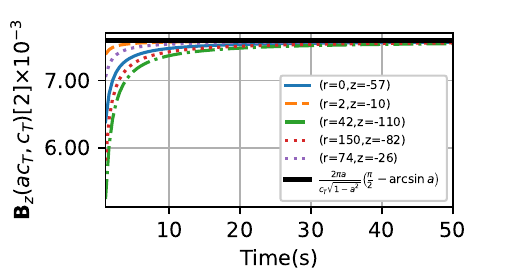}}
		\subfigure[2nd-order term of H-G-T]{\includegraphics[scale=0.8]{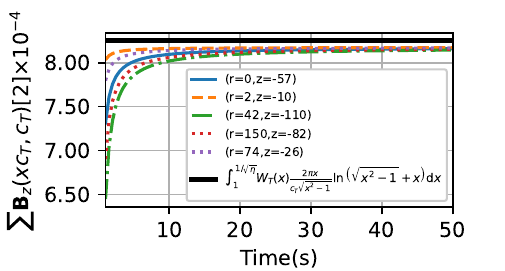}}
		\caption{The attenuation of $\mathbf{B}_z(p,c_T)[2]$.}\label{9th.-Jan.-2024-P005}
	\end{figure}

	In this section, we discuss about how does each wave vary with lapse. If a wave ends in a finite time $t<t_{end}$, the wave is said to have coda.\\
	\begin{equation}
		t_{end}(\mathbf{x}):=\sup_{\mathbf{u}(\tau,\mathbf{x})\neq0}\tau<\infty,\quad\mathbf{x}\in\mathbb{R}^2\times(-\infty,0).
	\end{equation}
	In this study, it is observed that all direct waves (D-L, D-T, D-A) exhibit codas, whereas surface waves (S-R, S-G) and Huygens waves (H-R-L, H-R-T, H-G-L, H-G-T) do not. Waves without codas attenuate over time. The rate of attenuation is significant for detection purposes, as waves with lower attenuation rates are easier to detect. (The temporal decay quality typically correlates with spatial decay under a fixed wave mode.)\\
	\\
	The $t_{end}$ of D-L and D-T coincides with their $t_{arrival}$ since they manifest as imimpulses. D-A has a finite duration. Interestingly, these waves also have coda fronts similar to the definition of wave fronts, appearing as spherical shapes.\\
	\begin{equation}
		t^{D-L}_{end}=\frac{R}{c_L}\left(=t^{D-L}_{arrival}\right),\quad t^{D-T}_{end}=\frac{R}{c_T}\left(=t^{D-T}_{arrival}\right),\quad t^{D-A}_{end}=\frac{R}{c_T}\quad\left(t^{D-A}_{arrival}=\frac{R}{c_L}\right).
	\end{equation}
	The surface wave attenuates with the speed $1/t$.\\
	\begin{equation}
		t\to\infty,\quad\mathbf{S}(p)=\frac{{\rm H}(t-r/p)}{\sqrt{t^2-r^2/p^2}}\sim\frac{1}{t}.
	\end{equation}
	The attenuation of Huygens waves depends on the temporal change of $\mathscr{H}_0$ and $\mathscr{H}_1$. The convergence of the components of $\mathbf{B}_r(p,c)$ and $\mathbf{B}_z(p,c)$ as $t$ approaches infinity is contingent upon the following inequality..\\
	\begin{equation}
		\begin{aligned}
			t\to\infty,\quad \mathscr{H}_v(t+r/p,0,z;p,c;m,n)-O(1/t^{-m+n-1-v})&<\mathscr{H}_v(t,r,z;p,c;m,n)\\
			&<\mathscr{H}_v(t-r/p,0,z;p,c;m,n)+O(1/t^{-m+n-1-v}).
		\end{aligned}
		\label{26th.-Dec.-2023-E001}
	\end{equation}
	The \ref{26th.-Dec.-2023-E001} highlights an interesting observation: within any plane at a constant depth $z$, the fluctuation of every point will be constrained to an "acoustical taper". The limits of the right and left terms in \ref{26th.-Dec.-2023-E001} with specific parameters can be computed using the squeeze theorem.\\
	\begin{equation}
		\lim_{t\to\infty}\mathscr{H}_v(t-r/p,0,z;p,c;m,n)=\lim_{t\to\infty}\mathscr{H}_v(t-r/p,0,z;p,c;m,n)=\lim_{t\to\infty}\mathscr{H}_v(t,0,z;p,c;m,n).
	\end{equation}
	The situation of $v=0$ is given in \ref{C.3} and the situation of $v=1$ can be easily computed.\\
	\begin{equation}
		\lim_{t\to\infty}\mathscr{H}_1(t+r/p,0,z;p,c;m,n)=\lim_{t\to\infty}\mathscr{H}_1(t-r/p,0,z;p,c;m,n)=0.
	\end{equation}
	In \ref{2nd.-Dec.-2023-T001}, the limit of each component in Huygens wave mode and the attenuation speed are given, where $\mathbf{B}(p,c)[n]$ represents the n-th component of Huygens wave mode vector $\mathbf{B}(p,c)$ and the index $n$ is from 0 to 2.\\
	\begin{table}[h!]
		\begin{center}
			\begin{tabular}{cll}
				\hline\hline\noalign{\smallskip}
				\textbf{Components of wave modes} & \textbf{Limit} & \textbf{Attenuation speed}\\
				\hline
				$\mathbf{B}_r(p,c)$[0] & 0 & $t^{-1}$\\
				$\mathbf{B}_r(p,c)$[1] & 0 & $t^{-2}$\\
				$\mathbf{B}_r(p,c)$[2] & 0 & $t^{-3}$\\
				$\mathbf{B}_z(p,c_L)$[0] & 0 & $t^{-2}\ln t$\\
				$\mathbf{B}_z(p,c_L)$[1] & 0 & $t^{-1}\ln t$\\
				$\mathbf{B}_z(p,c_L)$[2] & 0 & $t^{-1}$\\
				$\mathbf{B}_z(p,c_T)$[0] & 0 & $t^{-2}\ln t$\\
				$\mathbf{B}_z(p,c_T)$[1] & 0 & $t^{-1}\ln t$\\
				$\mathbf{B}_z(ac_T,c_T)$[2] & $\frac{2\pi a}{c_T\sqrt{1-a^2}}\left(\frac{\pi}{2}-\arcsin a\right)$ & $t^{-1}\ln t$\\
				$\mathbf{B}_z(c(x),c_T)$[2] & $\frac{2\pi x}{c_T\sqrt{x^2-1}}\ln\left(\sqrt{x^2-1}+x\right)$ & $t^{-1}\ln t$\\
				\noalign{\smallskip}\hline
			\end{tabular}\caption{Convergence of Huygens waves in displacements when $t\to\infty$.}\label{2nd.-Dec.-2023-T001}
		\end{center}
	\end{table}\\
	In \ref{2nd.-Dec.-2023-T001}, $\mathbf{B}_z(p,c_T)[2]$ does not attenuate to zero as $t\to \infty$, which is consistent with the behavior expected in Time-Green functions. However, after convolutions with a source satisfying $\lim_{t\to\infty}F(t)=0$, the displacements still attenuate to zero.\\
	\begin{equation}
		\begin{aligned}
			\lim_{t\to\infty}\mathbf{B}_z(p,c_T)[2]*F^{(n)}(t)=\lim_{t\to\infty}\left(\mathbf{B}_z(p,c_T)[2]-Const.\right)*F^{(2)}(t)+\lim_{t\to\infty}Const.*F^{(2)}(t)=0,
		\end{aligned}
		\notag
	\end{equation}
	Another noteworthy observation is that the non-zero limit of $\mathbf{B}_z(p,c_T)[2]$ is a constant independent of $r$ and $z$. \ref{9th.-Jan.-2024-P005} shows the attenuation of $\mathbf{B}_z(ac_T,c_T)$[2] and $\mathbf{B}_z(c(x),c_T)$[2].\\
	\\
	\subsection{Spatial distribution of displacements}\label{4.3}
	In this section, the displacements activated by the Ricker wavelet are discussed. Specifically, the attenuation of the axial displacement $u_z$ with respect to depth $z$ and diameter $r$ are analyzed in \ref{4.3.1}, while the approximations of these displacements using elementary functions are detailed in \ref{4.3.2}. Additionally, the discussions regarding the radial displacement $u_r$ varying with diameter $r$ are presented in \ref{4.3.3}.\\
	\subsubsection{Axial displacement underground varying with depth and diameter}\label{4.3.1}
\begin{figure}[h!]
		\centering
		\subfigure[D-L part of $u_z$.]{\includegraphics[scale=0.8]{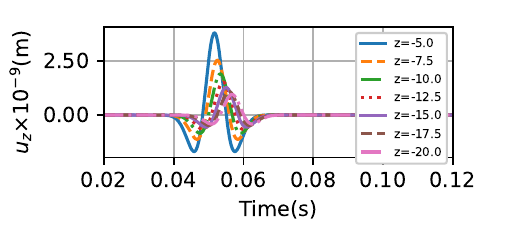}}
		\subfigure[Maximum of D-L varies with z.]{\includegraphics[scale=0.8]{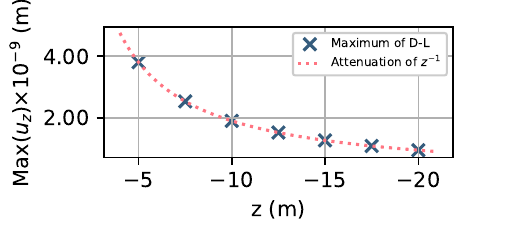}}\\
		\subfigure[D-A part of $u_z$.]{\includegraphics[scale=0.8]{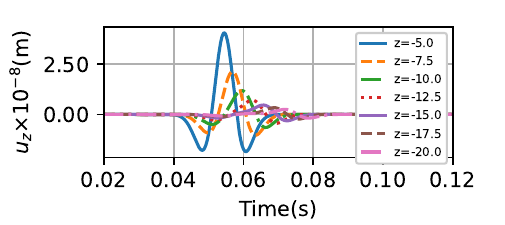}}
		\subfigure[Maximum of D-A varies with z.]{\includegraphics[scale=0.8]{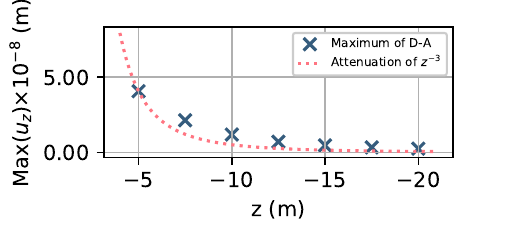}}\\
		\subfigure[H-R-L part of $u_z$.]{\includegraphics[scale=0.8]{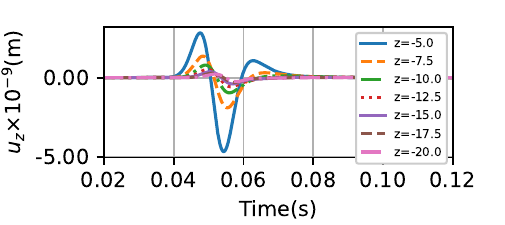}}
		\subfigure[Maximum of H-R-L varies with z.]{\includegraphics[scale=0.8]{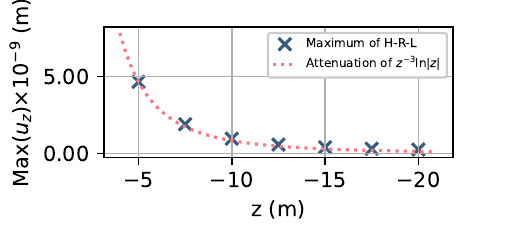}}\\
		\subfigure[H-R-T part of $u_z$.]{\includegraphics[scale=0.8]{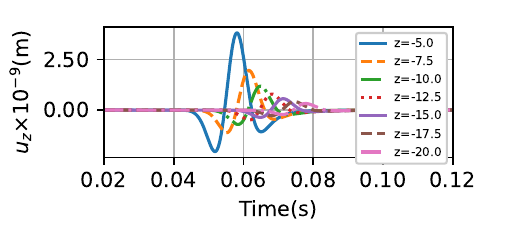}}
		\subfigure[Maximum of H-R-T varies with z.]{\includegraphics[scale=0.8]{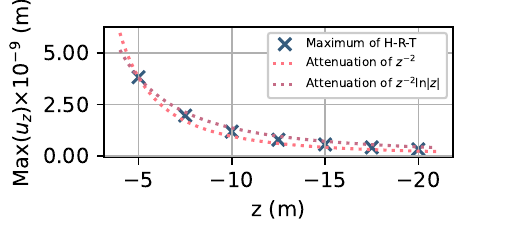}}\\
		\subfigure[H-G-L part of $u_z$.]{\includegraphics[scale=0.8]{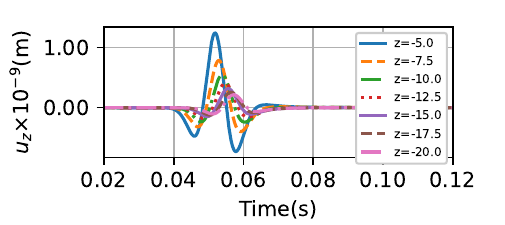}}
		\subfigure[Maximum of H-G-L varies with z.]{\includegraphics[scale=0.8]{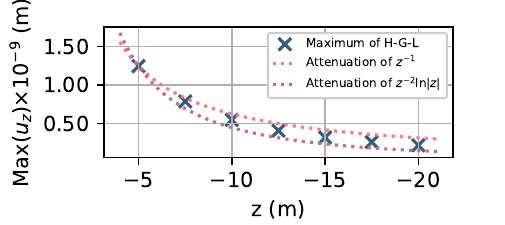}}\\
		\subfigure[H-G-T part of $u_z$.]{\includegraphics[scale=0.8]{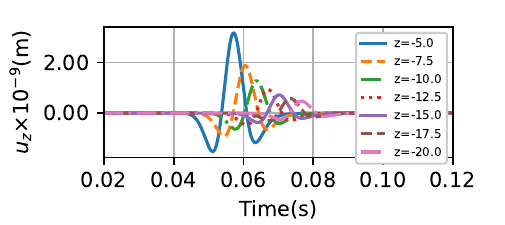}}
		\subfigure[Maximum of H-G-T varies with z.]{\includegraphics[scale=0.8]{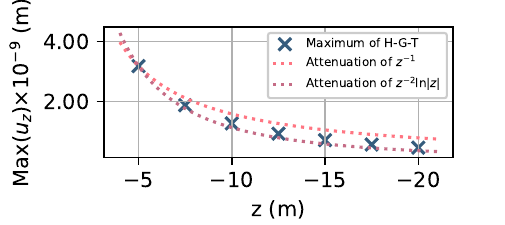}}\\
		\caption{The attenuation of $u_z$ with $z$.}\label{F4.14}
	\end{figure}

\begin{figure}[h!]
		\centering
		\subfigure[D-L part of $u_z$.]{\includegraphics[scale=0.8]{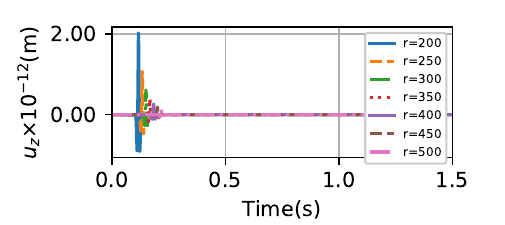}}
		\subfigure[Maximum of D-L varies with r.]{\includegraphics[scale=0.8]{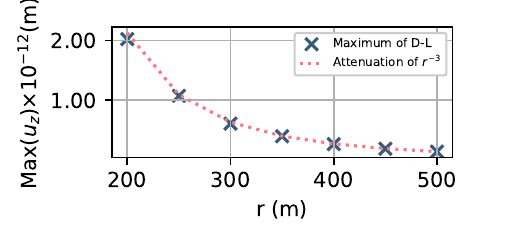}}\\
		\subfigure[D-T part of $u_z$.]{\includegraphics[scale=0.8]{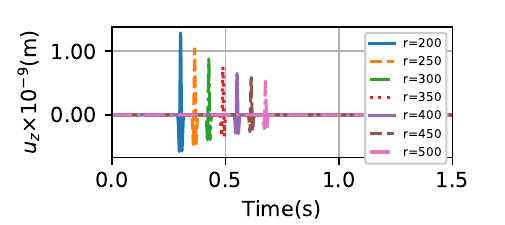}}
		\subfigure[Maximum of D-T varies with r.]{\includegraphics[scale=0.8]{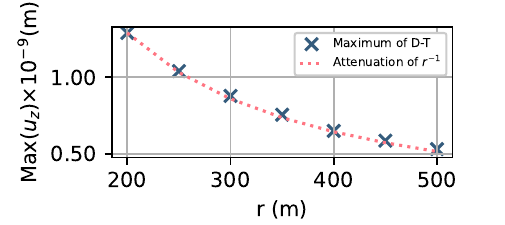}}\\
		\subfigure[H-R-T part of $u_z$.]{\includegraphics[scale=0.8]{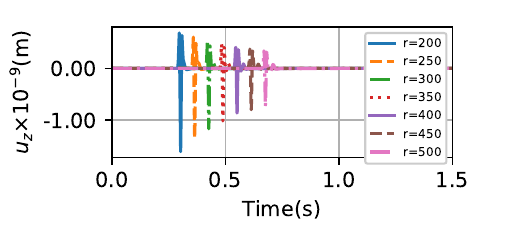}}
		\subfigure[Maximum of H-R-T varies with r.]{\includegraphics[scale=0.8]{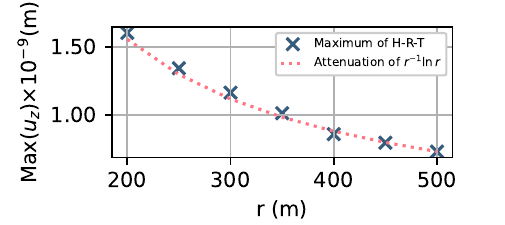}}\\
		\subfigure[H-G-L part of $u_z$.]{\includegraphics[scale=0.8]{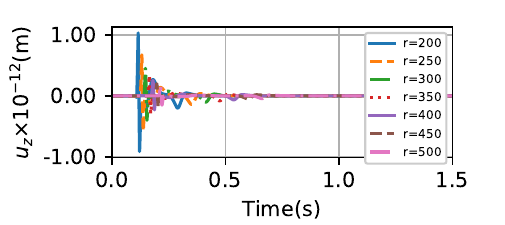}}
		\subfigure[Maximum of H-G-L varies with r.]{\includegraphics[scale=0.8]{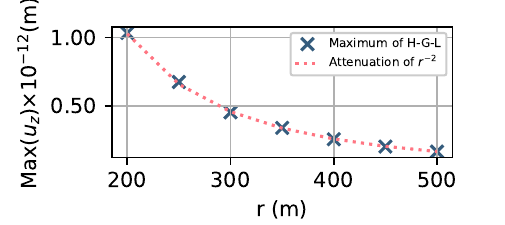}}\\
		\subfigure[H-G-T part of $u_z$.]{\includegraphics[scale=0.8]{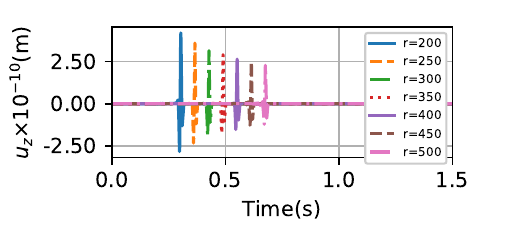}}
		\subfigure[Maximum of H-G-T varies with r.]{\includegraphics[scale=0.8]{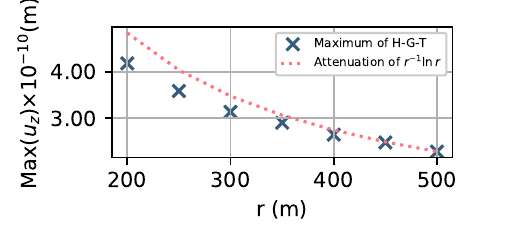}}\\
		\caption{The attenuation of $u_z$ (D-L, D-T, H-R-T, H-G-L, H-G-T) with $r$.}\label{F4.15}
	\end{figure}

\begin{figure}[h!]
		\centering
		\subfigure[D-A part of $u_z$.]{\includegraphics[scale=0.8]{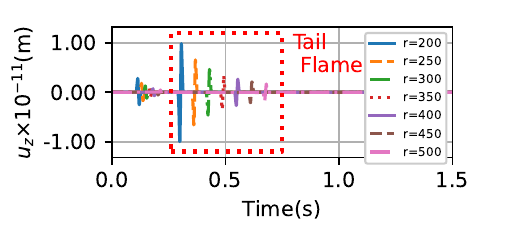}}
		\subfigure[H-R-L part of $u_z$.]{\includegraphics[scale=0.8]{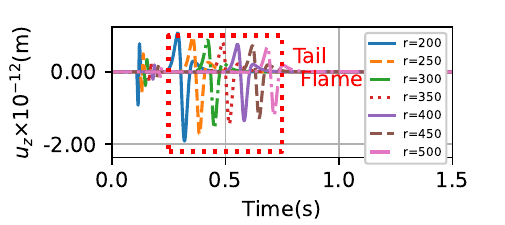}}\\
		\subfigure[Maximum of D-A (head wave) varies with r.]{\includegraphics[scale=0.8]{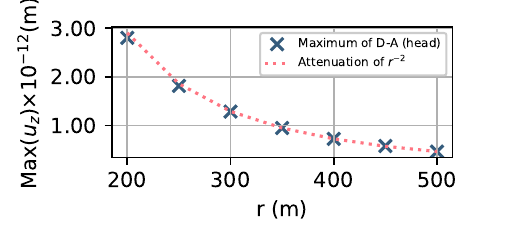}}
		\subfigure[Maximum of H-R-L (head wave) varies with r.]{\includegraphics[scale=0.8]{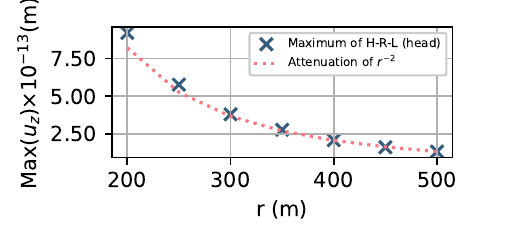}}\\
		\subfigure[Maximum of D-A (tail flame) varies with r.]{\includegraphics[scale=0.8]{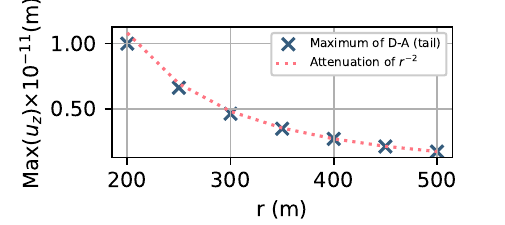}}
		\subfigure[Maximum of H-R-L (tail flame) varies with r.]{\includegraphics[scale=0.8]{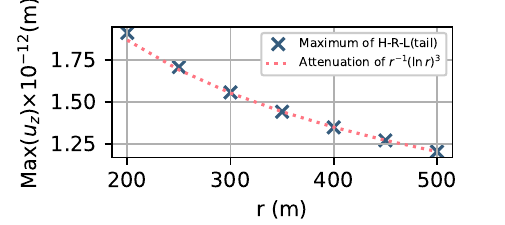}}
		\caption{The attenuation of $u_z$ (D-A, H-R-L) with $r$.}\label{F4.16}
	\end{figure}

\begin{figure}[h!]
		\centering
		\subfigure[D-A mode.]{\includegraphics[scale=0.8]{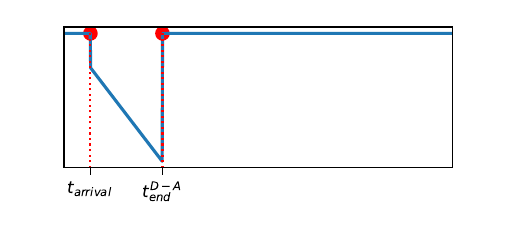}}
		\subfigure[H-R-L mode.]{\includegraphics[scale=0.8]{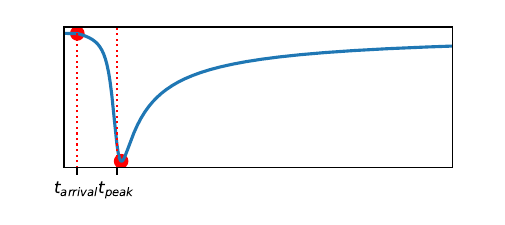}}
		\caption{Green's functions of D-A mode and H-R-L mode.}\label{F4.17}
	\end{figure}

	This section aims to study the axial displacement $u_z$ varying with spatial coordinates. According to the results in \ref{26th.-Dec.-2023-E001}, $u_z$ in a plane with a fixed depth $z$ is controlled by the displacement on axis. So how does $u_z$ vary with depth is analyzed firstly and the evolution between $u_z$ and diameter $r$ is similar with the evolution between $u_z$ and $t$ in a fixed plane.\\
	\begin{figure}[h!]
		\centering
		\includegraphics[scale=0.8]{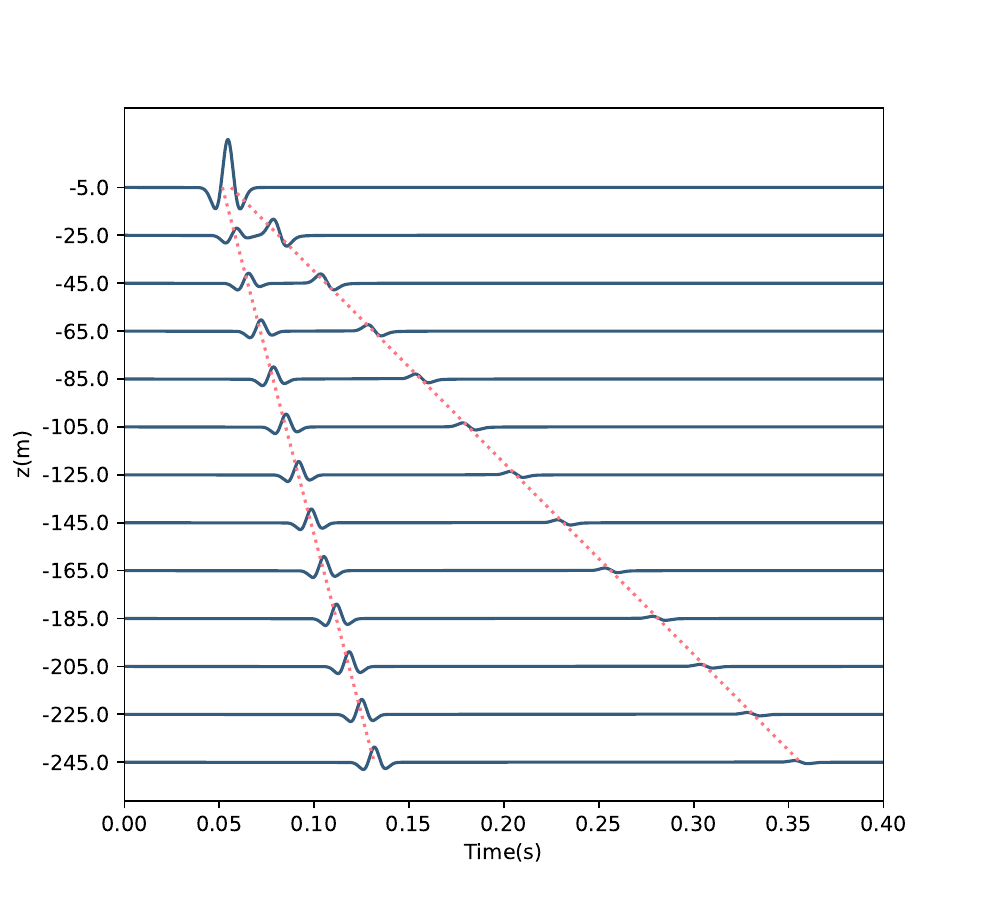}
		\caption{Array waveforms on the axis of depth from -5 m to -245 m.}\label{5th.-Jan-2024-P001}
	\end{figure}\\
	In \ref{5th.-Jan-2024-P001}, thirteen recievers are setted every 20 m from 5 m underground. The two wave fronts of longitudinal wave (the sum of waves propagating with $c_L$) and transverse wave (the sum of waves propagating with $c_T$) are clearly visible. The transverse wave is larger than the longitudinal wave near the surface and attenuate fastly with the increase of depth. Because of the very fast attenuation, an uneven scale is used to describe the ampitude of displacement. The attenuation of $u_z$ in each wave mode is drawed in \ref{F4.14} seprately and the conclusion is shown in \ref{5th.-Jan.-2024-T001}.\\
	\\
	In \ref{F4.14}, there is no direct transverse wave in $u_z$ on the axis. According to \ref{5th.-Jan.-2024-T001}, we find that the D-A has the fastest attenuation speed $|z|^{-3}$ and plays the main role in the longitudinal wave and the transverse wave near the surface. D-L has the slowest attenuation speed $|z|^{-1}$ and is the main part of longitudinal wave. The Huygens waves activated by Rayleigh wave (H-R-L, H-R-T) have fast attenuation speeds and are not tangible in the whole field. Finally, the Huygens waves activated by wave group (H-G-L, H-G-T) have slow attenuation speed (but faster than D-L), and H-G-L attenuates a bit slower than H-G-T. Therefore, with the increase of depth, the longitudinal wave has a better property of detection than the transverse wave.\\
	\begin{table}[h!]
		\begin{center}
			\begin{tabular}{cc}
				\hline\hline\noalign{\smallskip}
				\textbf{Wave mode} & \textbf{Attenuation speed } (estimation)\\
				\hline
				D-L & $|z|^{-1}$\\
				D-A & $|z|^{-3}$\\
				H-R-L & $|z|^{-3}\ln|z|$\\
				H-R-T & between $|z|^{-2}\ln|z|$ and $|z|^{-2}$\\
				H-G-L & between $|z|^{-1}$ and $|z|^{-2}\ln|z|$\\
				H-G-T & close to $|z|^{-2}\ln|z|$\\
				\noalign{\smallskip}\hline
			\end{tabular}\caption{The displacemnts on the axis of different wave modes attenuate with the increase of depth.}\label{5th.-Jan.-2024-T001}
		\end{center}
	\end{table}\\
	Then, consider about the fluction in the far-field of a plane. The attenuation of $u_z$ in each wave mode is drawed in \ref{F4.15} and \ref{F4.16} seprately, and particularly, the D-A and the H-R-L can be divided into two parts visually. The conclusion is shown in \ref{5th.-Jan.-2024-T001}. Seven recievers are set from 200 m (diameter $r$) every 50 m in the plane 30 m deep underground ($z=-30\text{ m}$). In \ref{F4.15}, with the increase of diameter, the direct transverse wave (D-T) and the Huygens waves carried by transverse wave (H-R-T, H-G-T) have a very slow attenuation, while the longitudinal waves (D-L, H-G-L) attenuate fastly.\\
	\\
	In \ref{F4.16}, except the head waves after the arrival time, there are another waves existing. We call this phenomenon as {\it tail flame} (it is not coda). The cause of tail flame in D-A and H-R-L are different. The head wave and tail flame are not different kinds of waves, and they just look like two parts visually.\\
	\\
	The sepration of head wave and tail flame is just a visual effect of mutagenic points in Green's functions. As shown in \ref{F4.17}, there are two mutagenic points in D-A mode and H-R-L mode (we choose $\mathbf{B}_z(c_R,c_L)[2]$ to represent that). The head wave is generated by the first point, which is the delineation between stationary and fluction, and its visual arrival time is \\
	\begin{equation}
		T^{H.W.}_{visual}=t_{arrival}+T_s,
	\end{equation}
	where $T_s$ is the delay of source. The tail flame is generated by the second point. The cause of the second point can be different for different waves. For D-A, the second point is the coda, and for H-R-L, it is the peak of Time-Green function,\\
	\begin{subequations}
		\begin{align}
			\text{D-A: }&T^{T.F.}_{visual}=t^{D-A}_{end}+T_s,\\
			\text{H-R-L: }&T^{T.F.}_{visual}\approx t_{peak}(c_R,c_L)+T_s.
		\end{align}
	\end{subequations} 
	For H-R-T and H-G-T, as shown in \ref{23th-Nov.-2023-P001}, the Time-Green function do not have the second point. For H-G-L, the distance between the two mutagenic points is very close, so the tail flame is "eatten" by the head wave and is not visible. Only when there are plural mutagenic points with long enough distances, the visual effect of sepration happens.\\
	\\
	As conclude in \ref{7th.-Jan.-2024-T001}, in general, with the increase of diameter, the transverse wave has a better property of detection than the longitudinal wave, and particularly, the tail flame of H-R-L has the best property of detection due to the slowest attenuation.\\
	\begin{table}[h!]
		\begin{center}
			\begin{tabular}{cc}
				\hline\hline\noalign{\smallskip}
				\textbf{Wave mode} & \textbf{Attenuation speed} (estimation)\\
				\hline
				D-L& $r^{-3}$ \\
				D-T& $r^{-1}$ \\
				D-A & head wave, tail flame: $r^{-2}$ \\
				H-R-L & head wave: $r^{-2}$, tail flame: $r^{-1}(\ln r)^3$\\
				H-R-T& $r^{-1}\ln r$\\
				H-G-L& $r^{-2}$\\
				H-G-T& $r^{-1}\ln r$\\
				\noalign{\smallskip}\hline
			\end{tabular}\caption{The attenuation with diameter of the displacemnts in the far-field on a plane for different wave modes.}\label{7th.-Jan.-2024-T001}
		\end{center}
	\end{table}\\
	Next we discuss about the velocity of waves in the plane. We define the wave velocity in a fixed plane $\Sigma_{P.}:\mathbf{a}\cdot\mathbf{x}=const.$ firstly. With time increase, the 3D wave front $\Sigma_{W.F.}(t)$ (suppose it must be convex) will touch $\Sigma_{P.}$ at a point $p_o$ and denote $p_o$ as the origin of the 2D wave front $\Gamma_{W.F.}(t)$. The vector between $p_o$ and any $p$ on $\Gamma_{W.F.}(t)$ is denoted as $\mathbf{r}$ and the time derivative of $\mathbf{r}$ is defined as the plane wave velocity.\\
	\begin{equation}
		\mathbf{v}_P(t;\theta):=\frac{{\rm d}}{{\rm d}t}\mathbf{r}(t;\theta)\xlongequal{\Gamma_{W.F.}(t):g(r)=t}\mathbf{v}_P(g(r);\theta).
	\end{equation}
	Here, $\Sigma_{W.F.}(t)$ is in the form of \ref{6th.-Jan.-2024-E003a}, \ref{6th.-Jan.-2024-E003b} or \ref{6th.-Jan.-2024-E004}. $\Sigma_{P.}(t)$ is the plane that $x_3=z$ and $p_o$ is $(0,0,z)$. In this situation, all the $\Gamma_{W.F.}$ are circle and $\mathbf{v}_P$ can be represented by a scalar $v_r$ (shown in \ref{8th.-Jan.-2024-T001}).\\
	\begin{table}[h!]
		\begin{center}
			\begin{tabular}{r|ccc|ccc|c}
				\hline\hline\noalign{\smallskip}
				\textbf{Wave mode} & D-L & H-R-L & H-G-L & D-T & H-R-T & H-G-T, $r\leq R\sqrt{\eta}$ & H-G-T, $r> R\sqrt{\eta}$\\
				\hline
				\textbf{Plane wave velocity} & \multicolumn{3}{|c|}{$c_L\sqrt{1-(z/r)^2}$} & \multicolumn{3}{|c|}{$c_T\sqrt{1-(z/r)^2}$} & $c_L$\\
				\noalign{\smallskip}\hline
			\end{tabular}\caption{Plane wave velocity of each wave mode with independent variable $r$.}\label{8th.-Jan.-2024-T001}
		\end{center}
	\end{table}\\
	The variation makes H-G-T wave propagating with the velocity of longitudinal wave in the plain when $r$ is large enough.\\
	\subsubsection{Approximations of displacements underground near the axis}\label{4.3.2}
\begin{figure}[h!]
		\centering
		\subfigure[$r=1$ m and $z=-30$ m.]{\includegraphics[scale=0.8]{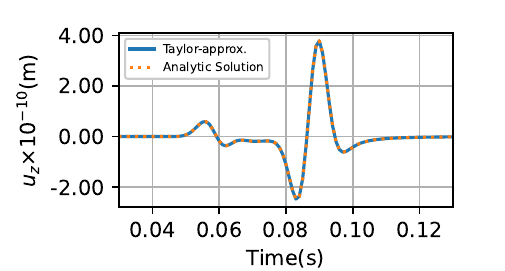}}
		\subfigure[$r=8$ m and $z=-30$ m.]{\includegraphics[scale=0.8]{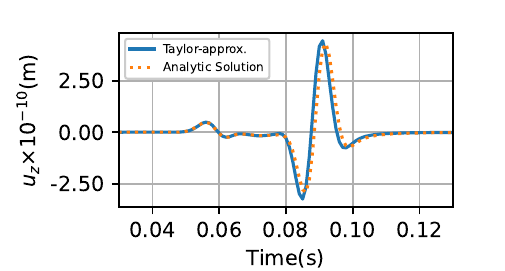}}
		\caption{The comparison between approximation and analytic solution.}\label{9th.-Jan.-2024-P003}
	\end{figure}

\begin{figure}[h!]
		\centering
		\subfigure[Axial displacements $u_z$.]{\includegraphics[scale=0.8]{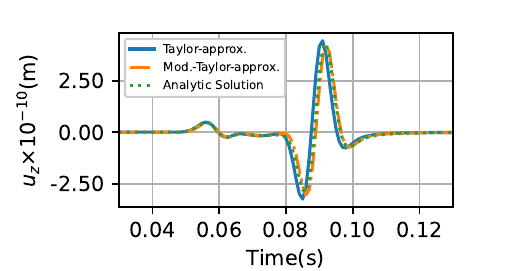}}
		\subfigure[Bias.]{\includegraphics[scale=0.8]{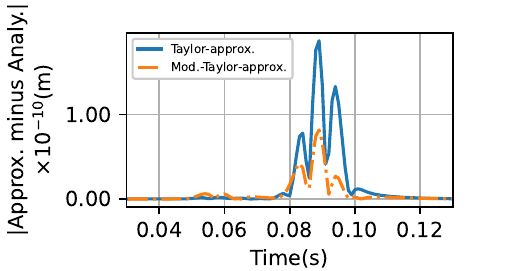}}\\
		\subfigure[Axial displacements $u_r$.]{\includegraphics[scale=0.8]{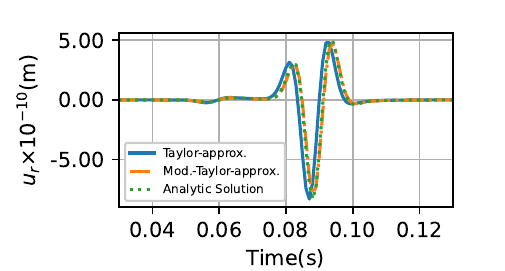}}
		\subfigure[Bias.]{\includegraphics[scale=0.8]{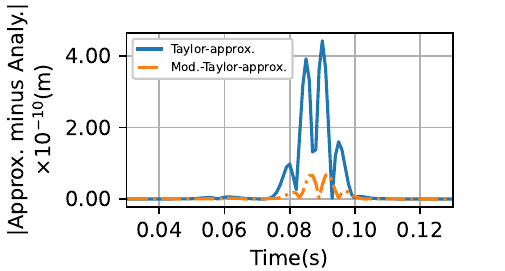}}
		\caption{The comparison between approximation and analytic solution.}\label{23th.-Feb.-2024-P002}
	\end{figure}

	In this section, the approximations of displacements near the axis are given and the concrete expressions are in \ref{C.4}. Beacuse $u_z^{Huygens}$ on axis can be expressed in elementary functions, Taylor expand (the displacements activated by a formal source can be expended into the form of polynomial and Piano residual) can be used to give the approximation of $u_z^{Huygens}$ around axis (when $r<|z|\sqrt{\eta/(1-\eta)}$).\\
	\begin{equation}
		u_z^{Huygens}= u_z^{Huygens}(r=0)+r\frac{\partial }{\partial r}u_z^{Huygens}(r=0)+\frac{r^2}{2}\frac{\partial^2}{\partial r^2}u_z^{Huygens}(r=0)+O(r^3),\notag
	\end{equation}
	where the value of partial derivative of $u_z^{Huygens}$ when $r=0$ is what we need to compute. In \ref{8th.-Jan.-2024-E001}, the first-order derivative of $\mathbf{B}_z(p,c)$ are all composed by $\mathscr{H}_1$ (see \ref{8th.-Jan.-2024-E002}) whose value at $r=0$ is zero.\\
	\begin{align}
		&\begin{aligned}
			&\frac{\partial }{\partial r}u_z^{Huygens}(r=0)=-\frac{1}{2\pi}\frac{\partial }{\partial r}\left(N_1\mathbf{B}_z(c_R,c_L)+N_2\mathbf{B}_z(c_R,c_T)+C_1\int_1^{1/\sqrt{\eta}}W_L(x)\mathbf{B}_z(xc_T,c_L){\rm d}x\right.\\
			&\left.+C_2\int_1^{1/\sqrt{\eta}}W_T(x)\mathbf{B}_z(xc_T,c_T){\rm d}x\right)*\mathbf{F}^{(T)}=0,
		\end{aligned}\label{8th.-Jan.-2024-E001}\\
		&\begin{aligned}
			&\frac{\partial^2}{\partial r^2}u_z^{Huygens}=\\
			&-\frac{1}{2\pi}\frac{\partial^2}{\partial r^2}\left(N_1\mathbf{B}_z(c_R,c_L)+N_2\mathbf{B}_z(c_R,c_T)+C_1\int_1^{1/\sqrt{\eta}}W_L(x)\mathbf{B}_z(xc_T,c_L){\rm d}x+C_2\int_1^{1/\sqrt{\eta}}W_T(x)\mathbf{B}_z(xc_T,c_T){\rm d}x\right)*\mathbf{F}^{(T)}\\
			&=-\frac{1}{2\pi}\left(N_1\tilde{\mathbf{B}}_z(c_R,c_L)+N_2\tilde{\mathbf{B}}_z(c_R,c_T)+C_1\int_1^{1/\sqrt{\eta}}W_L(x)\tilde{\mathbf{B}}_z(xc_T,c_L){\rm d}x+C_2\int_1^{1/\sqrt{\eta}}W_T(x)\tilde{\mathbf{B}}_z(xc_T,c_T){\rm d}x\right)*\tilde{\mathbf{F}}^{(T)},
		\end{aligned}\label{9th.-Jan.-2024-E001}
	\end{align}
	In \ref{9th.-Jan.-2024-E001}, the transformations from position to time partial derivates are uesd here and get $\tilde{\mathbf{B}}_z$ and $\tilde{\mathbf{F}}^{(T)}$ (the same idea to get Time-Green fuction: the position partial derivates of $\mathbf{B}_z$ are transfered to $\mathbf{F}^{(T)}$). The same to the above, $\tilde{\mathbf{B}}(p,c)[n]$ is the n-th component of vector.\\
	\begin{subequations}
		\begin{align}
			&\tilde{\mathbf{B}}_z(p,c):=\left[\tilde{\mathbf{B}}_z(p,c)[0],\tilde{\mathbf{B}}_z(p,c)[1],\tilde{\mathbf{B}}_z(p,c)[2],\tilde{\mathbf{B}}_z(p,c)[3],\tilde{\mathbf{B}}_z(p,c)[4]\right],\label{22th-Mar.-2024-E001a}\\
			&\tilde{\mathbf{F}}^{(T)}:=\left[F(t),F^{(1)}(t),F^{(2)}(t),F^{(3)}(t),F^{(4)}(t)\right]^{\textbf{T}}.
		\end{align}
	\end{subequations}
	The approximation of $u_z^{Huygens}$ is\\
	\begin{equation}
		\begin{aligned}
			u_z^{Taylor-approx.}&= u_z^{Huygens}(r=0)-\frac{r^2}{4\pi}\left(N_1\tilde{\mathbf{B}}_z(c_R,c_L;r=0)+C_1\int_1^{1/\sqrt{\eta}}W_L(x)\tilde{\mathbf{B}}_z(xc_T,c_L;r=0){\rm d}x\right.\\
			&\left.+N_2\tilde{\mathbf{B}}_z(c_R,c_T;r=0)+C_2\int_1^{1/\sqrt{\eta}}W_T(x)\tilde{\mathbf{B}}_z(xc_T,c_T;r=0){\rm d}x\right)*\tilde{\mathbf{F}}^{(T)}.
		\end{aligned}\label{23th.-Feb.-2024-E002}
	\end{equation}
	\ref{9th.-Jan.-2024-P003} shows the comparison between approximation and analytic solution at $r=1$ m and $z=-30$ m and $r=8$ m and $z=-30$ m.\\
	\\
	The way to get the approximations of $u_r^{Huygens}$ is similar with that of $u_z^{Huygens}$. The Taylor expand of $u_r^{Huygens}$ near $r=0$ is\\
	\begin{equation}
		u_r^{Huygens}=u_r^{Huygens}(r=0)+r\frac{\partial}{\partial r}u_r^{Huygens}(r=0)+O(r^2).
	\end{equation}
	The radial displacements on the axis is zero (\ref{3.2.2}), $u_r^{Huygens}(r=0)=0$.\\
	\begin{equation}
		\begin{aligned}
			&\frac{\partial}{\partial r}u_r^{Huygens}=\\
			&=\frac{1}{2\pi}\frac{\partial}{\partial r}\left(-N_1\mathbf{B}_r(c_R,c_L)+N_2\mathbf{B}_r(c_R,c_T)-C_1\int_1^{1/\sqrt{\eta}}W_L(x)\mathbf{B}_r(xc_T,c_L){\rm d}x+C_2\int_1^{1/\sqrt{\eta}}W_T(x)\mathbf{B}_r(xc_T,c_T){\rm d}x\right)*\mathbf{F}^{(T)}\\
			&=\frac{1}{2\pi}\left(-N_1\bar{\mathbf{B}}_r(c_R,c_L)+N_2\bar{\mathbf{B}}_r(c_R,c_T)-C_1\int_1^{1/\sqrt{\eta}}W_L(x)\bar{\mathbf{B}}_r(xc_T,c_L){\rm d}x+C_2\int_1^{1/\sqrt{\eta}}W_T(x)\bar{\mathbf{B}}_r(xc_T,c_T){\rm d}x\right)*\bar{\mathbf{F}}^{(T)}.\\
		\end{aligned}\label{24th.-Jan.-2024-E001}
	\end{equation}
	In \ref{24th.-Jan.-2024-E001}, the transformations from position to time partial derivates are uesd here and get $\bar{\mathbf{B}}_r$ and $\bar{\mathbf{F}}^{(T)}$ (the same idea to get Time-Green fuction: the position partial derivates of $\mathbf{B}_r$ are transfered to $\mathbf{F}^{(T)}$).\\
	\begin{subequations}
		\begin{align}
			&\bar{\mathbf{B}}_z(p,c):=\left[\bar{\mathbf{B}}_z(p,c)[0],\bar{\mathbf{B}}_z(p,c)[1],\bar{\mathbf{B}}_z(p,c)[2],\bar{\mathbf{B}}_z(p,c)[3]\right],\label{22th-Mar.-2024-E002a}\\
			&\bar{\mathbf{F}}^{(T)}:=\left[F(t),F^{(1)}(t),F^{(2)}(t),F^{(3)}(t)\right]^{\textbf{T}}.
		\end{align}
	\end{subequations}
	The approximation of $u_r^{Huygens}$ is\\
	\begin{equation}
		\begin{aligned}
			u_r^{Taylor-approx.}&= \frac{r}{2\pi}\left(-N_1\bar{\mathbf{B}}_r(c_R,c_L;r=0)-C_1\int_1^{1/\sqrt{\eta}}W_L(x)\bar{\mathbf{B}}_r(xc_T,c_L;r=0){\rm d}x\right.\\
			&\left.+N_2\bar{\mathbf{B}}_r(c_R,c_T;r=0)+C_2\int_1^{1/\sqrt{\eta}}W_T(x)\bar{\mathbf{B}}_r(xc_T,c_T;r=0){\rm d}x\right)*\bar{\mathbf{F}}^{(T)}.
		\end{aligned}\label{23th.-Feb.-2024-E003}
	\end{equation}
	\ref{9th.-Jan.-2024-P003} shows the comparison between approximation and analytic solution at $r=1$ m and $z=-30$ m and $r=8$ m and $z=-30$ m.\\
	\begin{figure}[h!]
		\centering
		\subfigure[$r=1$ m and $z=-30$ m.]{\includegraphics[scale=0.8]{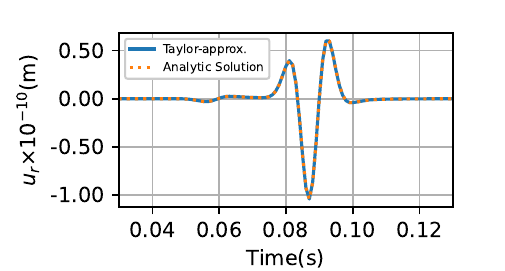}}
		\subfigure[$r=8$ m and $z=-30$ m.]{\includegraphics[scale=0.8]{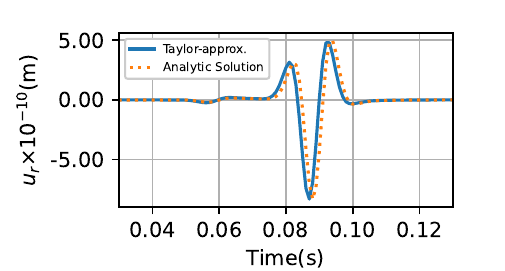}}
		\caption{The comparison between approximation and analytic solution.}\label{31th.-Jan.-2024-P001}
	\end{figure}\\
	From \ref{9th.-Jan.-2024-P003} and \ref{31th.-Jan.-2024-P001}, the arrival time of Taylor-approx. is much faster than that of analytic solutions with the increase of $r$. It is easy to find the reason of the stagnate from \ref{23th.-Feb.-2024-E002} and \ref{23th.-Feb.-2024-E003}. The Taylor-approx. has the same arrival time with the point $(0,z)$ on the axis. To handle this problem, we modify the approximations with transform $z=-R=-\sqrt{z^2+r^2}$.\\
	\begin{subequations}
		\begin{align}
			&\begin{aligned}
				u_z^{Mod.-Taylor-approx.}&= u_z^{Huygens}(r=0)-\frac{r^2}{4\pi}\left(N_1\tilde{\mathbf{B}}_z(c_R,c_L;t,0,-R)+C_1\int_1^{1/\sqrt{\eta}}W_L(x)\tilde{\mathbf{B}}_z(xc_T,c_L;t,0,-R){\rm d}x\right.\\
				&\left.+N_2\tilde{\mathbf{B}}_z(c_R,c_T;t,0,-R)+C_2\int_1^{1/\sqrt{\eta}}W_T(x)\tilde{\mathbf{B}}_z(xc_T,c_T;t,0,-R){\rm d}x\right)*\tilde{\mathbf{F}}^{(T)},
			\end{aligned}\label{23th.-Feb.-2024-E004a}\\
			&\begin{aligned}
				u_r^{Mod.-Taylor-approx.}&= \frac{r}{2\pi}\left(-N_1\bar{\mathbf{B}}_r(c_R,c_L;t,0,-R)-C_1\int_1^{1/\sqrt{\eta}}W_L(x)\bar{\mathbf{B}}_r(xc_T,c_L;t,0,-R){\rm d}x\right.\\
				&\left.+N_2\bar{\mathbf{B}}_r(c_R,c_T;t,0,-R)+C_2\int_1^{1/\sqrt{\eta}}W_T(x)\bar{\mathbf{B}}_r(xc_T,c_T;t,0,-R){\rm d}x\right)*\bar{\mathbf{F}}^{(T)}.
			\end{aligned}\label{23th.-Feb.-2024-E004b}
		\end{align}	
	\end{subequations}
	In \ref{23th.-Feb.-2024-P002}, the comparisons among analytic solutions, Taylor-approx. and Mod.-Taylor-apporx. at the point $(8,-30)$ are made. It seems that Mod.-Taylor-apporx. is more accurate than Taylor-apporx., but it is not mathematically rigorous.\\
	\\
	\subsubsection{Radial displacement on the surface varying with diameter}\label{4.3.3}
\begin{figure}[h!]
		\centering
		\includegraphics[scale=0.8]{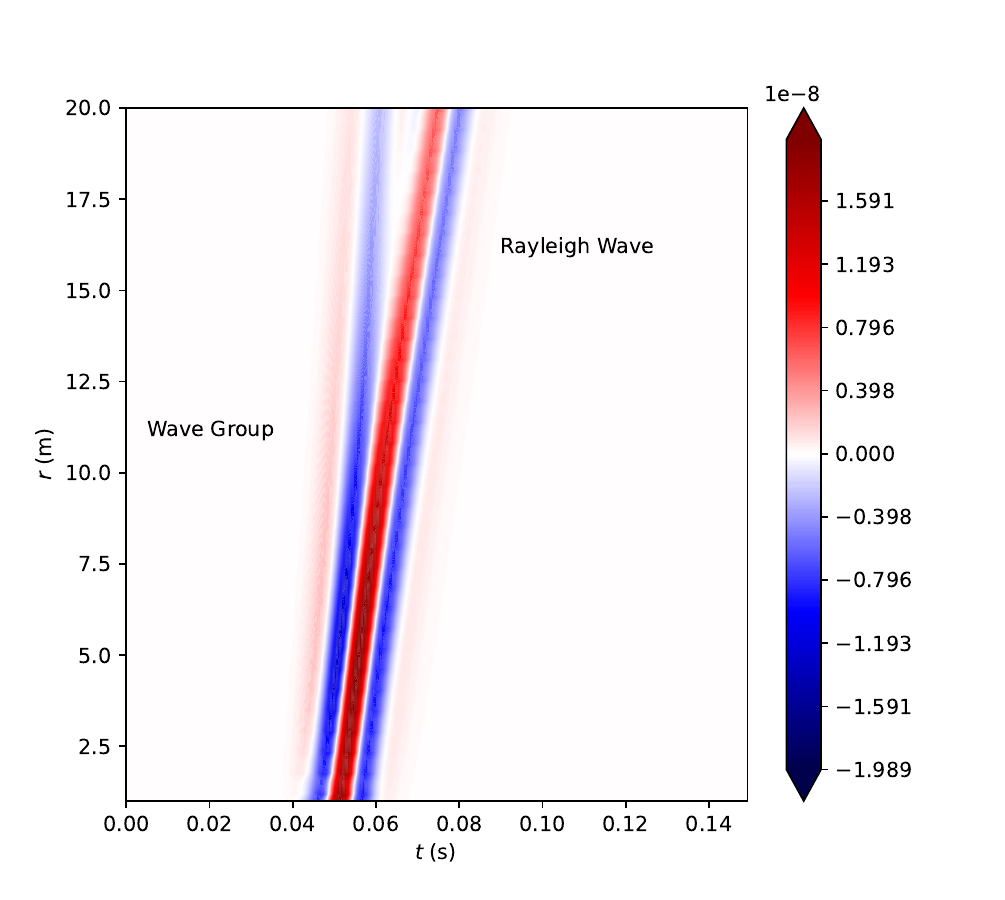}
		\caption{Map of $u_r^{sur.}$ in $t-r$ domain.}\label{9th.-Jan.-2024-P001}
	\end{figure}

\begin{figure}[h!]
		\centering
		\subfigure[Rayleigh wave.]{\includegraphics[scale=0.8]{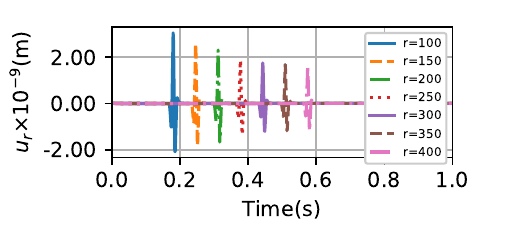}}
		\subfigure[Maximum varying with $r$.]{\includegraphics[scale=0.8]{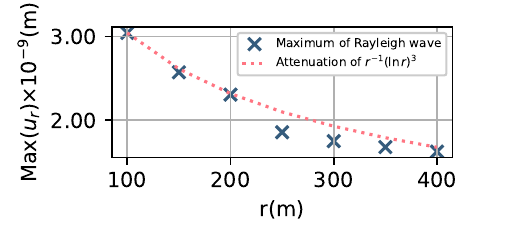}}\\
		\subfigure[Wave group.]{\includegraphics[scale=0.8]{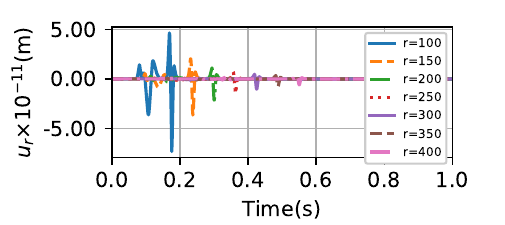}}
		\subfigure[Maximum varying with $r$.]{\includegraphics[scale=0.8]{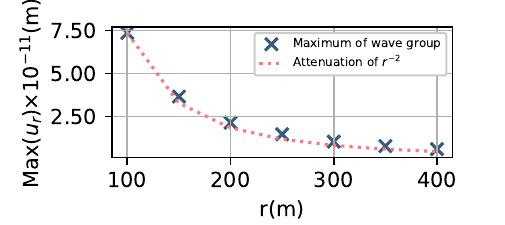}}
		\caption{The attenuation of $u_r^{sur.}$ with $r$.}\label{9th.-Jan.-2024-P002}
	\end{figure}

	This section aims to study the distribuation of $u_r$ on the free surface. As discussed above, $u_r^{sur.}$ has specificity that it shares similiar wave modes with surface waves which do not appear in the 3D domain. As shown in \ref{9th.-Jan.-2024-P001}, there are two kinds of wave in $u_r^{sur.}$: one is a wave group with velocity distributing in $(c_T,c_L)$ and the other is the wave propagating in the speed of Rayleigh wave.\\
	\\
	The radial Time-Green fuction on the surface $G^{(T)}_{1-r}$ and $G^{(T)}_{2-r}$ are composed by \ref{9th.-Jan.-2024-E003a} and \ref{9th.-Jan.-2024-E003b}. The attenuation of them determines that of $G^{(T)}_{1-r}$ and $G^{(T)}_{2-r}$.\\
	\begin{subequations}
		\begin{align}
			t\to\infty,\quad& S_1(t,r;p)=-\frac{1}{r}\sqrt{1-(\frac{r}{pt})^2}=-\frac{1}{r}-\frac{1}{2r}O(\frac{1}{t^2}),\\
			& S_2(t,r;p)=-\frac{1}{p}\arccos(\frac{r}{pt})=-\frac{1}{p}-\frac{1}{p}O(\frac{1}{t}).
		\end{align}
	\end{subequations}
	In \ref{9th.-Jan.-2024-P002}, we find that the amplitude of Rayleigh wave is much larger than that of wave group and its attenuation with diameter is much slower when the source is Ricker wavelet.\\ 
	\\
	\section{Conclusion and discussion}\label{5}
	Compared to previous works, this work has resulted in the following advances. The solutions to half-space given by this article are without the limit of Poisson's ratio and singularity in computation. A unified expression for the velocity of Rayleigh waves $c_R$ is given and the change of $c_R$ with medium is discussed in detail. The sepration of component-waves is made to explain the generation mechanisms of waves. The discussions about the attenuation of component-waves with time and space help us to select a collection of component-waves with good detection property.\\
	\\
	In this article, a novel method, the Huygens method (\ref{3}), is introduced, differing from the Cagniard-de Hoop method. The Huygens method is based on the solutions of wave equations in transform domain (\ref{2}) and can be divided into three steps: 1. Splitting the solutions in transform domain into D-term and P-term (\ref{3.1.1}); 2. Computing the inverters of D-term and P-term separately (\ref{3.1.2}, \ref{B.1}); 3. Computing the the 3D convolution of D-term and P-term (\ref{3.1.3}, \ref{C}). For a 3D fluctuation problem, points on the surface satisfying boundary conditions can be expressed by two independent space variables (denote them as $x_1$ and $x_2$) and the third space variable is a constant (denote it as $x_3$). The first step of Huygens method is separating $x_3$ from the soluition, while leaving a D-term totally depended on $x_1$ and $x_2$ and a P-term repressenting the mode of propagation in 3D domain. Physically, D-term is the distributed responsiveness due to the source, and P-term is a basic carrier via which the point source propagates. Similar to the Huygens-Fresnel principle, the 3D concolution in the third step can be viewed as that there is a distributed source, with D-term as its distribuation, splitted into infinite point sources, generating the wave which is the sum of P-term that each point source generates. In fact, the separation of wave can be done by the first step and third step, and the second step is not necessary but can greatly simplify the final expression. In general, there is not a common way to do the second step and it is difficult to do it in most of time. Fortunately, the second step is workable in this problem. Notably, to eliminate singularities of the Green's function, a new form (\ref{3.2.1}), Time-Green function is developped and used to express the results of this problem (\ref{28th-Sept.-2023-E001}). The transform from Green's function to Time-Green function is independent from the Huygens method. While the Cagniard-de Hoop method relies purely on mathematical techniques and is constrained by media properties, the Huygens method is developed from physical principles, offering clear insight into wave generation mechanisms and facilitating easy wave decomposition. Moreover, the Huygens method holds potential for a broad array of applications in various linear boundary problems.\\
	\\
	For the most concerned situation by seismology, displacements on the free surface, this article discussed the expressions and properties separately (\ref{3.2.2}, \ref{4.3.3}). The expressions of displacements on the surface can be greatly simplified (\ref{1st-Apr.-2024-E001}, \ref{1st-Apr.-2024-E002}), which makes the computation less time consuming. The discussions about the surface make a fact clear that "wave on the surface" do not equal to "surface wave". Here, surface wave is a kind of wave mode in 2D domain (almost invisible in 3D reality) and different from the wave that we observe on the surface, such as that $u_z(z=0)$ is not in the surface wave mode but Huygens wave mode (most of the waves on the surface are 3D waves generated by low dimensional waves but not direct low dimensional waves). Peculiarly, in the radial displacement on the surface $u_r(z=0)$, the surface wave is directly visible and $u_r(z=0)$ propagates with the surface wave mode. Mathematically, the reason of the appearance of surface wave is the degeneracy of weak function (\ref{28th-Sept.-2023-E002}). Physically, beacause the imimpulse in this problem is vertical ($z$-direction) and the radial displacement is orthogonal to that direction, the appearance of polarity makes P-term eliminated.\\
	\\
	This article mathematically defines three kinds of generation mechanisms (\ref{4.1}): direct wave, surface wave and Huygens wave. The direct wave in Lamb's problem shares the same form with the wave in full-space problem. The displacements caused by direct wave differs from the displacement in full-space by a factor of $2/(1-\eta)$, which means the half-space problem is not simplly cutting the full-space problem into half, but exists more complex effect, the Huygens effect. Traditionally, two kinds of surface waves are considered existing on the surface: a group of wave with velocities distributing in the interval $(c_L,c_T)$, and the Rayleigh wave with velocity defined by \ref{1st-Apr.-2024-E003}. The expression of the root of \ref{1st-Apr.-2024-E003} has been a segmented expression, until this article gives the Rayleigh wave's velocity in a unified formula (\ref{10th.Jan.-2024-E001}). Generally, the surface wave only exists on the low dimensional boundary and is invisible. From the computation of D-term (\ref{3.1.2}), we know that the wave group corresponds to the branch cut and the Rayleigh wave corresponds to the first-order pole. Specially, the first-order pole makes the determinant in \ref{15th-Sept.-2023-E011} equalling zero, which makes the dimension of the solution space diminishing. This makes the surface looking like a totally different medium. It is hopeful to explain this difference clearly by the definition of interior point and boundary point in algebra. Through wave decomposition, a novel wave mode, the Huygens wave, is identified, elucidating the intricate processes of acoustical reflection and refraction. The Huygens wave generated by the Huygens effect (\ref{1st-Apr.-2024-E004}). Physically, Huygens wave is the effect of low dimension to the three dimension. The surface waves excite the direct waves in 3D domain. Therefore, the wave journey of Huygens wave is a two-segment vector (\ref{22th-Mar.-2024-P004}). In Lamb's problem, there are four kinds of Huygens waves ($2\times2$, two kinds of surface waves and two kinds of direct waves) : the longitudinal wave excited by the Rayleigh wave (H-R-L), the transverse wave excited by the Rayleigh wave (H-R-T), the longitudinal wave excited by the wave group (H-G-L), and the transverse wave excited by the wave group (H-G-T). In common condition, the wavefronts of H-R-L, H-R-T and H-G-L are sphere. When the velocity of surface wave is larger than the velocity of direct wave, an aspherical wavefront is discovered in H-G-T. This phenomenon is called as variation (\ref{22th-Sept.-2023-E002}) and explained by the Fermat principle (\ref{4.2.1}). The variation also brings an interesting effect (\ref{8th.-Jan.-2024-T001}) that, on a plane with a fixed depth, waves of H-G-T mode propagate with the velocity of the longitudinal wave when $r$ is larger than a value about $z$, though the "carrier" of H-G-T is the transverse wave. We can view it simplly as the appearance of surface wave (wave group).\\
	\\
	Although the expressions of displacements are without singularities, the algorithmic complexity of the computation is still $O(n)$ or $O(n^2)$. For the further simplification of calculations, we discuss the properties of displacements near the axis and give the reliable approximations of the displacements near the axis (\ref{4.3.2}, \ref{C.5}), which is composed by the elementary functions and has an algorithmic complexity of $O(1)$ or $O(n)$.\\
	\\
	To enhance the practical engineering aspects of this work, this article discusses the attenuation of displacements with time and space, focusing on selecting component-waves with good detection property (\ref{4.3.1} and \ref{4.3.3}). As the article has demonstrated, for $u_z$ underground, in general, with the increase of depth, the longitudinal wave attenuates much slower than the transverse wave, and with the increase of diameter, situation is opposite. Whichever the condition is, the H-R-L always has a slow attenuation. For $u_r$ on the surface, with the increase of diameter, the Rayleigh wave has a slower attenuation than the wave group, which is useful in the far-field detection.\\
	\\
	The final noteworthy thing is phenomenon in \ref{F4.16}, tail flame. We observed that the theoretical separation of wave do not always equal  to what we see. Sometimes, in the waveform of one generation mechanism, there are one or more independent-looking waveforms following the head wave. Actually, as \ref{4.3.1} has explained, this phenomenon is caused by plural mutagenic points with long enough distances of Green's functions. This visual effect may make the observers and the image recognition algorithms confused and mistakening the tail flame as waves of different generation mechanisms. On the other hand, the existence of tail flame makes wave having more comprehensive properties in the complex conditions, such as H-R-L having a slow attenuating tail flame, whlie the other longitudinal waves attenuate fastly with the increase of diameter. So, more studies on the "visual separation of waves" are needed, which has the potential to be applied in the image recognition algorithms and deep learning in seismology and oil exploration.\\
	\newpage
	\appendix
	\labelformat{section}{{\color{red}{Appendix #1}}}
	\labelformat{subsection}{{\color{red}{Appendix #1}}}
	
	\section{Details of solutions in frequency and wave-number domain}\label{A}	
	\subsection{Applications of Laplace transformation and variable separation method}\label{A.1}
	This appendix is a supplement to \ref{2.2}, giving the detailed process from \ref{17th-Jan.-2024-E001a} and \ref{17th-Jan.-2024-E001b} to \ref{14th-Sept.-2023-E014a} and \ref{14th-Sept.-2023-E014b}.\\
	\\
	Performing Laplace transformation to \ref{17th-Jan.-2024-E001a} and \ref{17th-Jan.-2024-E001b}, we get\\
	\begin{subequations}
		\begin{align}
			(\frac{\partial^2}{\partial r^2}+\frac{1}{r}\frac{\partial}{\partial r}+\frac{\partial^2}{\partial z^2}-\frac{s^2}{c_L^2})\Phi&=0,
			\label{14th-Sept.-2023-E006a}\\
			(\frac{\partial^2}{\partial r^2}+\frac{1}{r}\frac{\partial}{\partial r}+\frac{\partial^2}{\partial z^2}-\frac{1}{r^2}-\frac{s^2}{c_T^2})\Psi&=0.
			\label{14th-Sept.-2023-E006b}
		\end{align}
	\end{subequations}
	In \ref{14th-Sept.-2023-E006a} and \ref{14th-Sept.-2023-E006b}, $\Phi$ and $\Psi$ are the Laplace transformations of $\varphi$ and $\psi$, and the Laplace transformations of their derivatives can be calculated by the differential property of Laplace transformation.\\
	\begin{subequations}
		\begin{align}
			\mathcal{L}\left[\frac{\partial^2\varphi(t,r,z)}{\partial t^2}\right]&=s^2\Phi(s,r,z)-s\varphi(0,r,z)-\frac{\partial\varphi(0,r,z)}{\partial t},
			\label{14th-Sept.-2023-E007a}\\
			\mathcal{L}\left[\frac{\partial^2\psi(t,r,z)}{\partial t^2}\right]&=s^2\Psi_{\theta}(s,r,z)-s\psi(0,r,z)-\frac{\partial\psi(0,r,z)}{\partial t}.
			\label{14th-Sept.-2023-E007b}
		\end{align}
	\end{subequations}
	According to the initial condition \ref{13th-Sept.-2023-E006}, we can get the initial condition of displacement potential.\\
	\begin{subequations}
		\begin{align}
			\varphi(0,r,z)=\frac{\partial\varphi(0,r,z)}{\partial t}&=0,
			\label{14th-Sept.-2023-E008a}\\
			\psi(0,r,z)=\frac{\partial\psi(0,r,z)}{\partial t}&=0.
			\label{14th-Sept.-2023-E008b}
		\end{align}
	\end{subequations}
	Assuming that $\Phi(s,r,z)$ and $\Psi(s,r,z)$ can be seprated by variables $r$ and $z$.\\
	\begin{subequations}
		\begin{align}
			\Phi(s,r,z)&=\Phi_r(r;s)\Phi_z(z;s),
			\label{14th-Sept.-2023-E009a}\\
			\Psi(s,r,z)&=\Psi_r(r;s)\Psi_z(z;s).
			\label{14th-Sept.-2023-E009b}
		\end{align}
	\end{subequations}
	\ref{14th-Sept.-2023-E006a} and \ref{14th-Sept.-2023-E006b} are seprated to two Bessel equations,
	\begin{subequations}
		\begin{align}
			\left(\frac{\partial^2}{\partial (kr)^2}+\frac{1}{(kr)}\frac{\partial}{\partial (kr)}+1\right)\Phi_r&=0,
			\label{14th-Sept.-2023-E010a}\\
			\left(\frac{\partial^2}{\partial (kr)^2}+\frac{1}{(kr)}\frac{\partial}{\partial (kr)}+1-\frac{1}{(kr)^2}\right)\Psi_r&=0.
			\label{14th-Sept.-2023-E010b}
		\end{align}
	\end{subequations}
	and two Euler equations.\\
	\begin{subequations}
		\begin{align}
			\frac{\partial^2\Phi_z}{\partial z^2}&=\frac{s^2}{c_L^2}+k^2,
			\label{14th-Sept.-2023-E011a}\\
			\frac{\partial^2\Psi_z}{\partial z^2}&=\frac{s^2}{c_T^2}+k^2.
			\label{14th-Sept.-2023-E011b}
		\end{align}
	\end{subequations}
	The general solutions of \ref{14th-Sept.-2023-E010a} and \ref{14th-Sept.-2023-E010b} are the following formula.\\
	\begin{subequations}
		\begin{align}
			\Phi_{r}&=A_{\Phi}(s,k){\rm J}_0(kr)+B_{\Phi}(s,k){\rm N}_0(kr),
			\label{14th-Sept.-2023-E012a}\\
			\Psi_{r}&=A_{\Psi}(s,k){\rm J}_1(kr)+B_{\Psi}(s,k){\rm N}_1(kr).
			\label{14th-Sept.-2023-E012b}
		\end{align}
	\end{subequations}
	The general solutions of \ref{14th-Sept.-2023-E011a} and \ref{14th-Sept.-2023-E011b} are the following formula.\\
	\begin{subequations}
		\begin{align}
			\Phi_{z}&=C_{\Phi}(s,k)\exp\left(z\sqrt{\frac{s^2}{c_L^2}+k^2}\right)+D_{\Phi}(s,k)\exp\left(-z\sqrt{\frac{s^2}{c_L^2}+k^2}\right),
			\label{14th-Sept.-2023-E013a}\\
			\Psi_{z}&=C_{\Psi}(s,k)\exp\left(z\sqrt{\frac{s^2}{c_T^2}+k^2}\right)+D_{\Psi}(s,k)\exp\left(-z\sqrt{\frac{s^2}{c_T^2}+k^2}\right).
			\label{14th-Sept.-2023-E013b}
		\end{align}
	\end{subequations}
	The combinations of \ref{14th-Sept.-2023-E012a} and \ref{14th-Sept.-2023-E013a}, \ref{14th-Sept.-2023-E012b} and \ref{14th-Sept.-2023-E013b} are the general solutions of homogeneous wave equation.\\
	\subsection{Branch analysis of $k_L$ and $k_T$}\label{A.2}
	This appendix is a supplement to \ref{2.2} and gives the choice of branch of functions in form of \ref{15th-Sept.-2023-E001} through the article, which simplifies \ref{17th-Jan.-2024-E001a} and \ref{17th-Jan.-2024-E001b} to \ref{15th-Sept.-2023-E005a} and \ref{15th-Sept.-2023-E005b}.\\
	\\
	Let $s=\sigma+\mathbf{i}\omega$, $\sigma$ is a positive real number, ${\rm Im}[\omega]=0$ is the situation what we need to consider. Start with the following function.\\
	\begin{equation}
		f(\omega)=\sqrt{\frac{1}{c^2}\left(\mathbf{i}\omega+\sigma\right)^2+k^2}.
		\label{15th-Sept.-2023-E001}
	\end{equation}
	On the complex plane, $f(\omega)$ has two branch point $p_1$ and $p_2$.\\
	\begin{equation}
		p_{1,2}=\pm ck+\mathbf{i}\sigma.
		\label{15th-Sept.-2023-E002}
	\end{equation}
	Rewrite \ref{15th-Sept.-2023-E001} in new form.\\
	\begin{equation}
		f(\omega)=\frac{1}{c}\sqrt{\left(p_1-\omega\right)\left(\omega-p_2\right)}.
		\label{15th-Sept.-2023-E003}
	\end{equation}
	Denote that $\omega_0$ is a point on the real axis, $\angle op_1\omega_0=\theta_1$ and $\angle op_2\omega_0=\theta_2$. According to the geometric relation (see \ref{15th-Sept.-2023-P001}), there are $-\angle op_2p_1<\theta_1<\pi-\angle op_2p_1$ and $\angle op_2p_1-\pi<\theta_2<\angle op_2p_1$. These relations show the total change of argument angle between orign and $\omega_0$ in square root function.\\
	\begin{equation}
		\left|\frac{\theta_1+\theta_2}{2}\right|<\frac{\pi}{2}.
		\label{15th-Sept.-2023-E004}
	\end{equation}
	\begin{figure}[h!]
		\centering
		\begin{tikzpicture}
			\draw[black,->,thick] (-4,0)--(4,0);
			\draw[black,->,thick] (0,-1)--(0,2);
			\draw[black] (-1.5,1)--(0,0);
			\draw[black] (1.5,1)--(0,0);
			\draw[black] (-1.5,1)--(2,0);
			\draw[black] (1.5,1)--(2,0);
			\filldraw[red] (2,0) circle (2pt);
			\filldraw[red] (0,0) circle (2pt);
			\node[above right] at (2,0) {$\omega_0$};
			\filldraw[blue] (1.5,1) circle (2pt);
			\node[above right] at (1.5,1) {$p_1$};
			\filldraw[blue] (-1.5,1) circle (2pt);
			\node[above left] at (-1.5,1) {$p_2$};
			\draw[blue,thick] (-1.5,1)--(1.5,1);
			\node[above right] at (-0.65,0.15) {$\theta_2$};
			\node[below] at (1.5,0.8) {$\theta_1$};
			\node[right] at (4,0) {${\rm Re}[\omega]$};
			\node[right] at (0,2) {${\rm Im}[\omega]$};
			\node[below left] at (0,0) {$o$};
		\end{tikzpicture}
		\caption{Analsis to the value of $f(\omega)$ on real axis.}
		\label{15th-Sept.-2023-P001}
	\end{figure}\\
	Through \ref{15th-Sept.-2023-E004}, the real part of $f(\omega_0)$ and $f(o)$ have the same sign. So, choose the branch that can make $f(o)=\left(c^2k^2+\sigma^2\right)^{\frac{1}{2}}>0$. It ensures that $f(\omega)$ has a positive real part when $\omega$ is a real number, and $\exp\left(zf(\omega)\right)$ converges, $\exp\left(-zf(\omega)\right)$ diverges when $z\to-\infty$. In this choice of branch, the coefficients of $\exp\left(-z\sqrt{\frac{s^2}{c^2}+k^2}\right)$ in \ref{14th-Sept.-2023-E013a} and \ref{14th-Sept.-2023-E013b} are confirmed to be zero.\\
	\newpage
	\section{Supplementary derivations in computation of D-term and P-term}\label{B}	
	\subsection{Split integral into D-term and P-term}\label{B.1}
	This appendix shows the works on the integral in form of \ref{16th-Sept.-2023-E005}, giving the transformation from the displacement potential functions in time domain (\ref{16th-Sept.-2023-E010a} and \ref{16th-Sept.-2023-E010b}) to the 3D convolution of D-term and P-term in \ref{3.1.1} (\ref{16th-Sept.-2023-E002a} and \ref{16th-Sept.-2023-E002b}).\\
	\begin{equation}
		I:=\frac{1}{4\pi^2\mathbf{i}}\int_{\sigma-\mathbf{i}\infty}^{\sigma+\mathbf{i}\infty}\int_{0}^{\infty}F(s,k)G(s,k)k{\rm J}_0(kr)e^{st}{\rm d}k{\rm d}s.
		\label{16th-Sept.-2023-E005}
	\end{equation}
	Denote the Fourier transformation and its inverse transformation as the following symbols.\\
	\begin{subequations}
		\begin{align}
			\mathcal{F}\left[f(\mathbf{x})\right]&=\int_{\Omega}f(\mathbf{x})e^{-\mathbf{i}\mathbf{k}\cdot\mathbf{x}}{\rm d}\Omega, \label{16th-Sept.-2023-E006a}\\
			\mathcal{F}^{-1}\left[F(\mathbf{k})\right]&=\frac{1}{(2\pi)^n}\int_{\Omega}F(\mathbf{k})e^{\mathbf{i}\mathbf{k}\cdot\mathbf{x}}{\rm d}\Omega. \label{16th-Sept.-2023-E006b}
		\end{align}
	\end{subequations}
	Perform a few transformations to the inner integral of the above equation.\\
	\begin{equation}
		\begin{aligned}
			\tilde{I}&:=\int_{0}^{\infty}F(k;s)G(k;s)k{\rm J}_0(kr){\rm d}k\\
			&=\frac{1}{2\pi}\int_0^{2\pi}\int_0^{\infty}F(k;s)G(k;s)ke^{\mathbf{i}kr\cos\theta}{\rm d}k{\rm d}\theta\\
			&=\frac{1}{2\pi}\iint_{\mathbb{R}^2}F(\sqrt{k_1^2+k_2^2};s)G(\sqrt{k_1^2+k_2^2};s)e^{\mathbf{i}kr\cos\theta}{\rm d}k_1{\rm d}k_2\\
			&=2\pi\mathcal{F}_{1,2}^{-1}\left[F(\sqrt{k_1^2+k_2^2};s)G(\sqrt{k_1^2+k_2^2};s)\right].
			\\
		\end{aligned}
		\notag
	\end{equation}
	Laplace transformation and Fourier transformation both have the convolution property.\\
	\begin{subequations}
		\begin{align}
			\mathcal{L}^{-1}\left[F(s)G(s)\right]&=\mathcal{L}^{-1}\left[F(s)\right]*\mathcal{L}^{-1}\left[G(s)\right],
			\label{16th-Sept.-2023-E007a}\\
			\mathcal{F}^{-1}\left[F(\mathbf{k})G(\mathbf{k})\right]&=\mathcal{F}^{-1}\left[F(\mathbf{k})\right]*\mathcal{F}^{-1}\left[G(\mathbf{k})\right].
			\label{16th-Sept.-2023-E007b}
		\end{align}
	\end{subequations}
	Thus, we can represent \ref{16th-Sept.-2023-E005} in the following form.\\
	\begin{equation}
		\begin{aligned}
			I&=\mathcal{L}^{-1}\left[\mathcal{F}_{1,2}^{-1}\left[F(\sqrt{k_1^2+k_2^2};s)G(\sqrt{k_1^2+k_2^2};s)\right]\right]\\
			&=\left<\mathcal{L}^{-1}\left[\mathcal{F}_{1,2}^{-1}\left[F(\sqrt{k_1^2+k_2^2};s)\right]\right],\mathcal{L}^{-1}\left[\mathcal{F}_{1,2}^{-1}\left[G(\sqrt{k_1^2+k_2^2};s)\right]\right]\right>.\\
			\label{16th-Sept.-2023-E008}
		\end{aligned}
	\end{equation}
	In \ref{2.1}, the pending coffecients of \ref{15th-Sept.-2023-E005a} and \ref{15th-Sept.-2023-E005b} are calculated, and the exact forms of them are comfirmed.\\
	\begin{subequations}
		\begin{align}
			\Phi&=\int_0^{\infty}\frac{k}{2\pi}\frac{\rho s^2+2\mu k^2}{(\rho s^2+2\mu k^2)^2-4\mu^2k^2k_Lk_T}
			{\rm J}_0(kr)e^{k_Lz}{\rm d}k,
			\label{16th-Sept.-2023-E009a}\\
			\Psi&=-\int_0^{\infty}\frac{k}{2\pi}\frac{2\mu kk_L}{(\rho s^2+2\mu k^2)^2-4\mu^2k^2k_Lk_T}{\rm J}_1(kr)e^{k_Tz}{\rm d}k.
			\label{16th-Sept.-2023-E009b}
		\end{align}
	\end{subequations}
	And $\varphi$ and $\psi$ are the integrals of form like \ref{16th-Sept.-2023-E005}, so they can be splited in the way shown in \ref{16th-Sept.-2023-E008}.\\
	\begin{subequations}
		\begin{align}
			&\begin{aligned}
				\varphi=\mathcal{L}^{-1}\left[\Phi\right]&=\frac{1}{4\pi^2\mathbf{i}}\int_{\sigma-\mathbf{i}\infty}^{\sigma+\mathbf{i}\infty}\int_{0}^{\infty}\frac{\rho s^2+2\mu k^2}{(\rho s^2+2\mu k^2)^2-4\mu^2k^2k_Lk_T}
				e^{k_Lz}k{\rm J}_0(kr)e^{st}{\rm d}k{\rm d}s\\
				&=\frac{\partial}{\partial z}\frac{1}{4\pi^2\mathbf{i}}\int_{\sigma-\mathbf{i}\infty}^{\sigma+\mathbf{i}\infty}\int_{0}^{\infty}\frac{\rho s^2+2\mu k^2}{(\rho s^2+2\mu k^2)^2-4\mu^2k^2k_Lk_T}\times\frac{e^{k_Lz}}{k_L}k{\rm J}_0(kr)e^{st}{\rm d}k{\rm d}s,
			\end{aligned}
			\label{16th-Sept.-2023-E010a}\\
			&\begin{aligned}
				\psi=\mathcal{L}^{-1}\left[\Psi\right]&=-\frac{1}{4\pi^2\mathbf{i}}\int_{\sigma-\mathbf{i}\infty}^{\sigma+\mathbf{i}\infty}\int_{0}^{\infty}\frac{2\mu k_Le^{k_Tz}}{(\rho s^2+2\mu k^2)^2-4\mu^2k^2k_Lk_T}k^2{\rm J}_1(kr)e^{st}{\rm d}k{\rm d}s\\
				&=\frac{\partial}{\partial r}\frac{1}{4\pi^2\mathbf{i}}\int_{\sigma-\mathbf{i}\infty}^{\sigma+\mathbf{i}\infty}\int_{0}^{\infty}\frac{2\mu k_Lk_T}{(\rho s^2+2\mu k^2)^2-4\mu^2k^2k_Lk_T}\times\frac{e^{k_Tz}}{k_T}k{\rm J}_0(kr)e^{st}{\rm d}k{\rm d}s.
			\end{aligned}
			\label{16th-Sept.-2023-E010b}
		\end{align}
	\end{subequations}
	In the above equations, the first part of quadrature function is distribution term mentioned in \ref{16th-Sept.-2023-E003a} and \ref{16th-Sept.-2023-E003b}. The second part is propagation term.\\
	\begin{subequations}
		\begin{align}
			P_L=\frac{1}{4\pi^2\mathbf{i}}\int_{\sigma-\mathbf{i}\infty}^{\sigma+\mathbf{i}\infty}\int_{0}^{\infty}\frac{e^{k_Lz}}{k_L}k{\rm J}_0(kr)e^{st}{\rm d}k{\rm d}s,
			\label{16th-Sept.-2023-E011a}\\
			P_T=\frac{1}{4\pi^2\mathbf{i}}\int_{\sigma-\mathbf{i}\infty}^{\sigma+\mathbf{i}\infty}\int_{0}^{\infty}\frac{e^{k_Tz}}{k_T}k{\rm J}_0(kr)e^{st}{\rm d}k{\rm d}s.
			\label{16th-Sept.-2023-E011b}
		\end{align}
	\end{subequations}
	Observe that,
	\begin{equation}
		\begin{aligned}
			\tilde{P}&:=\mathcal{F}_{1,2}\left[\mathcal{L}\left[\frac{1}{2\pi R}\delta(t-\frac{R}{c})\right]\right]=\iint_{\mathbb{R}^2}\int_{0}^{\infty}\frac{1}{2\pi R}\delta(t-\frac{R}{c})e^{-st-\mathbf{i}\mathbf{k}\cdot\mathbf{x}}{\rm d}t{\rm d}x_1{\rm d}x_2\\
			&=\int_0^{\infty}\int_{0}^{2\pi}\frac{r}{2\pi R}e^{-\frac{s}{c}R-\mathbf{i}kr\cos\theta}{\rm d}\theta{\rm d}r=\int_0^{\infty}\frac{r}{R}e^{-\frac{s}{c}R}{\rm J}_0(kr){\rm d}r\\
			&=\int_{|z|}^{\infty}e^{-\frac{s}{c}R}{\rm J}_0(k\sqrt{R^2-z^2}){\rm d}R=|z|\int_{1}^{\infty}e^{-\frac{|z|s}{c}x}{\rm J}_0(|z|k\sqrt{x^2-1}){\rm d}x\\
			&=\frac{1}{\sqrt{s^2/c^2+k^2}}\exp\left(-|z|\sqrt{s^2/c^2+k^2}\right).\\
		\end{aligned}
		\notag
	\end{equation}
	here, two integral formulas about Bessel functions are used and we get the expressions of P-terms (\ref{16th-Sept.-2023-E004a} and \ref{16th-Sept.-2023-E004b}).\\
	\begin{subequations}
		\begin{align}
			&\int_{0}^{2\pi}e^{-\mathbf{i}kr\cos\theta}{\rm d}\theta=2\pi{\rm J}_0(kr),
			\label{16th-Sept.-2023-E012a}\\
			&\int_{1}^{\infty}e^{-ax}{\rm J}_0(b\sqrt{x^2-1}){\rm d}x=\frac{1}{\sqrt{a^2+b^2}}\exp\left(-\sqrt{a^2+b^2}\right).
			\label{16th-Sept.-2023-E012b}
		\end{align}
	\end{subequations}
	At last, we finish the work and get the result in \ref{3.1.1}.\\
	\subsection{Rewriting of D-term and discussions on denominator polynomial}\label{B.2}
\begin{figure}[h!]
		\centering
		\subfigure[Real part] {\includegraphics[scale=0.8]{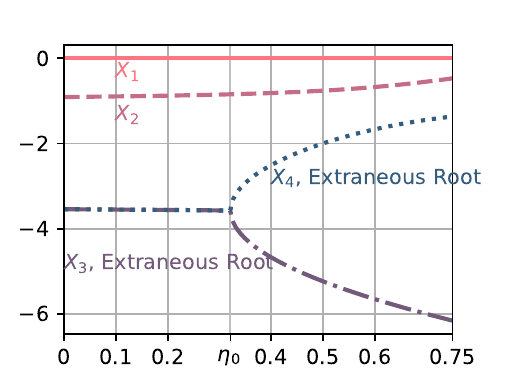}}
		\subfigure[Imaginary part] {\includegraphics[scale=0.8]{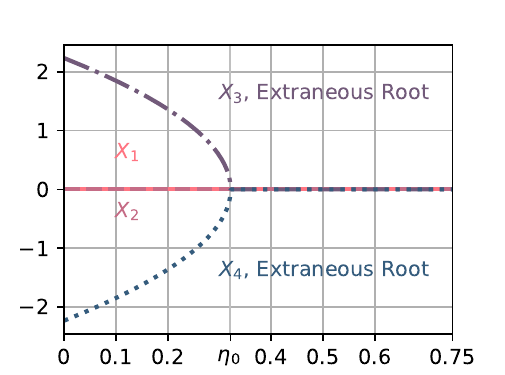}}
		\caption{$X$ varies with $\eta$.}
		\label{17th-Sept.-2023-P001}
	\end{figure}

	This appendix is a supplement to the step 1. \underline{\textbf{Change of integral-order}} in \ref{3.1.2}, while the zero points of denominator are given.\\
	\\
	In integral \ref{16th-Sept.-2023-E003a}, there is irrational term. Firstly, we rationalize the denominator.\\
	\begin{equation}
		\begin{aligned}
			L&=\frac{1}{4\pi^2\mathbf{i}}\int_{\sigma-\mathbf{i}\infty}^{\sigma+\mathbf{i}\infty}\int_{0}^{\infty}\frac{\rho s^2+2\mu k^2}{(\rho s^2+2\mu k^2)^2-4\mu^2k^2{\color{red}{k_Lk_T}}}k{\rm J}_0(kr)e^{st}{\rm d}k{\rm d}s\\
			&=\frac{1}{4\rho\pi^2\mathbf{i}}\int_{\sigma-\mathbf{i}\infty}^{\sigma+\mathbf{i}\infty}\int_{0}^{\infty}\frac{s^2+2c_T^2k^2}{F_1(s)}k{\rm J}_0(kr)e^{st}{\rm d}k{\rm d}s\\
			&=\frac{1}{4\rho\pi^2\mathbf{i}}\int_{\sigma-\mathbf{i}\infty}^{\sigma+\mathbf{i}\infty}\int_{0}^{\infty}\left(s^2+2c_T^2k^2\right)\frac{F_2(s)}{F(s)}k{\rm J}_0(kr)e^{st}{\rm d}k{\rm d}s\\
			k_Lk_T&=\sqrt{\left(\frac{s^2}{c_L^2}+k^2\right)\left(\frac{s^2}{c_T^2}+k^2\right)}.\\
		\end{aligned}	
		\notag
	\end{equation}
	The definition of functions are the same with \ref{17th-Sept.-2023-E002a}, \ref{17th-Sept.-2023-E002b} and \ref{17th-Sept.-2023-E002c}.\\
	\begin{align}
		&F_1(x):=(x^2+2c_T^2k^2)^2-4c_T^2k^2\sqrt{x^2+c_T^2k^2}\sqrt{\eta x^2+c_T^2k^2},
		\notag\\
		&F_2(x):=(x^2+2c_T^2k^2)^2+4c_T^2k^2\sqrt{x^2+c_T^2k^2}\sqrt{\eta x^2+c_T^2k^2}.
		\notag
	\end{align}
	Then, we analyse the roots of $F(x)$.\\
	\begin{equation}
		F(x):=F_1(x)F_2(x)=(x^2+2c_T^2k^2)^4-16c_T^4k^4(x^2+c_T^2k^2)(\eta x^2+c_T^2k^2).
		\label{17th-Sept.-2023-E004}
	\end{equation}
	The above equation is a quadratic polynomial of $X=x^2/(c_T^2k^2)$, and the roots can be calculated by root formula.\\
	\begin{subequations}
		\begin{align}
			X_1&=0,
			\label{17th-Sept.-2023-E005a}\\
			X_2&=\left(-\frac{8}{3}-\frac{1}{6}\frac{B}{A}+\frac{2}{3}A\right),
			\label{17th-Sept.-2023-E005b}\\
			X_3&=\left(-\frac{8}{3} + \frac{1+\mathbf{i}\sqrt{3}}{12}\frac{B}{A} - \frac{1-\mathbf{i}\sqrt{3}}{3}A\right),
			\label{17th-Sept.-2023-E005c}\\
			X_4&=\left(-\frac{8}{3} + \frac{1-\mathbf{i}\sqrt{3}}{12}\frac{B}{A} - \frac{1+\mathbf{i}\sqrt{3}}{3}A\right).
			\label{17th-Sept.-2023-E005d}
		\end{align}
	\end{subequations}
	The parameters in the above equations are below. $\eta$ is defined in \ref{17th-Sept.-2023-E003}.\\
	\begin{equation}
		\begin{aligned}
			A&=\left(17 - 45\eta + 3\sqrt{\Delta}\right)^{\frac{1}{3}},\\
			B&= 8-48\eta,\\
			\Delta&=-192\eta^3 + 321\eta^2 - 186\eta + 33.\\
		\end{aligned}
		\notag
	\end{equation}
	According to the choice of branch in \ref{2.2}, $X_1$ and $X_2$ are the real roots of $F_1(x)$ and $X_3$ and $X_4$ are the extraneous root. \ref{17th-Sept.-2023-P001} shows how the real part and imaginary part of the roots change with $\eta$.\\
	\\
	$X_1$ and $X_2$ are always real numbers. $X_3$ and $X_4$ are conjugate complex numbers when $0<\eta<\eta_0$, then turn to real numbers when $\eta_0\leq\eta<0.75$. The constant $\eta_0$ is a root of $\Delta=0$. There may be some interesting physical impunity relating to this constant, for example, it gives an exact standard of the so called soft stratum and hard stratum, and can explain some differences between these two kinds of stratum.\\
	\begin{equation}
		\eta_0=-\frac{(77293 + 7296\sqrt{114})^{1/3}}{192}+ \frac{455}{192(77293 + 7296\sqrt{114})^{1/3}} + \frac{107}{192}\approx0.321.
		\label{18th-Sept.-2023-E001}
	\end{equation}
	Based on the works above, $F(x)$ can be written as the form of \ref{17th-Sept.-2023-E002c}.\\
	\begin{equation}
		F(x)=x^2(x^2+a^2c_T^2k^2)(x^2-b^2c_T^2k^2)(x^2-c^2c_T^2k^2).
		\notag
	\end{equation}
	The parameters in the above equation are defined below. Specially, $ac_T<c_T$ will represent a special wave different from longitudinal and transverse waves. (In \ref{4.1.2}, we verify this wave is the Rayleigh wave, and $ac_T$ is the velocity of Rayleigh wave.)\\
	\begin{equation}
		a=\sqrt{-X_2}>0,\quad b=\sqrt{X_3},\quad c=\sqrt{X_4},\quad a^2b^2c^2=16(1-\eta)\text{ (Vieta's theorem)}.\\
		\notag
	\end{equation}
	The $L$-integral changes into the following form.\\
	\begin{equation}
		L=\frac{1}{4\rho\pi^2\mathbf{i}}\int_{\sigma-\mathbf{i}\infty}^{\sigma+\mathbf{i}\infty}\int_{0}^{\infty}k\left(s^2+2c_T^2k^2\right)\frac{F_2(s)}{F(s)}{\rm J}_0(kr)e^{st}{\rm d}k{\rm d}s.
		\label{18th-Sept.-2023-E003}
	\end{equation} 
	Then, if the rationality that the change of the order of the double integral is proved, we will get the final form \ref{17th-Sept.-2023-E001}. Here the Fubini's theorem is used.\\
	\begin{theorem}[Fubini's theorem]
		Let $P=P_1\times P_2$ be product measure. If $X$ is $\mathcal{B}_1\times\mathcal{B}_2$ measurable and is either non-negative or integrable with respect to $P$, then
		\begin{equation}
			\begin{aligned}
				\int_{\Omega_1\times\Omega_2}X{\rm d}P&=\int_{\Omega_1}\left[\int_{\Omega_2}X_{\omega_1}(\omega_2)P_2({\rm d}\omega_2)\right]P_1({\rm d}\omega_1)\\
				&=\int_{\Omega_2}\left[\int_{\Omega_1}X_{\omega_2}(\omega_1)P_1({\rm d}\omega_1)\right]P_2({\rm d}\omega_2).\\
			\end{aligned}
			\label{18th-Sept.-2023-E004}
		\end{equation}
	\end{theorem}
	\noindent We begin the proof with the boundedness of $f(k,s)$.\\
	\begin{equation}
		\begin{aligned}
			f(k,s)&:=\frac{k\left(s^2+2c_T^2k^2\right)}{F_1(s)}\\
			&=\frac{1}{k}\frac{\left(s^2/k^2+2c_T^2\right)}{(s^2/k^2+2c_T^2)^2-4c_T^2k^2\sqrt{s^2/k^2+c_T^2}\sqrt{\eta s^2/k^2+c_T^2}}.\\
		\end{aligned}
		\label{18th-Sept.-2023-E002}
	\end{equation}
	Briefly, we need to prove the following two integrals bounded.\\
	\begin{subequations}
		\begin{align}
			I_1&:=\int_0^{\infty}f(k,s){\rm J}_0(kr){\rm d}k,
			\label{18th-Sept.-2023-E005a}\\		I_2&:=\int_{\sigma-\mathbf{i}\infty}^{\sigma+\mathbf{i}\infty}f(k,s)e^{st}{\rm d}s.
			\label{18th-Sept.-2023-E005b}
		\end{align}
	\end{subequations}
	When $k\to\infty$, $|f(k,s)|\to0$. There must be a $K>0$, when $k>K$, $|f(k)|<C<\infty$, the inner integral of \ref{18th-Sept.-2023-E003} is converged.\\
	\begin{equation}
		\begin{aligned}
			|I_1|&\leq\left|\int_0^{K}f(k,s){\rm J}_0(kr){\rm d}k\right|+\left|\int_K^{\infty}f(k,s){\rm J}_0(kr){\rm d}k\right|<\left|\int_0^{K}f(k,s){\rm J}_0(kr){\rm d}k\right|+C\left|\int_K^{\infty}{\rm J}_0(kr){\rm d}k\right|\\
			&=\left|\int_0^{K}f(k,s){\rm J}_0(kr){\rm d}k\right|+\frac{C}{r}\left|1-{\rm J}_0(Kr)+\frac{Kr\pi}{2}\left({\rm J}_0(Kr){\rm H}^{(1)}_1(Kr)-{\rm J}_1(Kr){\rm H}^{(1)}_0(Kr)\right)\right|<\infty.\\
		\end{aligned}
		\notag
	\end{equation}
	Introduce the variable substitution $s=\sigma+\mathbf{i}\omega$.\\
	\begin{equation}
		I_2=\mathbf{i}e^{\sigma t}\int_{0}^{\infty}\left(f(k,\sigma+\mathbf{i}\omega)e^{\mathbf{i}\omega t}+f(k,\sigma-\mathbf{i}\omega)e^{-\mathbf{i}\omega t}\right){\rm d}\omega.\\
		\notag
	\end{equation}
	Since $|\omega|\to\infty$, $\left|f(k,\sigma+\mathbf{i}\omega)\right|\sim\frac{1}{|\omega|^2}$, $I_2$ is also converged. At last, we rewrite the $L$-integral into the target form.\\
	\begin{equation}
		|I_2|<2e^{\sigma t}\int_{0}^{\infty}\left|f(k,\sigma+\mathbf{i}\omega)\right|{\rm d}\omega<\infty.\\
		\notag
	\end{equation}
	\subsection{Calculation of frequency integral}\label{B.3}
	In this section, the detailed process to get the residues of poles and the branch-cut integral is given (step 2. \underline{\textbf{Path of $\omega$-integral}} in \ref{3.1.2}).\\
	\begin{figure}[h!]
		\begin{center}
			\begin{tikzpicture}
				\draw  [->](-6,0)--(6,0);
				\node[below left] at (6,0) {${\rm Re}[\omega]$};
				\draw  [->](0,-1)--(0,4);
				\node[below left] at (0,4) {${\rm Im}[\omega]$};
				\draw[blue,ultra thick,->] (-5,0)--(5,0);
				\draw[blue,ultra thick,->] (5,0)arc(0:180:5);
				\draw  [green](-4,2)--(-2.5,2);
				\draw  [green](4,2)--(2.5,2);
				\filldraw [red] (0,2) circle (2pt);
				\filldraw [red] (-1,2) circle (2pt);
				\filldraw [red] (1,2) circle (2pt);
				\filldraw [green] (-4,2) circle (2pt);
				\filldraw [green] (-2.5,2) circle (2pt);
				\filldraw [green] (4,2) circle (2pt);
				\filldraw [green] (2.5,2) circle (2pt);
				\draw[blue,ultra thick,->] (-0.3,2)arc(180:-180:0.3);
				\draw[blue,ultra thick,->] (-1.3,2)arc(180:-180:0.3);
				\draw[blue,ultra thick,->] (0.7,2)arc(180:-180:0.3);
				\draw[blue,ultra thick,<-] (-3.74,2.15)arc(30:330:0.3);
				\draw[blue,ultra thick,<-] (-2.76,1.85)arc(210:510:0.3);
				\draw[blue,ultra thick,->] (-3.74,2.15)--(-2.76,2.15);
				\draw[blue,ultra thick,<-] (-3.74,1.85)--(-2.76,1.85);
				\draw[blue,ultra thick,<-] (3.74,1.85)arc(210:510:0.3);
				\draw[blue,ultra thick,<-] (2.76,2.15)arc(30:330:0.3);
				\draw[blue,ultra thick,<-] (3.74,2.15)--(2.76,2.15);
				\draw[blue,ultra thick,->] (3.74,1.85)--(2.76,1.85);
				\node[below right] at (0,1.7) {$C_{pole}$};
				\node[below] at (-2.5,1.7) {$A_{Branch-point}$};
				\node[below] at (3.25,1.7) {$L_{Branch-cut}$};
				\node[above] at (0,4.5) {$A_{\infty}$};
				\node[above right] at (0,0) {$R.A.$};
			\end{tikzpicture}
		\end{center}
		\caption{The integral circuit that Cauchy's theorem is applied on.}
		\label{23th-Feb.-2024-P001}
	\end{figure}
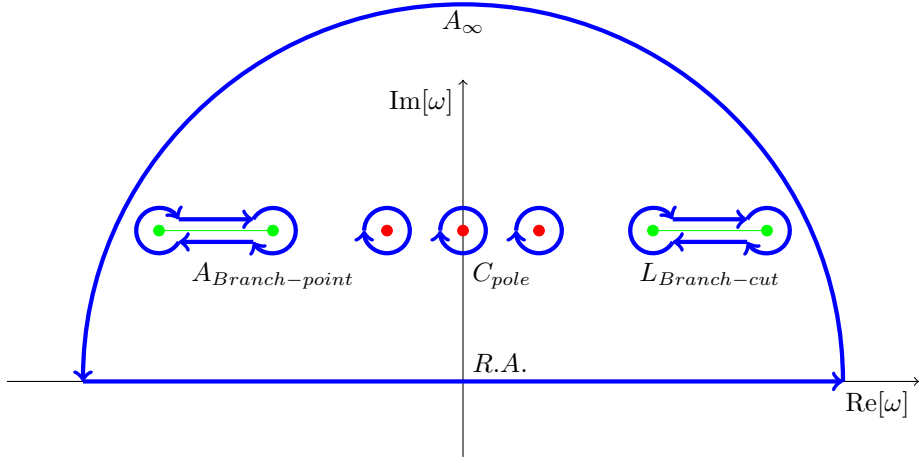\\
	The whole circuit $C$ is composed by the three circuits in \ref{23th-Feb.-2024-P001}: the outer counterclockwise semicircle composed $A_{\infty}$ and $R.A.$ (the integral we need); three small clockwise circles surround poles $C_{pole}$; two clockwise branch-cut circuits composed by small circular arcs surround branch-points $A_{Branch-point}$ and lines close to the shores of branch-cuts $L_{Branch-cut}$.\\
	\begin{subequations}
		\begin{align}
			&\oint_{C}=\int_{R.A.}+\int_{A_{\infty}}+\oint_{C_{pole}}+\int_{L_{Branch-cut}}+\int_{A_{Branch-point}}=0\text{, (via the Cauthy's theorem);}\\
			&\begin{aligned}
				\int_{A_{\infty}}=\lim_{R\to\infty}&\int_{\overset{\LARGE{\frown}}{(R+\mathbf{i}0^+)(-R+\mathbf{i}0^+)}}F(\omega)e^{\mathbf{i}\omega t}{\rm d}\omega\\
				&=\lim_{R\to\infty,\theta\in(0,\pi)}\pi R\left|F(\omega)\exp\left(Rt(-\sin\theta+\mathbf{i}\cos\theta)\right)\right|=0\text{, (via the Jordan's theorem);}
			\end{aligned}\\
			&\oint_{C_{pole}}=-2\pi\mathbf{i}\left(\left(Res[\mathbf{i}\sigma]+Res[-ac_Tk+\mathbf{i}\sigma]+Res[ac_Tk+\mathbf{i}\sigma]\right)\right)\text{, (via the residue theorem);}\\
			&\begin{aligned}
				\int_{A_{Branch-point}}=\lim_{\text{Arcs}\to\text{Points}}&\int_{\text{Arcs surround branch-points}}=0\text{, (via the Jordan's theorem);}
			\end{aligned}\\ 
			&\int_{R.A.}=\oint_{C}-\int_{A_{\infty}}-\oint_{C_{pole}}-\int_{L_{Branch-cut}}-\int_{A_{Branch-point}}=-\oint_{C_{pole}}-\int_{L_{Branch-cut}}.
		\end{align}
	\end{subequations}
	The residues of ploes are given below.\\
	\begin{equation}
		\begin{aligned}
			Res[\mathbf{i}\sigma]&=\lim_{\omega\to\mathbf{i}\sigma}\frac{{\rm d}}{{\rm d}\omega}\left(\left((\sigma+\mathbf{i}\omega)^2+2c_T^2k^2\right)\frac{(\omega-\mathbf{i}\sigma)^2F_2(\sigma+\mathbf{i}\omega)}{F(\sigma+\mathbf{i}\omega)}e^{\mathbf{i}\omega t}\right)=-\mathbf{i}e^{-\sigma t}\frac{16t}{a^2 b^2 c^2},\\
			Res[-ac_Tk+\mathbf{i}\sigma]&=\lim_{\omega\to-ac_Tk+\mathbf{i}\sigma}\left((\sigma+\mathbf{i}\omega)^2+2c_T^2k^2\right)\frac{(\omega+ac_Tk-\mathbf{i}\sigma)F_2(\sigma+\mathbf{i}\omega)}{F(\sigma+\mathbf{i}\omega)}e^{\mathbf{i}\omega t}\\
			&=-e^{-\sigma t}\frac{(2-a^2)^3}{a^3(a^2+b^2)(a^2+c^2)}\frac{e^{-\mathbf{i}ac_Tkt}}{c_Tk},\\
			Res[ac_Tk+\mathbf{i}\sigma]&=\lim_{\omega\to ac_Tk+\mathbf{i}\sigma}\left((\sigma+\mathbf{i}\omega)^2+2c_T^2k^2\right)\frac{(\omega-ac_Tk-\mathbf{i}\sigma)F_2(\sigma+\mathbf{i}\omega)}{F(\sigma+\mathbf{i}\omega)}e^{\mathbf{i}\omega t}\\
			&=e^{-\sigma t}\frac{(2-a^2)^3}{a^3(a^2+b^2)(a^2+c^2)}\frac{e^{\mathbf{i}ac_Tkt}}{c_Tk}.\\
		\end{aligned}
		\notag
	\end{equation}
	Next discussion is about $\int_{L_{Branch-cut}}$. Perform variable substitution $\zeta=\omega-\mathbf{i}\sigma$ to the part about branch cut.\\
	\begin{equation}
		\begin{aligned}
			I_{Branch-cut}&=\int_{L_{Branch-cut}}\left((\sigma+\mathbf{i}\omega)^2+2c_T^2k^2\right)\frac{F_2(\sigma+\mathbf{i}\omega)}{F(\sigma+\mathbf{i}\omega)}e^{\mathbf{i}\omega t}{\rm d}\omega\\
			&=e^{-\sigma t}\int_{L_{Branch-cut}}\frac{4c_T^2k^2\left(2c_T^2k^2-\zeta^2\right)\sqrt{c_T^2k^2-\zeta^2}\sqrt{c_T^2k^2-\eta\zeta^2}}{\zeta^2(\zeta^2-a^2c_T^2k^2)(\zeta^2+b^2c_T^2k^2)(\zeta^2+c^2c_T^2k^2)}e^{\mathbf{i}\zeta t}{\rm d}\zeta.\\
		\end{aligned}
		\label{19th-Sept.-2023-E005}
	\end{equation}
	The quadrature function has two branch cuts on the real aixs. The value of the quadrature function is comfirmed in the preceding choice of branch in \ref{A.2}.\\
	\begin{figure}[h!]
		\begin{center}
			\begin{tikzpicture}
				\draw  [black](-5,0)--(-0.2,0);
				\draw  [black](5,0)--(0.2,0);
				\filldraw [green] (-2,0) circle (2pt);
				\node[above] at (-2,0) {$-c_Tk$};
				\filldraw [green] (2,0) circle (2pt);
				\node[above] at (2,0) {$c_Tk$};
				\filldraw [green] (-4,0) circle (2pt);
				\node[above] at (-4,0) {$-c_Lk$};
				\filldraw [green] (4,0) circle (2pt);
				\node[above] at (4,0) {$c_Lk$};
				\draw[green,dashed,ultra thick] (-2,0)--(-4,0);
				\draw[green,dashed,ultra thick] (2,0)--(4,0);
				\node [left] at (-5,0) {$-P$};
				\node [above] at (-3,0) {$\mathbf{i}P$};
				\node [below] at (-3,0) {$-\mathbf{i}P$};
				\node [centered] at (0,0) {$P$};
				\node [right] at (5,0) {$-P$};
				\node [above] at (3,0) {$-\mathbf{i}P$};
				\node [below] at (3,0) {$\mathbf{i}P$};	
			\end{tikzpicture}
		\end{center}
		\caption{Value of the quadrature function on the 	real axis.}
		\label{19th-Sept.-2023-P002}
	\end{figure}
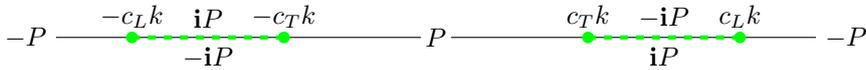\\
	In \ref{19th-Sept.-2023-P002} $P\geq0$ is the magnitude of the quadrature function. The integral $I_{Branch-cut}$ can be written in the following form.\\
	\begin{equation}
		\begin{aligned}
			I_{Branch-cut}=e^{-\sigma t}\left(\int_{-c_Lk+\mathbf{i}0^+}^{-c_Tk+\mathbf{i}0^+}+\int_{c_Tk+\mathbf{i}0^+}^{c_Lk+\mathbf{i}0^+}\right.&\left.+\int_{c_Lk+\mathbf{i}0^-}^{c_Tk+\mathbf{i}0^-}+\int_{-c_Tk+\mathbf{i}0^-}^{-c_Lk+\mathbf{i}0^-}\right)\\
			&\frac{4c_T^2k^2\left(2c_T^2k^2-\zeta^2\right)\sqrt{c_T^2k^2-\zeta^2}\sqrt{c_T^2k^2-\eta\zeta^2}}{\zeta^2(\zeta^2-a^2c_T^2k^2)(\zeta^2+b^2c_T^2k^2)(\zeta^2+c^2c_T^2k^2)}e^{\mathbf{i}\zeta t}{\rm d}\zeta.\\
		\end{aligned}
		\label{19th-Sept.-2023-E006}
	\end{equation}
	On the upper side of the branch cut, the quadrature function is a odd function. Furthermore, the integral along the upper side has the same value with the integral along the down side. Perform variable substitution $x=\frac{\zeta}{c_Tk}$ to \ref{19th-Sept.-2023-E006}.\\
	\begin{equation}
		\begin{aligned}
			I_{Branch-cut}&=2e^{-\sigma t}\left(\int_{-c_Tk+\mathbf{i}0^-}^{-c_Lk+\mathbf{i}0^-}+\int_{c_Tk+\mathbf{i}0^+}^{c_Lk+\mathbf{i}0^+}\right)\frac{4c_T^2k^2\left(2c_T^2k^2-\zeta^2\right)\sqrt{c_T^2k^2-\zeta^2}\sqrt{c_T^2k^2-\eta\zeta^2}}{\zeta^2(\zeta^2-a^2c_T^2k^2)(\zeta^2+b^2c_T^2k^2)(\zeta^2+c^2c_T^2k^2)}e^{\mathbf{i}\zeta t}{\rm d}\zeta\\
			&=-2e^{-\sigma t}\mathbf{i}\int_{c_Tk}^{c_Lk}\frac{4c_T^2k^2\left(2c_T^2k^2-\zeta^2\right)\sqrt{\zeta^2-c_T^2k^2}\sqrt{c_T^2k^2-\eta\zeta^2}}{\zeta^2(\zeta^2-a^2c_T^2k^2)(\zeta^2+b^2c_T^2k^2)(\zeta^2+c^2c_T^2k^2)}\left(e^{\mathbf{i}\zeta t}-e^{-\mathbf{i}\zeta t}\right){\rm d}\zeta\\
			&=\frac{16}{c_Tk}\int_{1}^{1/\sqrt{\eta}}\frac{\left(2-x^2\right)\sqrt{x^2-1}\sqrt{1-\eta x^2}}{x^2(x^2-a^2)(x^2+b^2)(x^2+c^2)}\sin (xc_Ttk){\rm d}x.\\
		\end{aligned}
		\notag
	\end{equation}
	Denote that, 
	\begin{equation}
		W_L(x)=\frac{\left(2-x^2\right)\sqrt{x^2-1}\sqrt{1-\eta x^2}}{x^3(x^2-a^2)(x^2+b^2)(x^2+c^2)},
		\notag
	\end{equation}
	the $I_{Branch-cut}$ can be writen as
	\begin{equation}
		I_{Branch-cut}=\frac{16}{c_T^2k}e^{-\sigma t}\int_{1}^{1/\sqrt{\eta}}W_L(x)xc_T\sin(xc_Ttk){\rm d}x.
	\end{equation}
	\newpage
	\section{Study on Huygens functions}\label{C}	
	\subsection{Elaborate expressions}\label{C.1}
	In this appendix, we calculate the Huygens functions $\mathscr{H}_v$ defined in \ref{3.1.3} (\ref{20th-Sept.-2023-E002b}) with $v=0,1$. \ref{22th-Sept.-2023-P002} shows the integral area changing with $\tau$. Based on \ref{22th-Sept.-2023-P002}, we divide the integral into two parts (zero value in other situations):
	\begin{itemize}
		\item[\textbf{a}.] $t_{arrival}(p,c)\leq t<t_{peak}(p,c)$;
		\item[\textbf{b}.] $t\geq t_{peak}(p,c).$
	\end{itemize}
	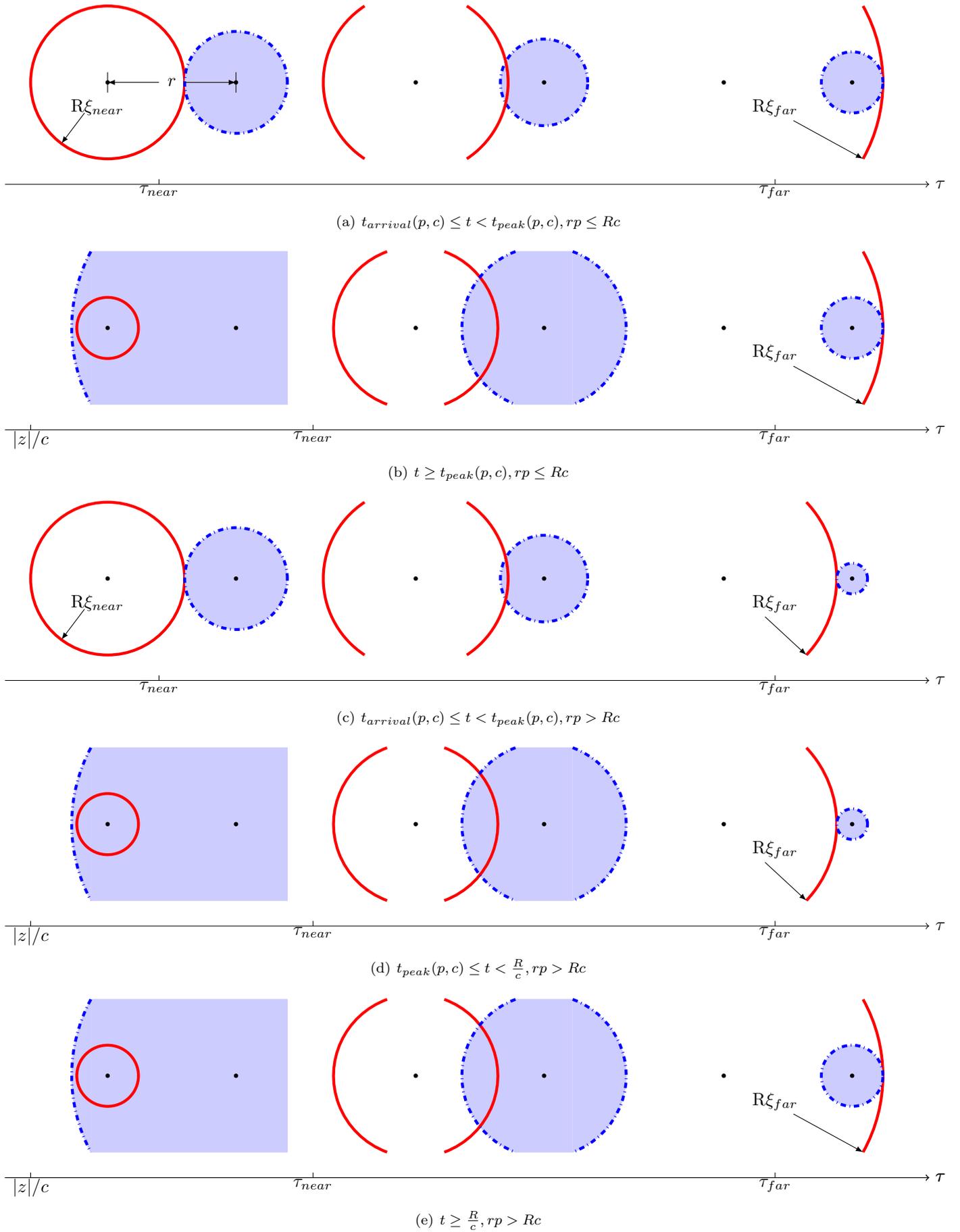
\begin{figure}[h!]
		\centering
		\subfigure[$t_{arrival}(p,c)\leq t<t_{peak}(p,c),rp\leq Rc$] 
		{\begin{tikzpicture}
				\draw[->](-9,0) --(9,0) node[right] {$\tau$};
				\draw (-6,0) --(-6,0.1) node[below]{$\tau_{near}$};
				\draw (6,0) --(6,0.1) node[below]{$\tau_{far}$};
				\draw[red,ultra thick] (-7,2) circle (1.5);
				\filldraw[dash dot,draw=blue,ultra thick,fill=blue!20] (-4.5,2) circle (1);
				\filldraw[fill=black] (-7,2) circle (1pt);
				\filldraw[fill=black] (-4.5,2) circle (1pt);
				\draw (-7,1.8) --(-7,2.2);
				\draw (-4.5,1.8) --(-4.5,2.2);
				\draw[-latex](-6,2) --(-7,2);
				\draw[-latex] (-5.5,2)--(-4.5,2);
				\node at (-5.75,2) {$r$};
				\draw[-latex](-7.45,1.4)--(-7.9,0.8);
				\node at(-7.2,1.52) {${\rm R}\xi_{near}$};
				\filldraw[dash dot,draw=blue,ultra thick,fill=blue!20] (1.5,2) circle (0.85);
				\draw[red,ultra thick] (-0.00501256,3.5) arc(56.4427:-56.4427:1.8);
				\draw[red,ultra thick] (-1.99499,3.5) arc(123.557:236.443:1.8);
				\filldraw[fill=black] (-1,2) circle (1pt);
				\filldraw[fill=black] (1.5,2) circle (1pt);
				\draw[red,ultra thick](7.71293,3.5) arc(28.9385:-28.9385:3.1);
				\filldraw[dash dot,draw=blue,ultra thick,fill=blue!20] (7.5,2) circle (0.6);
				\filldraw[fill=black] (5,2) circle (1pt);
				\filldraw[fill=black] (7.5,2) circle (1pt);
				\draw[-latex](6.35647,1.25)--(7.71293,0.5);
				\node at(6,1.5) {${\rm R}\xi_{far}$};
		\end{tikzpicture}}\\
		\subfigure[$t\geq t_{peak}(p,c),rp\leq Rc$] 
		{\begin{tikzpicture}
				\draw[->](-9,0) --(9,0) node[right] {$\tau$};
				\draw (-8.5,0) --(-8.5,0.1) node[below]{$|z|/c$};
				\draw (-3,0) --(-3,0.1) node[below]{$\tau_{near}$};
				\draw (6,0) --(6,0.1) node[below]{$\tau_{far}$};
				\filldraw[blue!20] (-7.32666,3.5)arc(152.047:207.953:3.2);
				\filldraw[blue!20](-7.32666,3.5)--(-3.5,3.5)--(-3.5,0.5)--(-7.32666,0.5)--(-7.32666,3.5);
				\draw[red,ultra thick] (-7,2) circle (0.6);
				\draw[dash dot,blue,ultra thick](-7.32666,3.5)arc(152.047:207.953:3.2);
				\filldraw[fill=black] (-7,2) circle (1pt);
				\filldraw[fill=black] (-4.5,2) circle (1pt);
				\filldraw[blue!20] (2.05678,3.5) arc(69.6359:-69.6359:1.6);
				\filldraw[blue!20] (0.94322,3.5) arc(110.364:249.636:1.6);
				\filldraw[blue!20] (2.05678,3.5)--(0.94322,3.5)--(0.94322,0.5)--(2.05678,0.5);
				\draw[red,ultra thick] (-0.443224,3.5) arc(69.6359:-69.6359:1.6);
				\draw[red,ultra thick] (-1.55678,3.5) arc(110.364:249.636:1.6);
				\draw[dash dot,blue,ultra thick] (2.05678,3.5) arc(69.6359:-69.6359:1.6);
				\draw[dash dot,blue,ultra thick] (0.94322,3.5) arc(110.364:249.636:1.6);
				\filldraw[fill=black] (-1,2) circle (1pt);
				\filldraw[fill=black] (1.5,2) circle (1pt);
				\draw[red,ultra thick](7.71293,3.5) arc(28.9385:-28.9385:3.1);
				\filldraw[dash dot,draw=blue,ultra thick,fill=blue!20] (7.5,2) circle (0.6);
				\filldraw[fill=black] (5,2) circle (1pt);
				\filldraw[fill=black] (7.5,2) circle (1pt);
				\draw[-latex](6.35647,1.25)--(7.71293,0.5);
				\node at(6,1.5) {${\rm R}\xi_{far}$};
		\end{tikzpicture}}\\
		\subfigure[$t_{arrival}(p,c)\leq t<t_{peak}(p,c),rp>Rc$] 
		{\begin{tikzpicture}
				\draw[->](-9,0) --(9,0) node[right] {$\tau$};
				\draw (-6,0) --(-6,0.1) node[below]{$\tau_{near}$};
				\draw (6,0) --(6,0.1) node[below]{$\tau_{far}$};
				\draw[red,ultra thick] (-7,2) circle (1.5);
				\filldraw[dash dot,draw=blue,ultra thick,fill=blue!20] (-4.5,2) circle (1);
				\filldraw[fill=black] (-7,2) circle (1pt);
				\filldraw[fill=black] (-4.5,2) circle (1pt);
				\draw[-latex](-7.45,1.4)--(-7.9,0.8);
				\node at(-7.2,1.52) {${\rm R}\xi_{near}$};
				\filldraw[dash dot,draw=blue,ultra thick,fill=blue!20] (1.5,2) circle (0.85);
				\draw[red,ultra thick] (-0.00501256,3.5) arc(56.4427:-56.4427:1.8);
				\draw[red,ultra thick] (-1.99499,3.5) arc(123.557:236.443:1.8);
				\filldraw[fill=black] (-1,2) circle (1pt);
				\filldraw[fill=black] (1.5,2) circle (1pt);
				\draw[red,ultra thick](6.60935,3.5) arc(42.9859:-42.9859:2.2);
				\filldraw[dash dot,draw=blue,ultra thick,fill=blue!20] (7.5,2) circle (0.3);
				\filldraw[fill=black] (5,2) circle (1pt);
				\filldraw[fill=black] (7.5,2) circle (1pt);
				\draw[-latex](5.80468,1.25)--(6.60935,0.5);
				\node at(6,1.5) {${\rm R}\xi_{far}$};
		\end{tikzpicture}}\\
		\subfigure[$t_{peak}(p,c)\leq t<\frac{R}{c},rp>Rc$] 
		{\begin{tikzpicture}
				\draw[->](-9,0) --(9,0) node[right] {$\tau$};
				\draw (-8.5,0) --(-8.5,0.1) node[below]{$|z|/c$};
				\draw (-3,0) --(-3,0.1) node[below]{$\tau_{near}$};
				\draw (6,0) --(6,0.1) node[below]{$\tau_{far}$};
				\filldraw[blue!20] (-7.32666,3.5)arc(152.047:207.953:3.2);
				\filldraw[blue!20](-7.32666,3.5)--(-3.5,3.5)--(-3.5,0.5)--(-7.32666,0.5)--(-7.32666,3.5);
				\draw[red,ultra thick] (-7,2) circle (0.6);
				\draw[dash dot,blue,ultra thick](-7.32666,3.5)arc(152.047:207.953:3.2);
				\filldraw[fill=black] (-7,2) circle (1pt);
				\filldraw[fill=black] (-4.5,2) circle (1pt);
				\filldraw[blue!20] (2.05678,3.5) arc(69.6359:-69.6359:1.6);
				\filldraw[blue!20] (0.94322,3.5) arc(110.364:249.636:1.6);
				\filldraw[blue!20] (2.05678,3.5)--(0.94322,3.5)--(0.94322,0.5)--(2.05678,0.5);	
				\draw[red,ultra thick] (-0.443224,3.5) arc(69.6359:-69.6359:1.6);
				\draw[red,ultra thick] (-1.55678,3.5) arc(110.364:249.636:1.6);
				\draw[dash dot,blue,ultra thick] (2.05678,3.5) arc(69.6359:-69.6359:1.6);
				\draw[dash dot,blue,ultra thick] (0.94322,3.5) arc(110.364:249.636:1.6);
				\filldraw[fill=black] (-1,2) circle (1pt);
				\filldraw[fill=black] (1.5,2) circle (1pt);
				\draw[red,ultra thick](6.60935,3.5) arc(42.9859:-42.9859:2.2);
				\filldraw[dash dot,draw=blue,ultra thick,fill=blue!20] (7.5,2) circle (0.3);
				\filldraw[fill=black] (5,2) circle (1pt);
				\filldraw[fill=black] (7.5,2) circle (1pt);
				\draw[-latex](5.80468,1.25)--(6.60935,0.5);
				\node at(6,1.5) {${\rm R}\xi_{far}$};
		\end{tikzpicture}}\\
		\subfigure[$t\geq\frac{R}{c},rp>Rc$] 
		{\begin{tikzpicture}
				\draw[->](-9,0) --(9,0) node[right] {$\tau$};
				\draw[->](-9,0) --(9,0) node[right] {$\tau$};
				\draw (-8.5,0) --(-8.5,0.1) node[below]{$|z|/c$};
				\draw (-3,0) --(-3,0.1) node[below]{$\tau_{near}$};
				\draw (6,0) --(6,0.1) node[below]{$\tau_{far}$};
				\filldraw[blue!20] (-7.32666,3.5)arc(152.047:207.953:3.2);
				\filldraw[blue!20](-7.32666,3.5)--(-3.5,3.5)--(-3.5,0.5)--(-7.32666,0.5)--(-7.32666,3.5);
				\draw[red,ultra thick] (-7,2) circle (0.6);
				\draw[dash dot,blue,ultra thick](-7.32666,3.5)arc(152.047:207.953:3.2);
				\filldraw[fill=black] (-7,2) circle (1pt);
				\filldraw[fill=black] (-4.5,2) circle (1pt);
				\filldraw[blue!20] (2.05678,3.5) arc(69.6359:-69.6359:1.6);
				\filldraw[blue!20] (0.94322,3.5) arc(110.364:249.636:1.6);
				\filldraw[blue!20] (2.05678,3.5)--(0.94322,3.5)--(0.94322,0.5)--(2.05678,0.5);
				\draw[red,ultra thick] (-0.443224,3.5) arc(69.6359:-69.6359:1.6);
				\draw[red,ultra thick] (-1.55678,3.5) arc(110.364:249.636:1.6);
				\draw[dash dot,blue,ultra thick] (2.05678,3.5) arc(69.6359:-69.6359:1.6);
				\draw[dash dot,blue,ultra thick] (0.94322,3.5) arc(110.364:249.636:1.6);
				\filldraw[fill=black] (-1,2) circle (1pt);
				\filldraw[fill=black] (1.5,2) circle (1pt);
				\draw[red,ultra thick](7.71293,3.5) arc(28.9385:-28.9385:3.1);
				\filldraw[dash dot,draw=blue,ultra thick,fill=blue!20] (7.5,2) circle (0.6);
				\filldraw[fill=black] (5,2) circle (1pt);
				\filldraw[fill=black] (7.5,2) circle (1pt);
				\draw[-latex](6.35647,1.25)--(7.71293,0.5);
				\node at(6,1.5) {${\rm R}\xi_{far}$};
		\end{tikzpicture}}\\
		\caption{Slices with a fixed $\tau$ of the integral area.}
		\label{22th-Sept.-2023-P002}
	\end{figure}
	In situation \textbf{a}, \\
	\begin{subequations}
		\begin{align}
			&\begin{aligned}
				\mathscr{H}_0(t,r,z;&p,c;m,n)\\
				&=\int_{\tau_{near}}^{\tau_{far}}\int_{-\theta(\tau)}^{\theta(\tau)}\int_{0}^{\infty}\frac{\xi^{m+1}}{(\sqrt{\xi^2+z^2})^{n}\sqrt{(t-\tau)^2-|\mathbf{r}-\bm{\xi}|^2/p^2}}\delta(\tau-\frac{\sqrt{\xi^2+z^2}}{c}){\rm d}\xi{\rm d}\theta{\rm d}\tau\\
				&=2pc^{2-n}\int_{\tau_{near}}^{\tau_{far}}\frac{\left(\sqrt{c^2\tau^2-z^2}\right)^{m}}{tau^{n-1}}\int_{0}^{\theta(\tau)}\frac{1}{\sqrt{2r\sqrt{c^2\tau^2-z^2}\cos\theta+p^2(t-\tau)^2-(r^2+c^2\tau^2-z^2)}}{\rm d}\theta{\rm d}\tau,
			\end{aligned}\\
			&\begin{aligned}
				\mathscr{H}_1(t,r,z;&p,c;m,n)\\
				&=\int_{\tau_{near}}^{\tau_{far}}\int_{-\theta(\tau)}^{\theta(\tau)}\int_{0}^{\infty}\frac{\xi^{m+1}\cos\theta}{(\sqrt{\xi^2+z^2})^{n}\sqrt{(t-\tau)^2-|\mathbf{r}-\bm{\xi}|^2/p^2}}\delta(\tau-\frac{\sqrt{\xi^2+z^2}}{c}){\rm d}\xi{\rm d}\theta{\rm d}\tau\\
				&=2pc^{2-n}\int_{\tau_{near}}^{\tau_{far}}\frac{\left(\sqrt{c^2\tau^2-z^2}\right)^{m}}{\tau^{n-1}}\int_{0}^{\theta(\tau)}\frac{\cos\theta}{\sqrt{2r\sqrt{c^2\tau^2-z^2}\cos\theta+p^2(t-\tau)^2-(r^2+c^2\tau^2-z^2)}}{\rm d}\theta{\rm d}\tau.\\	
			\end{aligned}
		\end{align}
		\label{9th-Nov.-2023-E001}
	\end{subequations}
	where\\
	\begin{equation}
		\cos\theta(\tau)=-\frac{p^2(t-\tau)^2-(r^2+c^2\tau^2-z^2)}{2r\sqrt{c^2\tau^2-z^2}}.
	\end{equation}
	Introduce two integral equations:\\
	\begin{subequations}
		\begin{align}
			&\int_{0}^{\arccos\left(-\frac{a}{b}\right)}\frac{{\rm d}x}{\sqrt{a+b\cos x}}=\sqrt{\frac{2}{b}}\mathbf{K}(\sqrt{\frac{a+b}{2b}}),\quad b\geq|a|>0,\\
			&\int_{0}^{\arccos\left(-\frac{a}{b}\right)}\frac{\cos\theta}{\sqrt{a+b\cos x}}{\rm d}x=\sqrt{\frac{2}{b}}\left(2\mathbf{E}(\sqrt{\frac{a+b}{2b}})-\mathbf{K}(\sqrt{\frac{a+b}{2b}})\right),\quad b\geq|a|>0.
		\end{align}
		\label{9th-Nov.-2023-E002}
	\end{subequations}
	where $\mathbf{K}(k)$ and $\mathbf{E}(k)$ are the first and the second elliptic integral.\\
	\begin{subequations}
		\begin{align}
			&\mathbf{K}(k)=\int_0^1\frac{{\rm d}t}{\sqrt{(1-t^2)(1-k^2t^2)}}>0,\quad 0<k<1,\\
			&\mathbf{E}(k)=\int_0^1\sqrt{\frac{1-k^2t^2}{1-t^2}}{\rm d}t>0,\quad 0<k<1.
		\end{align}
	\end{subequations}
	\ref{9th-Nov.-2023-E001} can be calculated by \ref{9th-Nov.-2023-E002},\\
	\begin{subequations}
		\begin{align}
			&\mathscr{H}_0(t,r,z;p,c;m,n)=\frac{2pc^{2-n}}{\sqrt{r}}\int_{\tau_{near}}^{\tau_{far}}\frac{\left(c^2\tau^2-z^2\right)^{m/2-1/4}}{\tau^{n-1}}\mathbf{K}(X){\rm d}\tau,\\
			&\mathscr{H}_1(t,r,z;p,c;m,n)=\frac{2pc^{2-n}}{\sqrt{r}}\int_{\tau_{near}}^{\tau_{far}}\frac{\left(c^2\tau^2-z^2\right)^{m/2-1/4}}{\tau^{n-1}}\left(2\mathbf{E}(X)-\mathbf{K}(X)\right){\rm d}\tau,
		\end{align}
	\end{subequations}
	where,\\
	\begin{equation}
		X(\tau;t,r,z,p,c):=\sqrt{\frac{p^2(t-\tau)^2-(r-\sqrt{c^2\tau^2-z^2})^2}{4r\sqrt{c^2\tau^2-z^2}}}.
	\end{equation}
	In situation \textbf{b},\\
	\begin{subequations}
		\begin{align}
			&\begin{aligned}
				\mathscr{H}_0(t,r,z;&p,c;m,n)\\
				&=\int_{|z|/c}^{\tau_{near}}\int_{-\pi}^{\pi}\int_{0}^{\infty}\frac{\xi^{m+1}}{(\sqrt{\xi^2+z^2})^{n}\sqrt{(t-\tau)^2-|\mathbf{r}-\bm{\xi}|^2/p^2}}\delta(\tau-\frac{\sqrt{\xi^2+z^2}}{c}){\rm d}\xi{\rm d}\theta{\rm d}\tau\\
				&+\int_{\tau_{near}}^{\tau_{far}}\int_{-\theta(\tau)}^{\theta(\tau)}\int_{0}^{\infty}\frac{\xi^{m+1}}{(\sqrt{\xi^2+z^2})^{n}\sqrt{(t-\tau)^2-|\mathbf{r}-\bm{\xi}|^2/p^2}}\delta(\tau-\frac{\sqrt{\xi^2+z^2}}{c}){\rm d}\xi{\rm d}\theta{\rm d}\tau\\
				&=2pc^{2-n}\int_{|z|/c}^{\tau_{near}}\frac{\left(\sqrt{c^2\tau^2-z^2}\right)^{m}}{\tau^{n-1}}\int_{0}^{\pi}\frac{1}{\sqrt{2r\sqrt{c^2\tau^2-z^2}\cos\theta+p^2(t-\tau)^2-(r^2+c^2\tau^2-z^2)}}{\rm d}\theta{\rm d}\tau\\
				&+2pc^{2-n}\int_{\tau_{near}}^{\tau_{far}}\frac{\left(\sqrt{c^2\tau^2-z^2}\right)^{m}}{\tau^{n-1}}\int_{0}^{\theta(\tau)}\frac{1}{\sqrt{2r\sqrt{c^2\tau^2-z^2}\cos\theta+p^2(t-\tau)^2-(r^2+c^2\tau^2-z^2)}}{\rm d}\theta{\rm d}\tau	,
			\end{aligned}\\
			&\begin{aligned}
				\mathscr{H}_1(t,r,z;&p,c;m,n)\\
				&=\int_{|z|/c}^{\tau_{near}}\int_{-\pi}^{\pi}\int_{0}^{\infty}\frac{\xi^{m+1}\cos\theta}{(\sqrt{\xi^2+z^2})^{n}\sqrt{(t-\tau)^2-|\mathbf{r}-\bm{\xi}|^2/p^2}}\delta(\tau-\frac{\sqrt{\xi^2+z^2}}{c}){\rm d}\xi{\rm d}\theta{\rm d}\tau\\
				&+\int_{\tau_{near}}^{\tau_{far}}\int_{-\theta(\tau)}^{\theta(\tau)}\int_{0}^{\infty}\frac{\xi^{m+1}\cos\theta}{(\sqrt{\xi^2+z^2})^{n}\sqrt{(t-\tau)^2-|\mathbf{r}-\bm{\xi}|^2/p^2}}\delta(\tau-\frac{\sqrt{\xi^2+z^2}}{c}){\rm d}\xi{\rm d}\theta{\rm d}\tau\\
				&=2pc^{2-n}\int_{|z|/c}^{\tau_{near}}\frac{\left(\sqrt{c^2\tau^2-z^2}\right)^{m}}{\tau^{n-1}}\int_{0}^{\pi}\frac{\cos\theta}{\sqrt{2r\sqrt{c^2\tau^2-z^2}\cos\theta+p^2(t-\tau)^2-(r^2+c^2\tau^2-z^2)}}{\rm d}\theta{\rm d}\tau\\
				&+2pc^{2-n}\int_{\tau_{near}}^{\tau_{far}}\frac{\left(\sqrt{c^2\tau^2-z^2}\right)^{m}}{\tau^{n-1}}\int_{0}^{\theta(\tau)}\frac{\cos\theta}{\sqrt{2r\sqrt{c^2\tau^2-z^2}\cos\theta+p^2(t-\tau)^2-(r^2+c^2\tau^2-z^2)}}{\rm d}\theta{\rm d}\tau	.
			\end{aligned}
		\end{align}
		\label{9th-Nov.-2023-E003}
	\end{subequations}
	Introduce another two integral equations,\\
	\begin{subequations}
		\begin{align}
			&\int_0^{\pi}\frac{{\rm d}x}{\sqrt{a+b\cos x}}=\frac{2}{\sqrt{a+b}}\mathbf{K}(\sqrt{\frac{2b}{a+b}}),\quad a>b>0,\\
			&\int_0^{\pi}\frac{\cos\theta}{\sqrt{a+b\cos x}}{\rm d}x=\frac{2}{b\sqrt{a+b}}\left((a+b)\mathbf{E}(\sqrt{\frac{2b}{a+b}})-a\mathbf{K}(\sqrt{\frac{2b}{a+b}})\right),\quad a>b>0.
		\end{align}
		\label{9th-Nov.-2023-E004}
	\end{subequations}
	\ref{9th-Nov.-2023-E003} can be calculated by \ref{9th-Nov.-2023-E002} and \ref{9th-Nov.-2023-E004}.\\
	\begin{subequations}
		\begin{align}
			&\begin{aligned}
				\mathscr{H}_0(t,r,z;p,c;m,n)&=\frac{2pc^{2-n}}{\sqrt{r}}\int_{\tau_{near}}^{\tau_{far}}\frac{\left(c^2\tau^2-z^2\right)^{m/2-1/4}}{\tau^{n-1}}\mathbf{K}(X){\rm d}\tau\\
				&+2pc^{2-n}\int_{|z|/c}^{\tau_{near}}\frac{\left(c^2\tau^2-z^2\right)^{m/2}}{\tau^{n-1}}\frac{2}{\sqrt{p^2(t-\tau)^2-(r-\sqrt{c^2\tau^2-z^2})^2}}\mathbf{K}(\frac{1}{X}){\rm d}\tau,
			\end{aligned}\\
			&\begin{aligned}
				\mathscr{H}_1(t,r,z;p,c;&m,n)=\frac{2pc^{2-n}}{\sqrt{r}}\int_{\tau_{near}}^{\tau_{far}}\frac{\left(c^2\tau^2-z^2\right)^{m/2-1/4}}{\tau^{n-1}}\left(2\mathbf{E}(X)-\mathbf{K}(X)\right){\rm d}\tau\\
				+\frac{2pc^{2-n}}{r}&\int_{|z|/c}^{\tau_{near}}\frac{\left(c^2\tau^2-z^2\right)^{m/2-1/2}}{\tau^{n-1}}\\
				&\left(\sqrt{p^2(t-\tau)^2-(r-\sqrt{c^2\tau^2-z^2})^2}\mathbf{E}(\frac{1}{X})-\frac{p^2(t-\tau)^2-(r^2+c^2\tau^2-z^2)}{\sqrt{p^2(t-\tau)^2-(r-\sqrt{c^2\tau^2-z^2})^2}}\mathbf{K}(\frac{1}{X})\right){\rm d}\tau.
			\end{aligned}
		\end{align}
		\label{11th-Mar.-2024-E001}
	\end{subequations}
	$\tau_{near}$ and $\tau_{far}$ in the above discussions are the intersections on the plane $\tau>0$ between two lines
	\begin{subequations}
		\begin{align}
			&L_1(\tau,\xi):p(t-\tau)=r-\xi,\\
			&L_2(\tau,\xi):p(t-\tau)=\xi-r.
		\end{align}
	\end{subequations}
	and a hyperbola:
	\begin{equation}
		\Gamma(\tau,\xi):c\tau=\sqrt{\xi^2+z^2},
	\end{equation}
	which are in the exact forms\\
	\begin{subequations}
		\begin{align}
			&\tau_{near}=\frac{1}{c^2-p^2}\left(\sqrt{p^2(pt-r)^2+(c^2-p^2)((pt-r)^2+z^2)}-p(pt-r)\right),\\
			&\tau_{far}=\left\{\begin{array}{ll}
				\frac{1}{p^2-c^2}\left(\sqrt{p^2(pt-r)^2+(c^2-p^2)((pt-r)^2+z^2)}+p(pt-r)\right) & ,\quad pr>cR\quad \text{and}\quad t_{arrival}<t<\frac{R}{c}\\
				\frac{1}{c^2-p^2}\left(\sqrt{p^2(pt+r)^2+(c^2-p^2)((pt+r)^2+z^2)}-p(pt+r)\right) & ,\quad \text{else}.\\
			\end{array}\right.
		\end{align}
	\end{subequations}
	\subsection{Transformations from position to time partial derivates}\label{C.2}
	This appendix is the base of the transform from Green's function to Time-Green function, which is explained in \ref{3.2.1}. The equations provided by this appendix are used in \ref{3.2} to get the analytic solutions, and in \ref{4.3.2} to get the approximations. The transformations in this appendix are depended on the denominations of $(r,t)$ and $(z,t)$ and the property of the derivative transmition of convolution.\\
	\begin{align}
		&\begin{aligned}
			\frac{\partial}{\partial z}\mathscr{H}_0(t,r,z;p,c;0,1)&=\left<\frac{{\rm H}(t-\frac{r}{p})}{\sqrt{t^2-r^2/p^2}},\frac{\partial}{\partial z}\frac{1}{R}\delta(t-\frac{R}{c})\right>\\
			&=-\frac{z}{c}\frac{\partial}{\partial t}\mathscr{H}_0(t,r,z;p,c;0,2)-z\mathscr{H}_0(t,r,z;p,c;0,3);\\
		\end{aligned}\\
		&\begin{aligned}
			&\frac{\partial}{\partial r}\mathscr{H}_0(t,r,z;p,c;m,n)=\left<\frac{{\rm H}(t-\frac{r}{p})}{\sqrt{t^2-r^2/p^2}},\frac{\partial}{\partial x_1}\frac{r^{m-1}}{R^{n}}\delta(t-\frac{R}{c})\right>\text{ (in axisymmetric case only)}\\
			&=-\frac{1}{c}\frac{\partial}{\partial t}\mathscr{H}_1(t,r,z;p,c;m+1,n+1)+m\mathscr{H}_1(t,r,z;p,c;m-1,n)-n\mathscr{H}_1(t,r,z;p,c;m+1,n+2);\\
		\end{aligned}\label{8th.-Jan.-2024-E002}\\
		&\begin{aligned}
			&\frac{\partial^2}{\partial z\partial r}\mathscr{H}_0(t,r,z;p,c;0,1)=\left<\frac{{\rm H}(t-\frac{r}{p})}{\sqrt{t^2-r^2/p^2}},\frac{\partial^2}{\partial z\partial x_1}\frac{1}{R}\delta(t-\frac{R}{c})\right>\text{ (in axisymmetric case only.)}\\
			&=\frac{z}{c^2}\frac{\partial^2}{\partial t^2}\mathscr{H}_1(t,r,z;p,c;1,3)+\frac{3z}{c}\frac{\partial}{\partial t}\mathscr{H}_1(t,r,z;p,c;1,4)+3z\mathscr{H}_1(t,r,z;p,c;1,5);\\
		\end{aligned}\\
		&\begin{aligned}
			&\frac{\partial^2}{\partial z^2}\mathscr{H}_0(t,r,z;p,c;0,1)=\left<\frac{{\rm H}(t-\frac{r}{p})}{\sqrt{t^2-r^2/p^2}},\frac{\partial^2}{\partial z^2}\frac{1}{R}\delta(t-\frac{R}{c})\right>\\
			&=\frac{z^2}{c^2}\frac{\partial^2}{\partial t^2}\mathscr{H}_0(t,r,z;p,c;0,3)+\frac{\partial}{\partial t}\left(\frac{3z^2}{c}\mathscr{H}_0(t,r,z;p,c;0,4)-\frac{1}{c}\mathscr{H}_0(t,r,z;p,c;0,2)\right)\\
			&+3z^2\mathscr{H}_0(t,r,z;p,c;0,5)-\mathscr{H}_0(t,r,z;p,c;0,3);\\
		\end{aligned}\\
		&\begin{aligned}
			&\nabla_r^2\mathscr{H}_0(t,r,z;p,c;m,n)=\left<\frac{{\rm H}(t-\frac{r}{p})}{\sqrt{t^2-r^2/p^2}},\nabla_r^2\frac{r^{m}}{R^{n}}\delta(t-\frac{R}{c})\right>\\
			&=\frac{1}{c^2}\frac{\partial^2}{\partial t^2}\mathscr{H}_0(t,r,z;p,c;m+2,n+2)\\
			&+\frac{\partial}{\partial t}\left(\frac{n+2}{c}\mathscr{H}_0(t,r,z;p,c;m+2,n+3)-\frac{m+2}{c}\mathscr{H}_0(t,r,z;p,c;m,n+1)\right)\\
			&+m^2\mathscr{H}_0(t,r,z;p,c;m-2,n)+n(n+2)\mathscr{H}_0(t,r,z;p,c;m+2,n+4)\\
			&-2(m+1)n\mathscr{H}_0(t,r,z;p,c;m,n+2);\\
		\end{aligned}\\
		&\begin{aligned}
			&\frac{\partial^2}{\partial r^2}\mathscr{H}_0(t,r,z;p,c;m,n)=\frac{1}{c^2}\frac{\partial^2}{\partial t^2}\mathscr{H}_2(;;m+2,n+2)\\
			&+\frac{1}{c}\frac{\partial}{\partial t}\left((2n+1)\mathscr{H}_2(;;m+2,n+3)-2m\mathscr{H}_2(;;m,n+1)-\mathscr{H}_0(;;m,n+1)\right)\\
			&+m\mathscr{H}_0(;;m-2,n)-n\mathscr{H}_0(;;m,n+2)+m(m-2)\mathscr{H}_2(;;0,m-2,n)\\
			&-2mn\mathscr{H}_2(;;0,m,n+2)+n(n+2)\mathscr{H}_2(;;0,m+2,n+4).\\
		\end{aligned}\label{8th.-Jan.-2024-E005}
	\end{align}
	\subsection{Simpified expressions of receivers on the axis}\label{C.3}
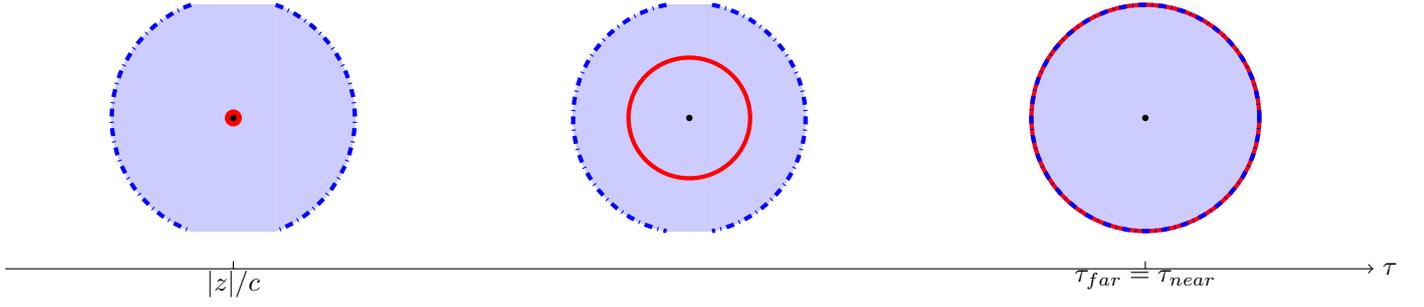
\begin{figure}[h!]
		\centering
		\begin{tikzpicture}
			\draw[->](-9,0) --(9,0) node[right] {$\tau$};
			\draw (-6,0) --(-6,0.1) node[below]{$|z|/c$};
			\draw (6,0) --(6,0.1) node[below]{$\tau_{far}=\tau_{near}$};
			\filldraw[blue!20] (-5.44322,3.5) arc(69.6359:-69.6359:1.6);
			\filldraw[blue!20] (-6.55678,3.5) arc(110.364:249.636:1.6);
			\filldraw[blue!20] (-5.44322,3.5)--(-6.55678,3.5)--(-6.55678,0.5)--(-5.44322,0.5);
			\draw[dash dot,blue,ultra thick] (-5.44322,3.5) arc(69.6359:-69.6359:1.6);
			\draw[dash dot,blue,ultra thick] (-6.55678,3.5) arc(110.364:249.636:1.6);
			\filldraw[red] (-6,2) circle (3pt);
			\filldraw[black] (-6,2) circle (1pt);
			\filldraw[blue!20] (0.3,3.5) arc(78.6901:-78.6901:1.529706);
			\filldraw[blue!20] (-0.3,3.5) arc(101.31:258.69:1.529706);
			\filldraw[blue!20] (0.3,3.5)--(-0.3,3.5)--(-0.3,0.5)--(0.3,0.5);
			\draw[dash dot,blue,ultra thick] (0.3,3.5) arc(78.6901:-78.6901:1.529706);
			\draw[dash dot,blue,ultra thick] (-0.3,3.5) arc(101.31:258.69:1.529706);
			\draw[red,ultra thick] (0,2) circle (0.8);
			\filldraw[black] (0,2) circle (1pt);
			\filldraw[blue!20] (6,2) circle (1.5);
			\draw[red,ultra thick] (6,2) circle (1.5);
			\draw[dash dot,blue,ultra thick] (6,2) circle (1.5);
			\filldraw[black] (6,2) circle (1pt);
		\end{tikzpicture}
		\caption{Slices with a fixed $\tau$ of the integral area when $r=0$.}\label{22th-Mar.-2024-P008}
	\end{figure}

	The equations in this appendix are used in \ref{3.2.2} to simplify the displacements on the axis, and in \ref{4.3.2} to calculate the approximations. When $r=0$, the calculations of $\mathscr{H}_0(;p,c;m=2l,n)$ are going under the axisymmetry. Consider geometrically, the hyperbolic surface and the cone are directly opposite. The area of integral is shown in \ref{22th-Mar.-2024-P008}.\\
	\\
	The expressions of $\mathscr{H}_0(;p,c;2l,n)$ can be normally simplified to
	\begin{equation}
		\begin{aligned}
			\mathscr{H}_0(;p,c;2l,n)&=\int_{|z|/c}^{T}\int_{-\pi}^{\pi}\int_{0}^{\infty}\frac{\xi^{2l+1}}{(\sqrt{\xi^2+z^2})^{n}\sqrt{(t-\tau)^2-\xi^2/p^2}}\delta(\tau-\frac{\sqrt{\xi^2+z^2}}{c}){\rm d}\xi{\rm d}\theta{\rm d}\tau\\
			&=2\pi pc^{2-n}{\rm H}(t-\frac{|z|}{c})\int_{|z|/c}^{T}\frac{(c^2\tau^2-z^2)^l}{\tau^{n-1}\sqrt{(p^2-c^2)\tau^2-2p^2t\tau+p^2t^2+z^2}}{\rm d}\tau\\
			&=2\pi pc^{2-n}{\rm H}(t-\frac{|z|}{c})\sum_{i=0}^l\tbinom{l}{i}(-z^2)^{l-i}c^{2i}\int_{|z|/c}^{T}\frac{1}{\tau^{n-2i-1}\sqrt{(p^2-c^2)\tau^2-2p^2t\tau+p^2t^2+z^2}}{\rm d}\tau,
		\end{aligned}\label{9th-Jan.-2024-E001}
	\end{equation}
	where 
	\begin{equation}
		\begin{aligned}
			&T=\frac{1}{c^2-p^2}\left(\sqrt{c^2p^2t^2+(c^2-p^2)z^2}-p^2t\right)\\
		\end{aligned}
	\end{equation}
	The integral in the form of\\
	\begin{equation}
		I_n(x;a_0>0,a_1,a_2):=\int\frac{1}{x^n\sqrt{a_0+a_1x+a_2x^2}}{\rm d}x
	\end{equation}
	can be expressed into elementary functions.\\
	\begin{equation}
		\begin{aligned}
			I_0&=\left\{\begin{array}{ll}
				-\frac{1}{\sqrt{-a_2}}\arcsin\frac{2a_2x+a_1}{\sqrt{-q}} & ,\quad a_2<0\\
				\frac{1}{\sqrt{a_2}}\ln\left(2\sqrt{a_2Y}+2a_2x+a_1\right) & ,\quad a_2>0\\			
			\end{array}\right.,\\
			I_1&=-\frac{1}{\sqrt{a_0}}\ln\frac{2\sqrt{a_0Y}+a_1x+2a_0}{x},\\			
			n\geq2,\quad I_n&=-\frac{\sqrt{Y}}{(n-1)a_0x^{n-1}}-\frac{(2n-3)a_1}{2(n-1)a_0}I_{n-1}-\frac{(n-2)a_2}{(n-1)a_0}I_{n-2} .
		\end{aligned}
		\notag
	\end{equation}
	where
	\begin{equation}
		\begin{aligned}
			Y&=a_0+a_1x+a_2x^2,\quad Y(T;p^2t^2-z^2,-2p^2t,p^2-c^2)=0\\
			q&=4a_0a_2-a_1^2.
		\end{aligned}
		\notag
	\end{equation}
	$I_n$ can be calculated by recurrence formula.\\
	\begin{equation}
		\begin{aligned}
			I_2&=-\frac{1}{a_0}\left(\frac{\sqrt{Y}}{x}+\frac{a_1}{2}I_1\right)\\
			I_3&=\frac{\sqrt{Y}}{a_0^2x^2}\left(a_1x-\frac{2a_0+a_1x}{4}\right)+\frac{2a_1^2-q}{8a_0^2}I_1\\
			...
		\end{aligned}
		\notag
	\end{equation}
	The expressions of $G^{(T)}_{0-z}$, $G^{(T)}_{1-z}$ and $G^{(T)}_{2-z}$, and the formulas used in \ref{4.3.2} are all composed by these following functions.\\
	\begin{equation}
		\begin{aligned}
			\mathscr{H}_0(;p,c;2l,n)&=\left.2\pi pc^{2-n}{\rm H}(t-\frac{|z|}{c})\sum_{i=0}^l\tbinom{l}{i}(-z^2)^{l-i}c^{2i}I_{n-2i-1}(\tau;p^2t^2-z^2,-2p^2t,p^2-c^2)\right|^T_{|z|/c}.
		\end{aligned}\label{23th.-Feb.-2024-E001}
	\end{equation}
	Further, with $t\to\infty$, some of the limits of Huygens functions are given below. These limits are used to get \ref{2nd.-Dec.-2023-T001} in \ref{4.2.3}.\\
	\begin{subequations}
		\begin{align}
			&\begin{aligned}
				\lim_{t\to\infty}\mathscr{H}_0(t,0,z;p,c;0,2)&=2\pi\lim_{t\to\infty}\frac{\ln t}{t}=0
			\end{aligned}\\
			&\begin{aligned}
				\lim_{t\to\infty}\mathscr{H}_0(t,0,z;p,c;0,3)&=2\pi\lim_{t\to\infty}\left(\frac{1}{|z|}\frac{1}{t}+\frac{1}{c}\frac{\ln t}{t^2}\right)=0
			\end{aligned}\\
			&\begin{aligned}
				\lim_{t\to\infty}\mathscr{H}_0(t,0,z;p,c;0,4)&=\pi\lim_{t\to\infty}\left(\frac{1}{|z|^2}\frac{1}{t}+\frac{2}{c|z|}\frac{1}{t^2}+\frac{2p^2+c^2}{p^2c^2}\frac{\ln t}{t^3}\right)=0
			\end{aligned}\\
			&\begin{aligned}
				\lim_{t\to\infty}\mathscr{H}_0(t,0,z;p,c;0,5)&=2\pi\lim_{t\to\infty}\left(\frac{1}{3|z|^3}\frac{1}{t}+\frac{1}{2c|z|^2}\frac{1}{t^2}+\frac{2c^2+3p^2}{3p^2c^2|z|}\frac{1}{t^3}+\frac{3c^2+2p^2}{2p^2}\frac{\ln t}{t^4}\right)=0
			\end{aligned}\\
			p<c_T,\quad&\begin{aligned}
				\lim_{t\to\infty}\mathscr{H}_0(t,0,z;p,c_T;2,3)&=\frac{2\pi pc_T}{\sqrt{c_T^2-p^2}}\left(\frac{\pi}{2}-\arcsin\frac{p}{c_T}\right)-2\pi\lim_{t\to\infty}\frac{|z|^2}{c_T}\frac{\ln t}{t}
			\end{aligned}\\
			p>c_T,\quad&\begin{aligned}
				\lim_{t\to\infty}\mathscr{H}_0(t,0,z;p,c_T;2,3)&=\frac{2\pi pc_T}{\sqrt{p^2-c_T^2}}\ln\left(\sqrt{\frac{p^2}{c_T^2}-1}+\frac{p}{c_T}\right)-2\pi\lim_{t\to\infty}\frac{|z|^2}{c_T}\frac{\ln t}{t}
			\end{aligned}\\
			&\begin{aligned}
				\lim_{t\to\infty}\mathscr{H}_0(t,0,z;p,c_T;2,4)&=2\pi\lim_{t\to\infty}\frac{\ln t}{t}=0
			\end{aligned}\\
			&\begin{aligned}
				\lim_{t\to\infty}\mathscr{H}_0(t,0,z;p,c_T;2,5)&=2\pi\lim_{t\to\infty}\left(\frac{2}{3|z|}\frac{1}{t}+\frac{1}{c_T}\frac{\ln t}{t^2}\right)=0
			\end{aligned}
		\end{align}
	\end{subequations}
	\subsection{Changed forms on the free surface}\label{C.4}
	This appendix is a supplement to \ref{3.2.1}, giving the process to get the displacements on the free surface.\\
	\\
	Firstly, the process of how \ref{30th-Sept.-2023-E001} changes to Time-Green functions is shown in this section. In the terms of \ref{30th-Sept.-2023-E001}, ${\rm H}(t^2-r^2/p^2)$ can replace ${\rm H}(t-r/p)$, beacuse $t+r/p>0$ establishes all the time. Then the position partial derivative $\frac{\partial}{\partial r}$ can be changed into the form of
	\begin{equation}
		\frac{\partial}{\partial r}=-\frac{r}{p^2t}\frac{\partial}{\partial t}.\label{3rd-Oct.-2023-E001}
	\end{equation}\\
	Later, observe that\\
	\begin{equation}
		\begin{aligned}
			\frac{\partial}{\partial r}\frac{{\rm H}(t-r/p)}{\sqrt{t^2-r^2/p^2}}*F(t)&=-\frac{r}{p^2}\int_0^{t}\frac{F(t-\tau)}{\tau}\frac{\partial}{\partial\tau}\frac{{\rm H}(\tau-r/p)}{\sqrt{\tau^2-r^2/p^2}}{\rm d}\tau\\
			&=-\frac{r}{p^2}{\rm H}(t-\frac{r}{p})\int_{r/p}^t\frac{F(t-\tau)}{\tau^2\sqrt{\tau^2-r^2/p^2}}{\rm d}\tau-\frac{r}{p^2}{\rm H}(t-\frac{r}{p})\int_{r/p}^t\frac{F'(t-\tau)}{\tau\sqrt{\tau^2-r^2/p^2}}{\rm d}\tau\\
			&{\color{red}{-\frac{rF(0)}{p^2t}\frac{{\rm H}(t-\frac{r}{p})}{\sqrt{t^2-r^2/p^2}}}}\\
			&=-\frac{1}{r}{\rm H}(t-\frac{r}{p})\int_{r/p}^t\frac{\sqrt{\tau^2-r^2/p^2}}{\tau}F'(t-\tau){\rm d}\tau-\frac{1}{p}{\rm H}(t-\frac{r}{p})\int_{r/p}^t\arccos(\frac{r}{p\tau})F''(t-\tau){\rm d}\tau\\
			&{\color{red}{-\left[\left(\frac{r}{p^2\sqrt{t^2-r^2/p^2}}+\frac{\sqrt{t^2-r^2/p^2}}{r}\right)\frac{F(0)}{t}+\frac{1}{p}\arccos(\frac{r}{pt})F'(0)\right]{\rm H}(t-\frac{r}{p})}}.
		\end{aligned}
		\notag
	\end{equation}
	Denote that\\
	\begin{subequations}
		\begin{align}
			&S_1(t,r;p)=-\frac{\sqrt{t^2-r^2/p^2}}{rt}{\rm H}(t-\frac{r}{p})\\
			&S_2(t,r;p)=-\frac{1}{p}\arccos(\frac{r}{pt}){\rm H}(t-\frac{r}{p})\\
			&ET(t,r;p)=-\left[\left(\frac{r}{p^2\sqrt{t^2-r^2/p^2}}+\frac{\sqrt{t^2-r^2/p^2}}{r}\right)\frac{F(0)}{t}+\frac{1}{p}\arccos(\frac{r}{pt})F'(0)\right]{\rm H}(t-\frac{r}{p}).
		\end{align}
	\end{subequations}
	Finally, we get the Time-Green functions and extra term for Informal sound sources functions.\\
	\\
	Secondly, consider the axial displacement \ref{15th-Sept.-2023-E014b}. When $z=0$,\\
	\begin{equation}
		\begin{aligned}
			G_z&=\frac{\rho}{4\pi^2\mathbf{i}}\int_{\sigma-\mathbf{i}\infty}^{\sigma+\mathbf{i}\infty}\int_{0}^{\infty}\frac{\rho s^2k_L}{(\rho s^2+2\mu k^2)^2-4\mu^2k^2k_Lk_T}k{\rm J}_0(kr)e^{st}{\rm d}k{\rm d}s\\
			&=\frac{\rho}{4\pi^2\mathbf{i}}\int_{\sigma-\mathbf{i}\infty}^{\sigma+\mathbf{i}\infty}\int_{0}^{\infty}\left[\frac{\rho s^2k_Lk_T}{(\rho s^2+2\mu k^2)^2-4\mu^2k^2k_Lk_T}\times\frac{1}{k_T}\right]k{\rm J}_0(kr)e^{st}{\rm d}k{\rm d}s\\
			&=\frac{\rho}{c_T^2}\frac{\partial^2}{\partial t^2}\left<\left[\frac{e^{\sigma t}}{4\rho\pi^2}\int_{0}^{\infty}\int_{-\infty}^{\infty}\frac{\rho s^2k_Lk_T}{(\rho s^2+2\mu k^2)^2-4\mu^2k^2k_Lk_T}e^{\mathbf{i}\omega t}{\rm d}\omega k{\rm J}_0(kr){\rm d}k\right],\frac{\delta(t-r/c_T)}{2\pi r}\right>\\
			&=\frac{\rho}{c_T^2}\frac{\partial^2}{\partial t^2}\left<\frac{1}{2}T\text{( distribuation term)},\frac{\delta(t-r/c_T)}{2\pi r}\right>\\
			&=\frac{M}{4\pi c_T^2r}\delta(t-\frac{r}{c_T})-\frac{1}{4\pi c_T^2}\frac{\partial^2}{\partial t^2}\left(N_2\mathscr{H}_0(t,r,0;ac_T,c_T;0,1)+C_2\int_1^{1/\sqrt{\eta}}W_T(x)\mathscr{H}_0(t,r,0;xc_T,c_T;0,1){\rm d}x\right).
		\end{aligned}
		\label{30th-Oct.-2023-E001}
	\end{equation}
	Here, differential properties of integral transforms are used to get the last equation of \ref{30th-Oct.-2023-E001}. Imitate method in \ref{3.1} to calculate \ref{30th-Oct.-2023-E001}, and the Time-Green functions can be written as\\
	\begin{subequations}
		\begin{align}
			z=0,\quad&G^{(T)}_{0-z}=\frac{M}{4\pi c_T^2r}\delta(t-\frac{r}{c_T}),\\
			&G^{(T)}_{1-z}=0,\\
			&G^{(T)}_{2-z}=-\frac{1}{4\pi c_T^2}\left(N_2\mathscr{H}_0(t,r,0;ac_T,c_T;0,1)+C_2\int_1^{1/\sqrt{\eta}}W_T(x)\mathscr{H}_0(t,r,0;xc_T,c_T;0,1){\rm d}x\right).
		\end{align}
	\end{subequations}
	\subsection{Exact expressions of approximations}\label{C.5}
	This appendix gives the exact expressions of the components in \ref{22th-Mar.-2024-E001a} and \ref{22th-Mar.-2024-E002a}. The components of $\tilde{\mathbf{B}}_z(p,c)$ can be calculated by \ref{8th.-Jan.-2024-E005}. Its value at $r=0$ is given by \ref{23th.-Feb.-2024-E001}.\\
	\begin{subequations}
		\begin{align}
			&\tilde{\mathbf{B}}_z(p,c_L;r=0)[0]=15z^2\left[7\mathscr{H}_0(;2,9)-2\mathscr{H}_0(;0,7)\right]-3\left[5\mathscr{H}_0(;2,7)-2\mathscr{H}_0(;0,5)\right],\\
			&\tilde{\mathbf{B}}_z(p,c_L;r=0)[1]=\frac{3z^2}{c_L}\left[31\mathscr{H}_0(;2,8)-10\mathscr{H}_0(;0,6)\right]-\frac{1}{c_L}\left[13\mathscr{H}_0(;2,6)-6\mathscr{H}_0(;0,4)\right],\\
			&\tilde{\mathbf{B}}_z(p,c_L;r=0)[2]=\frac{3z^2}{c_L^2}\left[13\mathscr{H}_0(;2,7)-4\mathscr{H}_0(;0,5)\right]-\frac{1}{c_L^2}\left[5\mathscr{H}_0(;2,5)-2\mathscr{H}_0(;0,3)\right],\\
			&\tilde{\mathbf{B}}_z(p,c_L;r=0)[3]=\frac{z^2}{c_L^3}\left[8\mathscr{H}_0(;2,6)-3\mathscr{H}_0(;0,4)\right],\\
			&\tilde{\mathbf{B}}_z(p,c_L;r=0)[4]=\frac{z^2}{c_L^4}\mathscr{H}_0(;2,5),\\
			&\tilde{\mathbf{B}}_z(p,c_T;r=0)[0]=3\left[8\mathscr{H}_0(;0,5)+35\mathscr{H}_0(;4,9)-40\mathscr{H}_0(;2,7)\right],\\
			&\tilde{\mathbf{B}}_z(p,c_T;r=0)[1]=\frac{3}{c_T}\left[39\mathscr{H}_0(;4,8)-44\mathscr{H}_0(;2,6)+8\mathscr{H}_0(;0,4)\right],\\
			&\tilde{\mathbf{B}}_z(p,c_T;r=0)[2]=\frac{4}{c_T^2}\left[9\mathscr{H}_0(;4,7)-4\mathscr{H}_0(;2,5)+2\mathscr{H}_0(;0,3)\right],\\
			&\tilde{\mathbf{B}}_z(p,c_T;r=0)[3]=\frac{2}{c_T^3}\left[4\mathscr{H}_0(;4,6)-3\mathscr{H}_0(;2,4)\right],\\
			&\tilde{\mathbf{B}}_z(p,c_T;r=0)[4]=\frac{1}{c_T^4}\mathscr{H}_0(;4,5).
		\end{align}
	\end{subequations}
	The components of $\tilde{\mathbf{B}}_r(p,c)$ can be calculated in the same way.\\
	\begin{subequations}
		\begin{align}
			&\bar{\mathbf{B}}_r(p,c;r=0)[0]=3z\mathscr{H}_0(;0,5)-15z\mathscr{H}_2(;2,7),\\
			&\bar{\mathbf{B}}_r(p,c;r=0)[1]=\frac{3z}{c}\mathscr{H}_0(;,4)-\frac{15z}{c}\mathscr{H}_2(;2,6),\\
			&\bar{\mathbf{B}}_r(p,c;r=0)[2]=\frac{z}{c^2}\mathscr{H}_0(;0,3)-\frac{6z}{c^2}\mathscr{H}_2(;2,5),\\
			&\bar{\mathbf{B}}_r(p,c;r=0)[3]=-\frac{z}{c^3}\mathscr{H}_2(;2,4).
		\end{align}
	\end{subequations}
	\newpage
	\bibliographystyle{plain}
	\bibliography{ref}
\end{document}